\documentclass[fleqn,usenatbib]{mnras}

\usepackage{newtxtext,newtxmath}

\usepackage[T1]{fontenc}

\DeclareRobustCommand{\VAN}[3]{#2}
\let\VANthebibliography\thebibliography
\def\thebibliography{\DeclareRobustCommand{\VAN}[3]{##3}\VANthebibliography}

\usepackage{graphicx}	
\usepackage{amsmath}	
\usepackage{tabularx}
\usepackage{afterpage}
\usepackage{pdflscape}   
\usepackage{booktabs}    
\usepackage{caption}
\usepackage{subcaption}
\usepackage{rotating}

\newcommand{\Teff}{\mbox{$T_{\mathrm{eff}}$}}
\newcommand{\logg}{\mbox{$\log{g}$}}
\newcommand{\htohe}{\mbox{$\log(\mathrm{H/He})$}}

\newcommand{\obgt}{\mbox{O$_{\mathrm{bgt}}$}}
\newcommand{\ztohe}[1]{$\log(\mathrm{#1/He})$}
\newcommand{\ztox}[1]{$\log(\mathrm{#1/H[e]})$}

\newcommand{\Lines}[3]{\Ion{#1}{#2}\,#3\,\AA}
\newcommand{\Ion}[2]{#1{\,\sc#2}}

\newcommand{\Msun}{\mbox{$\mathrm{M}_{\odot}$}}

\title[Heavily metal-rich white dwarfs in DESI DR1]{First planetesimals from DESI DR1: 12 highly metal-rich white dwarfs}

\author[Paula Izquierdo]{
Paula Izquierdo,$^{1}$\thanks{E-mail: Paula.Izquierdo-Sanchez@warwick.ac.uk}
Andrew Swan,$^{1}$
Boris T. G\"ansicke,$^{1}$
Jamie T. Williams,$^{1}$
Detlev Koester,$^{2}$
\newauthor
Nicola P. Gentile-Fusillo,$^{3}$
Christopher J. Manser,$^{1}$
Laura K. Rogers,$^{4}$
D.~Aguado,$^{5}$
J.~Aguilar,$^{6}$
S.~Ahlen,$^{7}$
\newauthor
C.~Allende Prieto,$^{5,8}$
D.~Bianchi,$^{9,10}$
D.~Brooks,$^{11}$
T.~Claybaugh,$^{5}$
A.~de la Macorra,$^{12}$
A.~ Dey,$^{13}$
P.~Doel,$^{11}$
\newauthor
J.~E.~Forero-Romero,$^{14,15}$
E.~Gaztañaga,$^{16,17}$
S.~Gontcho A Gontcho,$^{5}$
G.~Guti\'errez,$^{18}$
D.~Joyce,$^{13}$
T.~Kisner,$^{5}$
\newauthor
S.E.~Koposov,$^{19,20,21}$
A.~Kremin,$^{5}$
M.~Landriau,$^{5}$
L.~Le~Guillou,$^{22}$
T.~S.~Li,$^{23}$
M.~Manera,$^{24}$
A.~Meisner,$^{13}$
\newauthor
R.~Miquel,$^{24,25}$
J.~Moustakas,$^{26}$
J.~Najita,$^{13}$
W.~J.~Percival,$^{27,28,29}$
F.~Prada,$^{30}$
I.~P\'erez-R\`afols,$^{31}$
G.~Rossi,$^{32}$
\newauthor
E.~Sanchez,$^{33}$
D.~Schlegel,$^{5}$
M.~Schubnell,$^{34}$
D.~Sprayberry,$^{13}$,
G.~Tarl\'{e},$^{34}$
B.~A.~Weaver,$^{13}$
R.~Zhou,$^{5}$
H.~Zou,$^{35}$
}

\date{Accepted 2026 July 14. Received 2026 July 09; in original form 2026 April 16}

\pubyear{\the\year{2026}}

\begin{document}
\label{firstpage}
\pagerange{\pageref{firstpage}--\pageref{lastpage}}
\maketitle

\begin{abstract}
Metal-enriched white dwarfs provide a unique insight into the composition of exoplanet interiors. These stars accrete the debris of disrupted planetary bodies, and hence, measuring the stellar parameters and photospheric abundances yields the bulk compositions of the parent bodies. At present, over 1750 debris-accreting white dwarfs are known, but just a few dozen are sufficiently enriched to allow a detailed abundance study. Here we report the analysis of 12 highly metal-enriched white dwarfs observed within the Data Release~1 of the Dark Energy Spectroscopic Instrument (DESI). We characterised their stellar parameters and photospheric metal abundances and we identified between three and ten different elements in their optical spectra, including most of the rock-forming species: O, Mg, Si, Ca and Fe. We conclude that the accreted bodies broadly resemble compositions found within the inner Solar System such as primitive meteorites, processed material or planetary cores. Six of the systems allowed a more thorough analysis: four of the parent bodies are composed of dry rock-forming elements; and two of them of something akin to a water-rich planetesimal. Thus, this study establishes DESI as a potent survey for identifying metal-rich targets, yielding reliable compositions of accreted exoplanetary material.
\end{abstract}





\section{Introduction}

The field of exoplanets keeps growing and shedding light onto the diversity of other worlds, but measuring the composition of exoplanets remains a challenging quest. Conventional studies of planetary systems around main-sequence stars involve the use of radial velocity and transit photometry to derive planetary bulk densities \citep{charbonneau00,mandel02,seager10}, which are then compared to interior models to infer likely compositions \citep[see e.g.][]{fortney07, dorn17}. This technique has several shortcomings \citep{zeng08}, but the main concern is that it often yields degenerate solutions, and thus cannot uniquely identify bulk compositions. 

Nonetheless, the composition of exoplanets can be inferred from the study of white dwarfs. These are the end product of main-sequence stars with masses less than $8-10\,\Msun$, which in this stage are sustained against gravity by the pressure of degenerate electron gas. White dwarfs are typically Earth-sized with masses of $\simeq0.6\,\Msun$, with correspondingly high surface gravities ($\simeq10^{8}\mathrm{cm\,s^{-2}}$). As a result, elements heavier than helium sink out of the outer convective layers on relatively short timescales \citep[$\lesssim10^{6}$ years;][]{koester09-1, cunningham19}. As a consequence, white dwarfs are expected to have pristine hydrogen or helium spectra.

However, observations show that between $20-50$ per cent of white dwarfs are metal-enriched \citep{Zuckerman03,zuckerman10-1,Koester14,manser24,ouldrouis24}. The origin of these metals is attributed to accreted planetary bodies, as evidenced by manifold observational signatures: white dwarfs displaying rapid-evolving transits from disintegrating bodies \citep[see e.g.][]{vanderburg15,gansicke16-1,guidry21,aungwerojwit24}, detection of dust and/or gas debris discs \citep[see e.g.][]{gaensicke06-1,farihi09,gentile21b,guidry24}, X-ray emission from the accretion of such material \citep{cunningham22}, and the chemical composition of the accreted material being consistent with bulk-Earth/primordial chondrites or volatile-rich bodies \citep[see e.g.][]{klein10,dufour10,gaensicke12,farihietal13-1,rogers24b,sneha25}, akin to the chemical composition of Solar system bodies. The detailed analysis of the spectra of white dwarfs accreting planetary debris allows to quantitatively measure the photospheric metal abundances, which in turn are a proxy for the composition of the parent body. Currently, there are over 1750 known metal-enriched white dwarfs, of which 1480 only show Ca \citep{williams24}, reflecting the extreme strength of the  Ca H and K resonance lines rather than anomalously large Ca abundances. In contrast, only 86 systems exhibit more than five metals, allowing a more detailed study of the composition of the accreted bodies \citep[see e.g.][]{xu16-1,hoskin20,johnson22,lebourdais24}. Elements of particular interest are those that trace the core (Fe, Ni), mantle (Mg, Si) and crust (Na, Ca, Al, Ti) fractions of the accreted planetary bodies, as well as their volatile and water content (C, N, O, S).

The Dark Energy Spectroscopic Instrument (DESI) aims to constrain the effect of dark energy in the expansion of the Universe by acquiring spectra of over 100 million objects \citep{desi16b,DESI_must3,DESI_must4,schafly23_desi, desi_2025_ii}. As part of the DESI Data Release~1 \citep[DR1;][]{DESI_Y1, desi_2024_vii}, about 64\,000 white dwarf candidates were observed, which will be presented as the DESI DR1 white dwarf catalogue (Swan et al. in prep.). So far, it contains about $1700$ white dwarfs displaying metal lines (below 25\,000\,K)~--~81 per cent of which are new identifications based on the updated database of metal-enriched white dwarfs \citep{williams24}.

Here we present a pilot study of 12 of those white dwarfs selected based on their large number of metal lines in their spectra, spanning a range of effective temperatures and atmospheric compositions that is representative of the DESI DR1 white dwarf sample. We present the spectroscopic and photometric data in Section~\ref{sec:sample}, and describe the methodology used to model the stellar parameters and chemical compositions in Section~\ref{sec:stellar_mod}. We present the results and discuss the photospheric parameters of all white dwarfs and the exo-planetesimal compositions of the warmer ones in Section~\ref{sec:results}, along with a detailed explanation of the likely accretion of water-rich material and alternative accretion scenarios. Finally, the conclusions are presented in Section~\ref{sec:conclusions}. Additional material follows in Appendices~\ref{sec:appendix_phot} and \ref{sec:appendix_bestfits}, which encompasses the employed archival photometric data and all the best fits to the spectroscopic data.

\begin{table*}
\caption{White dwarf sample, including the WD\,J names from \citet{gentile19}, the short names used in this paper, the \textit{Gaia} $G$-band magnitude, the distance $D$ \citep[derived as $D$~(pc)~$=1000/\varpi$, where $\varpi$ is the parallax in mas;][]{gaia23}, the spectral classification from the DESI DR1 white dwarf catalogue (Swan et al. in prep.), the log of the X-Shooter spectroscopy and the signal-to-noise (S/N) ratios of the UVB and VIS X-shooter, and the number of DESI blue, red and near-infrared arms spectra (the last five columns). }
\label{tab:obslog}
\centering
\begin{tabular}{lccccccccccc}
\hline
Star & Short name  & \textit{Gaia G} & $D$ & Spectral & \multicolumn{4}{c}{X-shooter observations}  & \multicolumn{3}{c}{DESI} \\
     &  &  & (pc) & class & Date & Exposure time (s) & UVB & VIS & B & R & Z \\ 
\hline
WD\,J024258.64+042653.15    & 0242$+$0426 & 18.4 & $226 \pm 11$  & DBAZ &               &  &  &  & 23 & 26 & 19 \\
WD\,J025527.70+023714.08	& 0255$+$0237 & 17.2 & $68 \pm 1$    & DAZ  & 16/09/2025    & 4$\times$700/4$\times$650 & 69 & 47 & 34 & 37 & 34 \\
WD\,J045235.51-021446.07	& 0452$-$0214 & 17.8 & $141 \pm 2$   & DZAB &               &  &  &  & 29 & 26 & 27 \\
WD\,J085035.17+320804.29	& 0850$+$3208 & 17.2 & $115 \pm 1$   & DBAZ & 22--23/04/2025 & 6$\times$700/6$\times$650 & 56 & 19 & 50 & 46 & 43 \\
WD\,J092256.07+010310.15	& 0922$+$0103 & 16.5 & $32 \pm 1$    & DAZ  & 23/04/2025    & 2$\times$400/2$\times$350  & 43 & 39 & 46 & 55 & 54 \\
WD\,J125250.17+735216.84	& 1252$+$7352 & 18.8 & $148 \pm 3$   & DAZ  &               &  &  &  & 17 & 17 & 14 \\
WD\,J133305.34+325400.11    & 1333$+$3254 & 19.2 & $302 \pm 18$  & DBAZ & 22--23/04/2025 & 4$\times$1800/4$\times$1750 & 24 & 9 & 22 & 19 & 14\\
WD\,J133658.08$-$033733.08  & 1336$-$0337 & 19.0 & $87 \pm 1$    & DAZ  & 22--23/04/2025 & 4$\times$1800/4$\times$1750 & 33 & 28 & 25 & 35 & 35\\
WD\,J135217.76+032323.92	& 1352$+$0323 & 18.5 & $237 \pm 14$  & DZBA  & 22--23/04/2025 & 4$\times$1800/4$\times$1750 & 18 & 15 & 18 & 16 & 11 \\
WD\,J162646.91+313628.00	& 1626$+$3136 & 19.4 & $294 \pm 20$  & DBAZ &               &  &  &  & 24 & 22 & 18 \\
WD\,J175646.49+381616.30	& 1756$+$3816 & 19.0 & $170 \pm 6$   & DAZ  &               &  &  &  & 11 & 14 & 13 \\
WD\,J221414.67+092325.32	& 2214$+$0923 & 19.4 & $113 \pm 4$   & DZ   & 17--18/09/2025 & 4$\times$1800/4$\times$1750 & 5 & 21 & 5 & 7 & 7\\\hline
\label{tab:wd_observations}
\end{tabular}
\end{table*}

\newcommand{\mc}[1]{\multicolumn{2}{c}{#1}}

\section{White dwarf sample and data}
\label{sec:sample}

\subsection{Sample selection}

Based on our visual inspection of the $\simeq64\,000$ white dwarf candidates within DESI DR1 (Swan et al. in prep), we selected 12 strongly metal-enriched white dwarfs for the analysis presented here: nine of them are classified as D[BAZ]\footnote{He-dominated white dwarfs. The lower opacity of He compared to H results in a deeper view into their atmospheres, resulting in stronger metal lines in their spectra. Correspondingly, D[BAZ] often allow for a more detailed chemical analysis of the accreted parent body than metal-rich H-dominated white dwarfs, DAZ.} (B: He I transitions, A: Balmer lines, Z: metals, the sequence of the letters reflects the relative strength of these features); two as DAZ, and one as a DZ\footnote{White dwarfs showing only metal features in their spectra have typically cool He-dominated atmospheres.}. Details of these 12 systems are presented in Table~\ref{tab:wd_observations}.

\subsection{DESI DR1 spectroscopic data}

\begin{figure*}
        \includegraphics[width=0.9\hsize]{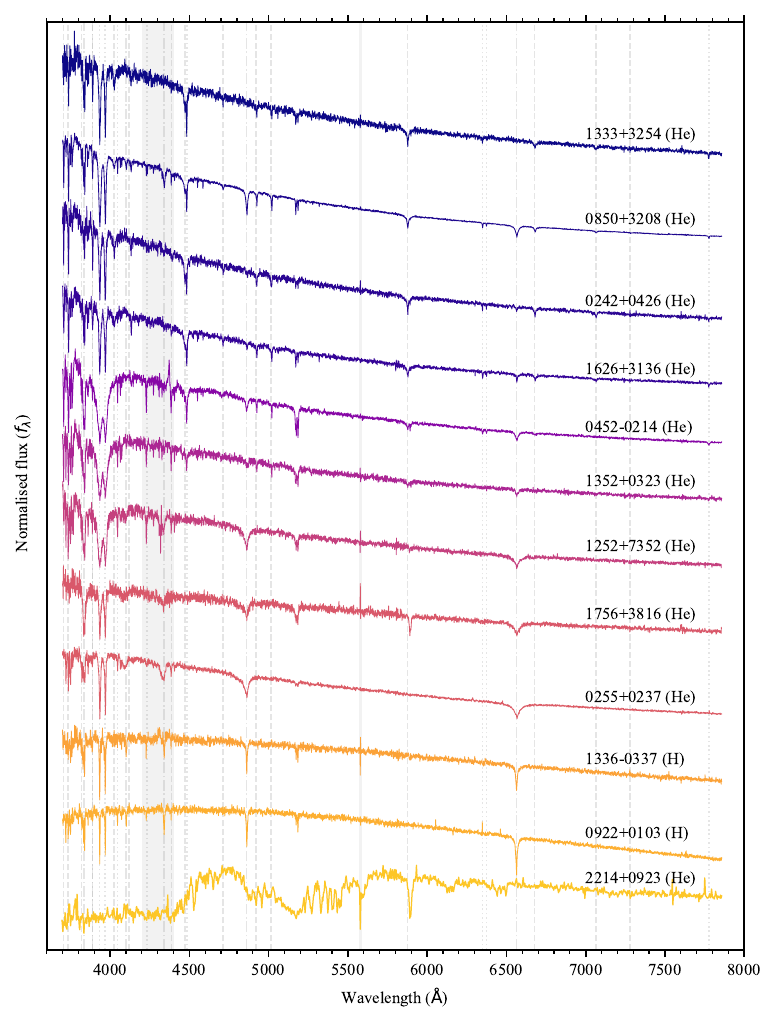}
    \caption{DESI DR1 coadded spectra of the 12 metal-enriched white dwarfs analysed, with the main photospheric element displayed in brackets. He, H and metal absorption lines are marked with dashed, dot-dashed and dotted grey vertical lines, respectively. The grey vertical region between 4200\,\AA\ < $\lambda$ < 4400\,\AA\ highlights where the spectrograph has calibration residuals \citep[see e.g.][]{manser24b}, while the region around
    $\lambda\simeq5580$\,\AA\ identifies where coadding between the blue and red arms causes artefacts. The effective temperature decreases from top to bottom. The spectra are offset vertically for display purposes.}
    \label{fig:DESIDR1_norm_spec}
\end{figure*}

DESI spectra are obtained with a three-arm spectrograph (blue, red, near-infrared) with a mean spectral resolution of {$\Delta\lambda\simeq1.8$\,\AA} at full-width of half maximum \citep{DESI_must2, guy23_desi}.  Several of the white dwarfs in our sample were observed more than once and Fig.~\ref{fig:DESIDR1_norm_spec} displays the uncertainty-weighted normalised co-adds using all the available individual exposures for each star. These spectra feature a wide mixture of H, He and metal lines. The average S/N ratios for the co-added DESI spectra in this sample are 25, 27 and 24, for the blue, red and near-infrared arms, respectively.

 \begin{figure*}
        \includegraphics[width=0.99\hsize]{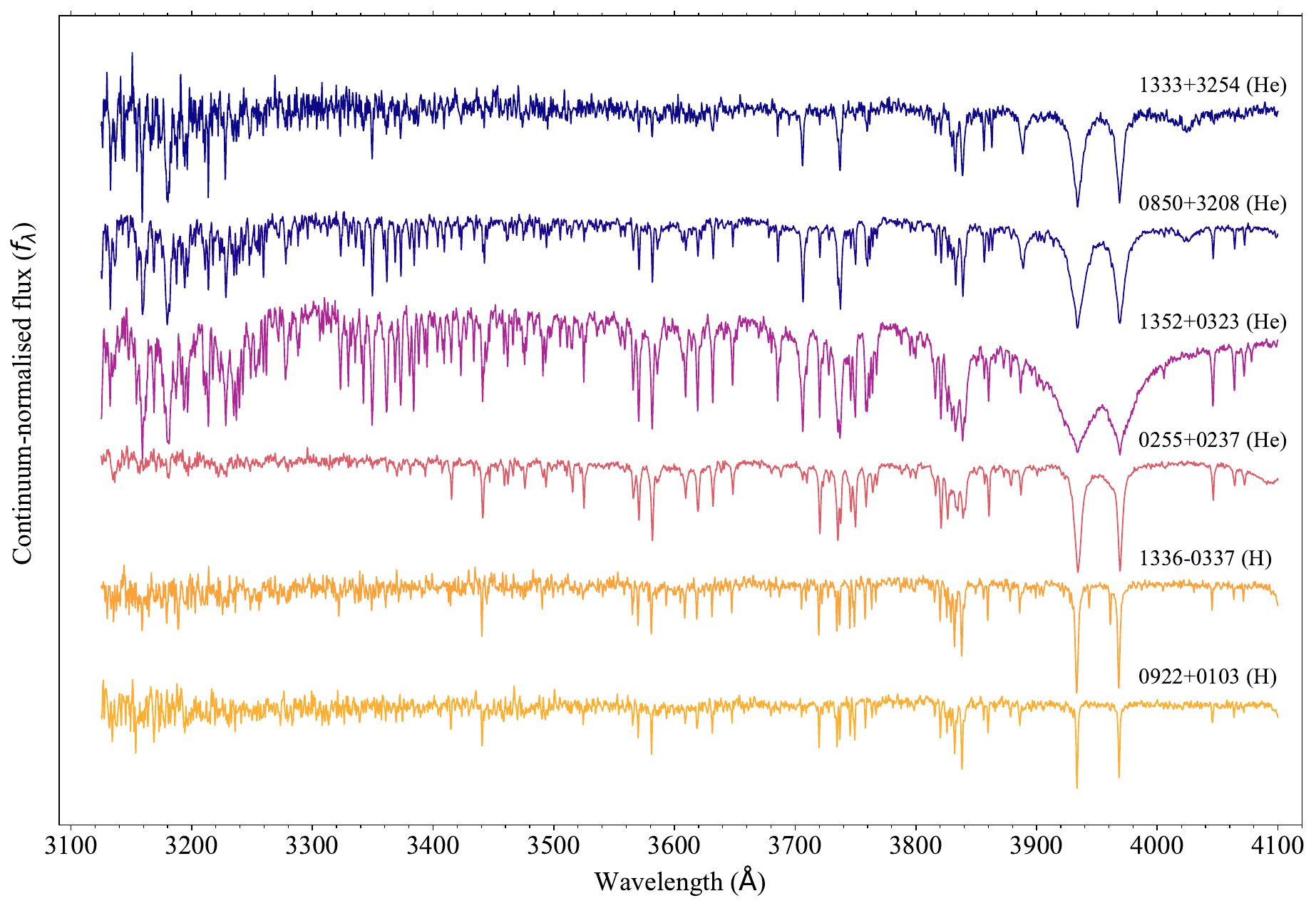}
    \caption{Bluest part of the continuum-normalised weighted average spectra of six metal-enriched white dwarfs obtained with X-shooter, with the main photospheric element given in brackets. Details of the observations can be found in Section~\ref{sec:xs_obs} and Table~\ref{tab:wd_observations}.}
    \label{fig:XS_norm_spec}
\end{figure*}

 \begin{figure*}
        \includegraphics[width=0.95\hsize]{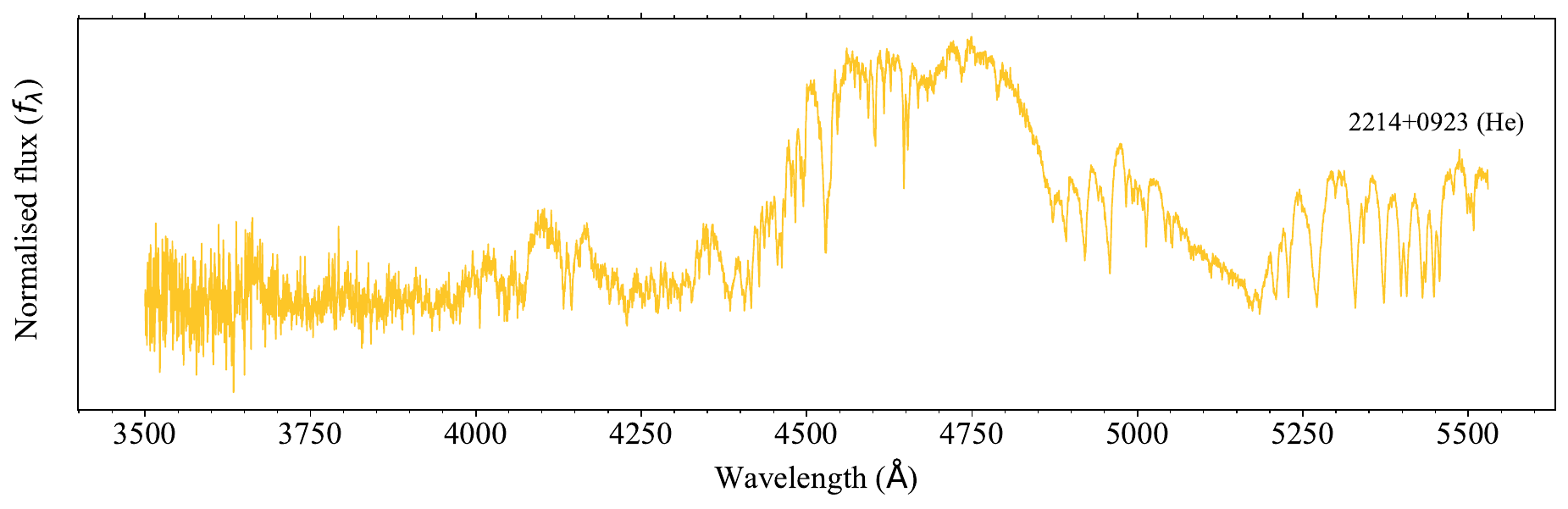}
    \caption{Normalised weighted average X-Shooter UVB-arm spectra of 2214+0923. Details of the observations can be found in Section~\ref{sec:xs_obs} and Table~\ref{tab:wd_observations}.}
    \label{fig:XS_norm_spec_J22}
\end{figure*}

\subsection{X-shooter spectroscopic follow-up}
\label{sec:xs_obs}

We secured additional intermediate resolution data for 0255+0237, 0850+3208, 0922+0103, 1333+3254, 1336$-$0337, 1352+0323 and 2214+0923 using the X-shooter spectrograph \citep{vernetetal11-1}, mounted on the UT3 Kueyen telescope of the 8.2-m Very Large Telescope at Cerro Paranal (Chile). The spectra were obtained using slit widths of 1.0, 0.9, and 0.9\,arcsec in the UVB, VIS, NIR arms, respectively, with resolving power $R=\lambda/\Delta\lambda$ of 5400, 8900, and 5600 in the UVB, VIS and NIR arms, respectively. The X-shooter data have a higher spectral resolution compared to DESI and additional coverage in the near-ultraviolet (UV) wavelength range 3100--3600\,\AA. This provides access to absorption lines of Ti, Cr, Mn, and Ni that have no strong features in the spectral range covered by the DESI data. The 3100--4100\,\AA\ range of the X-Shooter spectra are shown in Figs.~\ref{fig:XS_norm_spec} and \ref{fig:XS_norm_spec_J22}.

\subsection{Astrometric and photometric data}
\label{sec:phot_data}

We made use of archival photometry from SDSS \citep{SDSSpass}, Pan-STARRS \citep{PanSTARRS2}  and \textit{Gaia} DR3 \citep{gaia-edr3} for the photometric analysis of the white dwarfs. We also supplemented the photometric data with \textit{Gaia} DR3 parallaxes \citep{gaia-edr3}, as well as with extinction information from two independent all-sky maps \citep{lallement22, vergely22, edenhofer24}. The photometric data are listed in Table~\ref{tab:phot_points} and detailed information about the analysis can be found in Section~\ref{sec:photandspec}.

Additionally, we inspected the infrared (IR) photometry of the sources to look for possible surrounding disks of dust or debris, checking the archival available photometry from the Two Micron All‑Sky Survey \citep[2MASS;][]{skrutskieetal06-1}, the UKIRT Infrared Deep Sky Survey \citep[UKIDSS;][]{lawrence07}, and the Wide‑field Infrared Survey Explorer \citep[WISE;][]{wright10}.

\section{Stellar modelling}
\label{sec:stellar_mod}

\subsection{Model atmospheres}
\label{sec:model_atm}

The synthetic spectra used to model the spectroscopic and photometric data were computed with the latest version of the \cite{koester10-1} code. Convection zones (CVZ) of white dwarfs were treated with a 1D mixing-length (ML) prescription, using the ML2 parametrization and mean constant convective efficiency $\alpha=0.8$ \citep{tremblay13b,cukanovaiteetal19-1}.

We computed a generic grid of He+H+Z atmosphere models for the analysis of the D[BAZ] stars, spanning effective temperatures $\Teff = 7000$--10\,000\,K in 500\,K steps and $\Teff = 10\,000$--17\,000\,K in steps of 1000\,K; surface gravities $\logg = 7.5$--9.5 in 0.5\,dex steps, and H abundances $\htohe$ from $-8.0$ to $-1.0$ in steps of 1\,dex. We included the major constituents seen in CI chondrites, which are representative of primitive, unprocessed material with a similar composition to the Solar protoplanetary disc: N, O, Na, Mg, Al, Si, P, S, Cl, Ca, K, Sc, Ti, V, Cr, Mn, Fe, Co, Ni, Cu and Zn. The relative abundances of these metals were fixed to CI chondrite values with respect to Si \citep{lodders03-1}, and the overall metal content spanned \ztohe{Si} from $-10.8$ to $-5.3$ in steps of $0.5$\,dex. 

For the analysis of the DAZ stars ($1336-0337$ and $0922+0103$), we computed a general grid of H+Z atmospheres, spanning $\Teff = 5250$--7250\,K in steps of 250\,K and $\logg = 7.25$--8.75 in 0.25\,dex steps. The same metals as in the He+H+Z grid were included, with abundances fixed to CI chondritic ratios relative to Si, and the overall metal content varied $\log(\rm{Si/H})$ from $-9.4$ to $-3.4$ in steps of $0.5$\,dex. 

We generated a custom DZ grid to analyse $2214+0923$ using unified broadening profiles for \Lines{Na}{i}{5891.58, 5897.56}, \Lines{Mg}{i}{5168.76, 5174.13, 5185.05}, \Lines{Mg}{ii}{2796.352, 2803.531}, \Lines{Ca}{i}{4227.92} and \Lines{Ca}{ii}{3934.78, 3969.60}, adequate for the high density in the cool He atmosphere of this star. We used the preliminary \Teff\ and \logg\ published by \cite{gentile21} as starting values ($\Teff = 5030 \pm 300$\,K, and $\logg = 7.48 \pm 0.28$), computing a grid which spanned $\Teff = 5000$--6000\,K in steps of 250\,K and $\logg = 7.5$--8.0 in 0.25\,dex steps. The H abundance was fixed to $\htohe = -5.0$, as this is the lower boundary above which trace H starts to get detected \citep[mixed models are commonly adopted for cool He-dominated photospheres,][]{bergeron19}. We included the same metals as in the previous grids, with abundances fixed to CI chondrite ratios relative to Si, and the overall metal content spanning \ztohe{Si} from $-7.0$ to $-5.4$ in 0.25\,dex steps.  

These three generic grids were used as starting points to measure preliminary values of the photospheric parameters of the 12 white dwarfs. The spectroscopic analysis is described in detail in the following section and involves the computation of custom atmosphere grids for each star, adjusting for the specific metal content that each photosphere displays, once initial metal abundances are obtained.

\subsection{Photometric and spectroscopic fitting}
\label{sec:photandspec}
We measured the effective temperature $\Teff$ and surface gravity $\logg$ by modelling the archival photometry and parallaxes (Table~\ref{tab:phot_points}), and the photospheric compositions by fitting the DESI and X-shooter spectra. 

For the photometric analysis we followed the methodology described in Appendix B2 of \cite{manser24} with a few modifications. The archival photometry presented in Table~\ref{tab:phot_points} was modelled, fitting for \Teff, \logg, the distance $D$ and the extinction $E(B-V)$. Model fluxes were scaled by the star solid angle $\pi (R_{\mathrm{WD}}/D)^2$, with the radius of the white dwarf $R_{\mathrm{WD}}$ inferred from evolutionary models of \cite{althaus13}.  We supplemented the photometric data with the \textit{Gaia} DR3 parallax and reddened the synthetic data assuming an extinction ratio of $R_V = 3.1$.

The general spectroscopic fitting technique followed the same procedure as that presented in \citet{manser24}. We obtained an estimate of the overall metal content, $\log(\mathrm{Si/H[e]})$, by fitting the whole spectrum. We used windows 200--220\,\AA\ wide, normalising both the model and observed spectra within each window, and performed a joint fit of all the windows. For the D[BAZ] white dwarfs, we included an additional free parameter, \htohe, which we also obtained within this preliminary fit.

Once we had obtained a first estimation of \Teff, \logg, overall $\log(\mathrm{Si/H[e]})$ (and \htohe\  for the He-dominated white dwarfs), we proceeded with a more detailed analysis of the spectroscopic data (see the following section for a step-by-step explanation). In this new step (Stage 4 in Section~\ref{sec:iter}), we refined the estimation of the metal abundances. To do that, we analysed each metal individually, jointly fitting all its metal transitions within small windows of 10--60\,\AA\ wide (depending on the intrinsic width of the metal line). These smaller windows were normalised in the same aforementioned fashion. The metal lines fitted in this analysis are listed in Table~\ref{tab:metal_lines}.

The cool DZ 2214+0923 lacks a spectral continuum because of the numerous, strong, and extremely wide metal lines, and therefore requires a different approach \citep{hollands17}. For this star, we normalised the whole useful spectrum to preserve the overall shape, employing just one window. We normalised this window by dividing each data point by the mean flux, both in the data and the models, and then made a preliminary estimate of the metal content. As already mentioned in Section~\ref{sec:model_atm}, the H abundance was fixed to $\htohe = -5.0$, as this is the lower boundary above which trace hydrogen starts to get detected via H$\alpha$. For this system, the estimation of the abundances of individual metals was also done by fitting the whole spectrum. 

\subsection{Iterative analysis}
\label{sec:iter}

\begin{figure*}
        \includegraphics[width=0.95\textwidth]{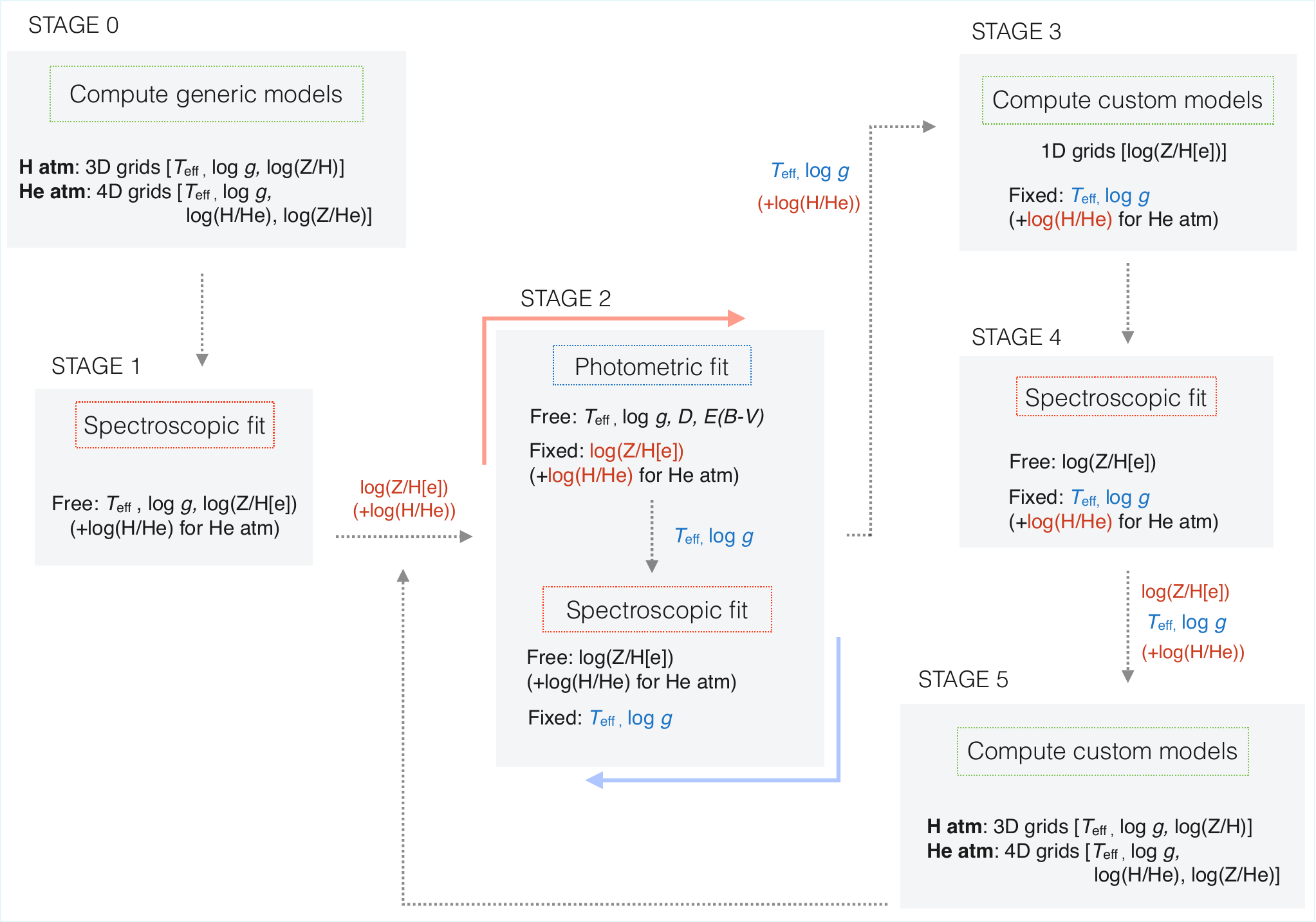}
    \caption{Flow chart of the iterative process used to measure the photospheric parameters of the white dwarfs. Parameters determined from a photometric (spectroscopic) fit are colour-coded in blue (red). Note the coloured arrows surrounding Stage 2 to mark an iterative stage in itself, where the spectroscopic values are fed to the photometric fitting, and vice versa, until convergence is achieved in this step (see the main text for details).}
    \label{fig:flow_chart}
\end{figure*}

We iterated between photometric and spectroscopic fittings to measure the \Teff, \logg, and chemical composition, generating fresh custom grids of synthetic spectra at each stage of the process. This whole workflow is illustrated in Fig.~\ref{fig:flow_chart}, and is divided in six steps: 

\begin{itemize}
\item \textbf{Stage 0}: We computed a generic grid of models with different \Teff, \logg, \ztox{Z} (and \htohe\  for He-dominated white dwarfs; see Section~\ref{sec:model_atm} for more details). 

\item \textbf{Stage 1}: The spectroscopic data were modelled, with up to four free parameters, \Teff, \logg, \ztox{Si}\ (and \htohe\ for He-dominated white dwarfs), to obtain an initial guess of their values. The inferred abundances were then fed to the following stage. 

\item \textbf{Stage 2}: This was an iterative step in itself. We first modelled the photometry, keeping the abundances (obtained from the previous step), which yielded \Teff\ and \logg. These parameters were then fed to a spectroscopic fit where \Teff\ and \logg\ were fixed to refine the chemical composition. This refined chemical composition was then fixed to model the photometric data again. This process wes repeated until convergence was achieved. 

\item \textbf{Stage 3}: We fixed the preliminary best-fitting values of \Teff, \logg\ (and \htohe\ for He-dominated photospheres) from Stage~2, and computed custom model grids for each star. Specifically, we generated a 1D grid for each element detected in the white dwarf under analysis, in which only the abundance of that element was varied, while the abundances of all other identified metals were fixed to the abundances from the previous step (CI chondrite ratios in the first iteration). For example, we computed seven 1D grids for $1333+3254$, one for each metal identified in their DESI and X-Shooter spectra (O, Mg, Si, Ca, Ti, Cr, Fe). 

\item \textbf{Stage 4}: We refined the abundances of all detected elements using the 1D grids computed in Stage~3. For the D[BA]Z and DAZ stars, in this step we just fitted the metal lines of one element at a time. Because of the strong line blending in the DZ 2214+0923 we fitted the whole spectrum. 

\item \textbf{Stage 5}: The best-fit set of \Teff, \logg, \ztox{Z} (and \htohe\ for He-dominated photospheres) was used to produce small custom grids in which the three (four for the D[BA]Z white dwarfs) parameters varied. We call these 3D and 4D model grids. Note that \ztox{Z} was updated to the specific abundances measured in each white dwarf, and hence this parameter encompassed all the abundances of all specific metals for that star in a single parameter. These custom grids typically spanned a few 100\,K, $\pm0.2$\,dex, and $\pm 1$\,dex around the best-fit values for \Teff, \logg, and \ztox{Z} obtained in Stage~4. For the D[BA]Z stars, \htohe\ typically spanned $\pm 0.5$\,dex around the best-fit value obtained in Stage~2.

\item \textbf{Go back to Stage 2}: Using these custom 3D / 4D grids the spectral analysis was repeated from Stage~2 onwards to fully account for the effect of the metals on the structure of the atmosphere. We repeated this procedure until the changes in \Teff, \logg, \ztox{Z} (and \htohe\ for He-dominated photospheres) from each iteration were within the statistical uncertainties provided by the photometric or spectroscopic fits.

\end{itemize}

We note that the final models contain the specific metals identified in the spectra of each white dwarf. However, we also included Si at CI chondrites abundances relative to Mg for the white dwarfs with untraceable Si: 0255+0237, 1336$-$0337, 0922+0103 and 2214+0923. This is standard practice for cool white dwarfs \citep[see e.g.][]{hollands17} since Si is one of the most common rock-forming element, has a low ionisation energy that adds free electrons contributing to the overall opacity, and its optical lines disappear below $7000$\, and $8000$\,K for H and He-dominated white dwarfs, respectively.

\begin{table}
\small
\caption{Spectral lines used in the determination of the metal chemical abundances.}
\vspace{0.2cm}
\begin{tabularx}{0.48\textwidth}{ll}
\hline\noalign{\smallskip}
Ion & Air wavelength (\AA)\\
\noalign{\smallskip}\hline
\noalign{\smallskip}
O\,{\sc i} & 7771.94, 7774.17, 7775.39, 8446.76\\
Na\,{\sc i} & 5889.95 , 5895.92\\
Al\,{\sc i} & 3944.01, 3961.52\\
Mg\,{\sc i} & 3829.36, 3832.30, 3838.29, 5172.68, 5183.60\\
Mg\,{\sc ii} & 4481.33, 7877.05, 7896.36\\
Si\,{\sc i} & 3905.52\\
Si\,{\sc ii} & 3853.66, 3856.02, 3862.60, 4128.07, 4130.89, 5055.98,\\ 
             & 6347.1, 6371.4\\
Ca\,{\sc i} & 4226.73 \\
Ca\,{\sc ii} & 3158.86, 3179.33, 3181.28, 3706.53, 3736.90, 3933.66,\\
 & 3968.47, 8498.02, 8542.09, 8662.14\\
Ti\,{\sc ii} & 3154.19, 3161.77, 3162.57, 3168.52, 3190.87, 3202.53,\\
& 3218.26, 3224.24, 3228.60, 3229.42, 3232.28, 3239.66,\\
& 3248.60, 3252.94, 3261.61, 3271.65, 3278.92, 3287.65,\\
& 3321.70, 3322.94, 3341.87, 3349.03, 3444.31, 3461.50,\\
& 3477.18, 3491.05, 3504.89, 3510.84, 3535.41, 3685.20,\\
& 3741.64, 3759.29, 3761.32, 3900.54, 4468.51, 4549.62 \\
Cr\,{\sc ii} & 3124.97, 3128.69, 3132.05, 3136.68, 3147.22, 3180.69,\\
& 3197.08, 3209.18, 3217.39, 3339.79, 3342.58, 3358.49,\\ 
& 3360.29, 3368.04, 3403.31, 3408.77, 3421.21, 3422.74,\\  
& 3433.29\\
Mn\,{\sc ii} & 3441.98, 3460.31, 3474.04, 3474.13, 3482.90 \\
Fe\,{\sc i} & 3190.82, 3249.50, 3440.99, 3490.57, 3565.38, 3570.10,\\  
            &  3581.19, 3608.86, 3618.77, 3631.46, 3719.93, 3734.86,\\
            & 3749.49, 3758.23, 3763.79, 3815.84, 3820.43, 3825.88,\\
            & 3827.82, 3834.22, 3840.44, 3856.37, 3859.91, 4383.54,\\ 
Fe\,{\sc ii} & 3105.17, 3114.32, 3118.29, 3131.10, 3154.20, 3167.86,\\
             & 3170.34, 3177.54, 3183.11, 3186.74, 3192.91, 3193.80,\\
             & 3196.07, 3210.45, 3213.31, 3227.74, 3247.18, 3255.87,\\
             & 3258.77, 3259.05, 3277.35, 3493.54, 4233.16, 4583.83,\\ 
             & 4923.92, 5018.44, 5169.03, 5275.99\\
Ni\,{\sc i} &  3414.76, 3524.54, 3619.39 \\
Ni\,{\sc ii} &  3513.99\\
\noalign{\smallskip}\hline
\end{tabularx}
\label{tab:metal_lines}
\end{table}

\subsection{Sampling technique}

We modelled both the spectroscopic and photometric data with the \texttt{pocoMC} package within \textsc{python} \citep{karamanis22-1,karamanis22-2}. This is a Bayesian inference algorithm designed to sample complex multimodal posterior distributions. We assumed a Gaussian likelihood function and employed 4096 effective and 2048 active particles, respectively, with a final 4096 required samples and 2048 for the evidence. 

Uniform priors for \Teff, \logg, \htohe\ and $\log({\mathrm{Z/H[e]}})$ were used. In order to account for uncertain parallaxes, an exponentially-decreasing prior was used for the distance, which is calibrated from simulated \textit{Gaia} data of a white dwarf population \citep{bailerjones18,rybizki20}. We also constructed a reddening prior from 3D extinction maps \citep{lallement22, vergely22, edenhofer24}, assuming $A_V = 3.1\times E(B-V)$, and using a Student's~$t$ distribution to accommodate local variations in the extinction that are not well-resolved in the maps.

\subsection{Inferring the composition of the accreted material}
\label{sec:phot_to_pb}

The outcome of the iterative analysis is the characterisation of the stellar parameters \Teff\ and \logg, and the photospheric abundances. We converted the photospheric abundances arising from the accretion of planetary debris into parent body compositions. Specifically, we used eq.(5) from  \cite{koester09-1}, evaluated at different times since accretion started/stopped, as is common practice \citep[see e.g.][]{izquierdo20}.

This technique assumes a simplistic scenario in which the mass transfer starts at some point, proceeds at a uniform rate, and then switches off when the mass reservoir is exhausted. In this scenario, a system is in one of the following  accretion states: \textit{increasing, steady} or \textit{decreasing}. In the increasing state, the material is accreted into a prior pristine CVZ and the photospheric abundances are approximately the same as in the parent body. In the steady state, the material is being accreted at the same rate as it sinks out of the photosphere. In this equilibrium between accretion and diffusion, the photospheric abundances are corrected by the sinking times of each metal and the parent body compositions are reliably inferred. Finally, in the decreasing state, accretion is no longer ongoing and the photospheric metal abundances decrease exponentially with time. As each element has a different diffusion time scale, an unambiguous interpretation of the parent body composition is not possible for systems in the decreasing state.

We probed for the state of accretion using a Bayesian model, with the methods described in full by \cite{swan23}. We modelled the accretion of Solar system objects onto each star at a constant rate, calculating the resulting photospheric abundances at some time after accretion ceased. Various meteorite samples were considered, as were mixtures of the core, mantle and crust of Earth or Mars. For the stars where O is detected, we used two versions of the models: one with an additional water component (`wet'), and one without (`dry'). For the He-atmosphere white dwarfs the intrinsic H/He abundance of the star \textit{before} the accretion episode was inferred as a nuisance parameter. While the result itself is uninformative, including that parameter in the analysis ensures that the quantity of H supplied by any accretion of water accretion does not exceed the observed H abundance.

We used Bayesian model averaging to infer the most likely phase of accretion and the water fraction in the accreted material, marginalising over all models to yield composition-agnostic posteriors for the duration of the accretion event. This procedure retains the underlying assumption that the exoplanetary material is broadly similar to Solar system objects, but otherwise avoids having to assume a specific composition. 

Additionally, we computed the minimum mass of the accreted body as the sum of the individual metal masses contained in the CVZ, which are proportional to the photospheric mass abundances $Z_{\mathrm{ph,m}}$:

\begin{equation}
\label{eq:m_z}
M_{z} = M_{\mathrm{CVZ}} \ Z_{\mathrm{ph,m}} .
\end{equation}

\section{Results and discussion}
\label{sec:results}

\subsection{Stellar parameters and photospheric metal abundances}
\label{sec:stellar_params}

Table~\ref{tab:stellar_params} presents the white dwarf stellar parameters~--~\Teff, \logg\ and \htohe~--~obtained by the joint modelling of photometric, astrometric, and spectroscopic data, employing detailed model spectra that accounted for the particularities of each white dwarf (see Section~\ref{sec:stellar_mod} for more details). The mass of the white dwarf ($M_{\mathrm{WD}}$) and cooling age were interpolated from the evolutionary models of \protect\cite{bedard20} and the mass of the convection zone $M_{\mathrm{CVZ}}$ was calculated by downward integration of the envelope equations until the Schwarzschild criterion was no longer fulfilled \citep{koester20}.

The quoted uncertainties are purely statistical and hence do not account for other systematic sources of uncertainty, such as the input microphysics in the computation of model spectra, different instruments with their individual limitations, and the use of particular spectral ranges and/or the S/N of the data. A detailed analysis of systematic uncertainties in the characterization of metal-enriched He-dominated white dwarfs was carried out by \citet{izquierdo23}, who found typical uncertainties of 500\,K, 0.05\,dex and 0.27\,dex for \Teff, \logg\ and \htohe, respectively. To date, we have not found an analogous analysis that estimates the uncertainties in the modelling of H-dominated metal-enriched white dwarfs.   

\begin{landscape}
\begin{table}
\centering
\caption{Stellar parameters of the 12 white dwarfs. $\Teff$ and $\logg$ measured from the modelling of photometric and astrometric data (Table~\ref{tab:phot_points}); $\htohe$ was obtained by modelling the H and He absorption lines in the DESI and X-shooter spectra; the mass of the white dwarf ($M_{\mathrm{WD}}$) and cooling age were interpolated from the evolutionary models of \protect\cite{bedard20}; $N_{\mathrm{Z}}$ is the total number of metals detected from the DESI and, if available, X-shooter spectra; and the mass of the convection zone $M_{\mathrm{CVZ}}$ was calculated by downward integration of the envelope equations until the Schwarzschild criterion was no longer fulfilled \citep{koester20}. The two white dwarfs marked with $^{\star}$ have H-dominated photospheres. Note the uncertainties are purely statistical.}
\label{tab:stellar_params}
\begin{tabular}{cccccccc}
\toprule
Star &  $\Teff$ (K) & $\logg$ &$\htohe$ & $M_{\mathrm{WD}}$ ($M_{\odot}$) & Cooling age (Gyr) & $N_{\mathrm{Z}}$ & $\log{M_{\mathrm{CVZ}} / M_{\mathrm{WD}}}$\\
\midrule
1333+3254 & 12830 $\pm$ 50 & 7.99 $\pm$ 0.01 &  $-6.3\pm0.25$ & $0.58 \pm 0.03$ & $0.33 \pm 0.04$ & 7 & $-4.99$ \\
0850+3208 & 12596 $\pm$ 120 & 8.15 $\pm$ 0.02 & $-3.98\pm0.01$ & $0.68 \pm 0.01$ & $0.46 \pm 0.01$ & 9 & $-5.39$\\
0242+0426 & 12290 $\pm$ 285 & 7.99 $\pm$ 0.06 & $-5.81\pm0.11$ & $0.58 \pm 0.03$ & $0.37 \pm 0.04$ & 6 & $-4.90$\\
1626+3136 & 12200 $\pm$ 230 &  8.14 $\pm$ 0.07 & $-4.91 \pm 0.04$  & $0.66 \pm 0.05$ & $0.50 \pm 0.06$ & 6 & $-5.31$ \\
0452$-$0214 & 10470 $\pm$ 170 &  7.92 $\pm$ 0.03 & $-4.22 \pm 0.02$ & $0.54 \pm 0.02$ & $0.50 \pm 0.02$ & 7 & $-4.68$\\
1352+0323 & 9615 $\pm$ 100 &  7.93 $\pm$ 0.07 & $-4.31 \pm 0.05$ & $0.54 \pm 0.03$ & $0.69 \pm 0.09$ & 10 & $-4.64$\\
1252+7352 & 8785 $\pm$ 110 &  8.17 $\pm$ 0.02 & $-2.60 \pm 0.03$  & $0.68 \pm 0.03$ & $1.49 \pm 0.09$ & 4 & $-5.47$\\
1756+3816 & 8100 $\pm$ 115 &  7.97 $\pm$ 0.04 & $-2.28 \pm 0.04$  & $0.55 \pm 0.03$ & $1.31 \pm 0.11$ & 4 & $-4.98$\\
0255+0237 & 7940 $\pm$ 40 & 8.03 $\pm$ 0.01 & $-2.10 \pm 0.01$  & $0.59 \pm 0.01$ & $1.56 \pm 0.02$ & 6 & $-5.16$ \\
1336$-$0337$^{\star}$ & 6215 $\pm$ 50 & 7.91 $\pm$ 0.03 & $-$  & $0.53 \pm 0.01$ & $1.93 \pm 0.07$ & 5 & $-7.32$\\
0922+0103$^{\star}$ & 5910 $\pm$ 25 & 7.89 $\pm$ 0.01 & $-$  & $0.53 \pm 0.01$ & $2.15 \pm 0.02$ & 3 & $-7.10$\\
2214+0923 & 5370 $\pm$ 25 & 7.84 $\pm$ 0.04 & $-5.0$ & $0.5 \pm 0.1$ & $3.26 \pm 0.47$ & 6 & $-4.87$\\
\bottomrule
\end{tabular}
\end{table}

\begin{table}
\setlength{\tabcolsep}{3.5pt}
\centering
\caption{Photospheric element abundances relative to the main photospheric element.  The displayed abundances were measured from the detected metal lines in the DESI and X-shooter spectra (first and second row of each star, respectively), and were obtained with \Teff, \logg\ and \htohe\ fixed to the values in Table~\ref{tab:stellar_params}. We note that the atmosphere models for the white dwarfs $0255+0237$, $1336-0337$, $0922+0103$ and $2214+0923$ included Si at CI chondrites abundances relative to Mg, however, those Si abundances are not included here as they were not inferred from detections (see Section~\ref{sec:stellar_mod} for more details). The two white dwarfs marked with $^{\star}$ have H-dominated photospheres. Note the uncertainties are purely statistical.}
\label{tab:metal_abs}
\begin{tabular}{lccccccccccc}
\toprule
Star & \ztox{O} & \ztox{Na} & \ztox{Mg} & \ztox{Al} & \ztox{Si} & \ztox{Ca} & \ztox{Ti} & \ztox{Cr} & \ztox{Mn} & \ztox{Fe} & \ztox{Ni} \\
\midrule
1333+3254 & $-4.85 \pm 0.09$ & $-$ & $-5.66 \pm 0.05$ & $-$ & $-6.00 \pm 0.07$ & $-7.09 \pm 0.03$ & $-$ & $-$ & $-$ & $-6.19 \pm 0.07$ & $-$ \\
         & $-4.68 \pm 0.13$  & $-$ & $-5.59 \pm 0.02$ & $-$ & $-5.92 \pm 0.02$ & $-7.07 \pm 0.02$ & $-8.94 \pm 0.04$ & $-8.15 \pm 0.09$ & $-$ & $-6.23 \pm 0.02$ & $-$ \\
0850+3208 & $-5.27\pm0.02$   &$-$& $-5.76\pm0.02$ & $-$ & $-5.86 \pm 0.04$ & $-6.87 \pm 0.01$ & $-8.48 \pm 0.02$ & $-$ & $-$ & $-5.78 \pm 0.02$ & $-$ \\
          & $-5.23\pm0.05$   &$-$& $-5.92\pm0.05$ & $-$ & $-5.96 \pm 0.04$ & $-6.94 \pm 0.02$ & $-8.51\pm0.01$  & $-7.82\pm0.03$ & $-8.58\pm0.03$ & $-5.79\pm 0.01$ & $-7.12 \pm 0.04$ \\
0242+0426 & $-5.12 \pm 0.08$ & $-$ & $-5.97 \pm 0.03$ & $-$ & $-5.88 \pm 0.05$ & $-7.35 \pm 0.02$ &  $-8.76 \pm 0.04$ &$-$ &$-$ & $-6.46 \pm 0.04$ & $-$ \\
1626+3136 & $-4.88 \pm 0.08$ &$-$& $-5.70 \pm 0.02$ & $-$ & $-5.56 \pm 0.03$ & $-7.32 \pm 0.02$ & $-8.84 \pm 0.11$ &$-$ &$-$ & $-5.90 \pm 0.02$ &$-$\\
0452-0214 & $-4.54 \pm 0.08$ & $-7.57 \pm 0.04$ & $-5.74 \pm 0.01$ & $-$ & $-6.00 \pm 0.03$ & $-7.40 \pm 0.02$ & $-8.65 \pm 0.04$ & $-$ &  $-$ & $-6.43 \pm 0.02$ &$-$ \\
1352+0323 & $-4.57 \pm 0.20$ &  $-8.17 \pm 0.20$ & $-6.35 \pm 0.03$ & $-$ & $-6.32 \pm 0.08$ & $-8.27 \pm 0.20$ & $-$ & $-$ & $-$ & $-6.83 \pm 0.03$ &  $-$\\
          & $-4.62 \pm 0.11$ &  $-8.21 \pm 0.11$ & $-6.25 \pm 0.01$ & $-$ & $-6.26 \pm 0.03$ & $-8.03 \pm 0.05$ & $-9.00 \pm 0.05$ & $-8.41 \pm 0.04$ & $-9.08 \pm 0.03$ & $-6.73 \pm 0.03$ &  $-8.13 \pm 0.03$\\
1252+7352   & $-$ & $-$ & $-6.65 \pm 0.05$  & $-$ & $-6.66 \pm 0.03$ & $-7.88 \pm 0.03$ & $-$ & $-$ & $-$ & $-6.95 \pm 0.20$ & $-$\\
1756+3816   & $-$ & $-7.44 \pm 0.05$ & $-6.78 \pm 0.04$  & $-$ & $-6.79 \pm 0.03$ & $-9.05 \pm 0.04$ & $-$ & $-$ & $-$ & $-$ & $-$\\
0255+0237   & $-$ & $-$ & $-7.61 \pm 0.05$  & $-$ & $-$ & $-9.29 \pm 0.02$ & $-$ & $-$ & $-$ & $-7.74 \pm 0.02$ & $-$\\
            & $-$ & $-$ & $-7.57 \pm 0.02$ & $-$  & $-$ & $-9.35 \pm 0.01$ & $-10.3 \pm 0.01$ & $-$ & $-10.48 \pm 0.02$ & $-7.72 \pm 0.02$ & $-8.84 \pm 0.02$\\
1336$-$0337$^{\star}$ & $-$ & $-8.85 \pm 0.25$ & $-6.92 \pm 0.05$  & $-8.32 \pm 0.13$ & $-$ & $-8.29 \pm 0.03$ & $-$ & $-$ & $-$ & $-7.61 \pm 0.06$ & $-$\\
            & $-$ & $-8.75 \pm 0.21$ & $-6.92 \pm 0.02$  & $-8.17 \pm 0.03$ & $-$ & $-8.32 \pm 0.01$ & $-$ & $-$ & $-$ & $-7.62 \pm 0.02$ & $-$\\
0922+0103$^{\star}$   & $-$ & $-$ & $-7.39\pm0.03$ & $-$ & $-$ & $-9.54 \pm 0.03$ & $-$ & $-$ & $-$ & $-8.30 \pm 0.02$ & $-$\\
            & $-$ & $-$ & $-7.41\pm0.01$ & $-$ & $-$ & $-9.53 \pm 0.01$ & $-$ & $-$ & $-$ & $-8.27 \pm 0.01$ & $-$\\

2214+0923   & $-$ & $-8.89\pm 0.20$  & $-6.18 \pm 0.30$ & $-$ & $-$ & $-7.84 \pm0.20$ & $-$ & $-$ & $-$ & $-6.06 \pm 0.20$ & $-$\\
            & $-$ & $-9.00 \pm 0.20$ & $-6.18 \pm 0.20$ & $-$ & $-$ & $-7.67 \pm 0.20$ & $-9.8 \pm 0.5$ & $-8.4 \pm 0.4$ & $-$ & $-6.18 \pm 0.20$ & $-$\\

\bottomrule
\end{tabular}
\end{table}
\end{landscape}

\begin{figure*}
        \includegraphics[width=\hsize]{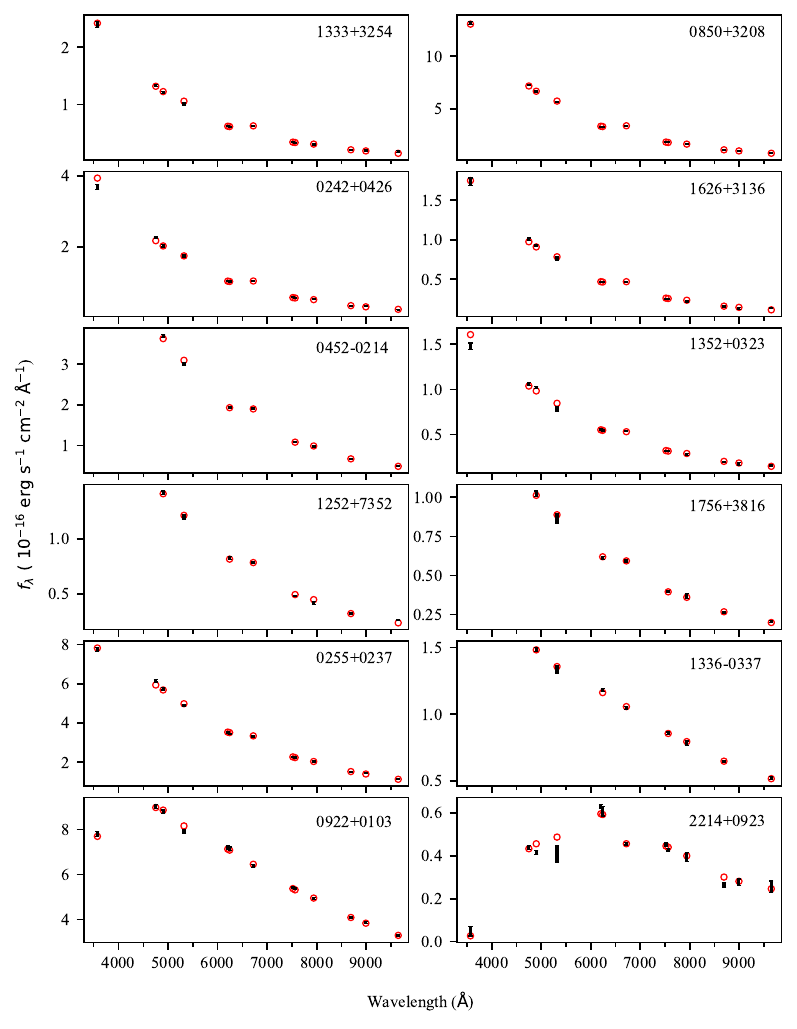}
    \caption{Spectral energy distribution (SED) from archival photometry (black markers; Appendix~\ref{sec:appendix_phot}) with the best photometric fit overplotted (red circles). The white dwarf SEDs are ordered by decreasing effective temperature, from top left to bottom right.}
    \label{fig:SEDs}
\end{figure*}

\begin{figure*}
        \includegraphics[width=0.94\textwidth]{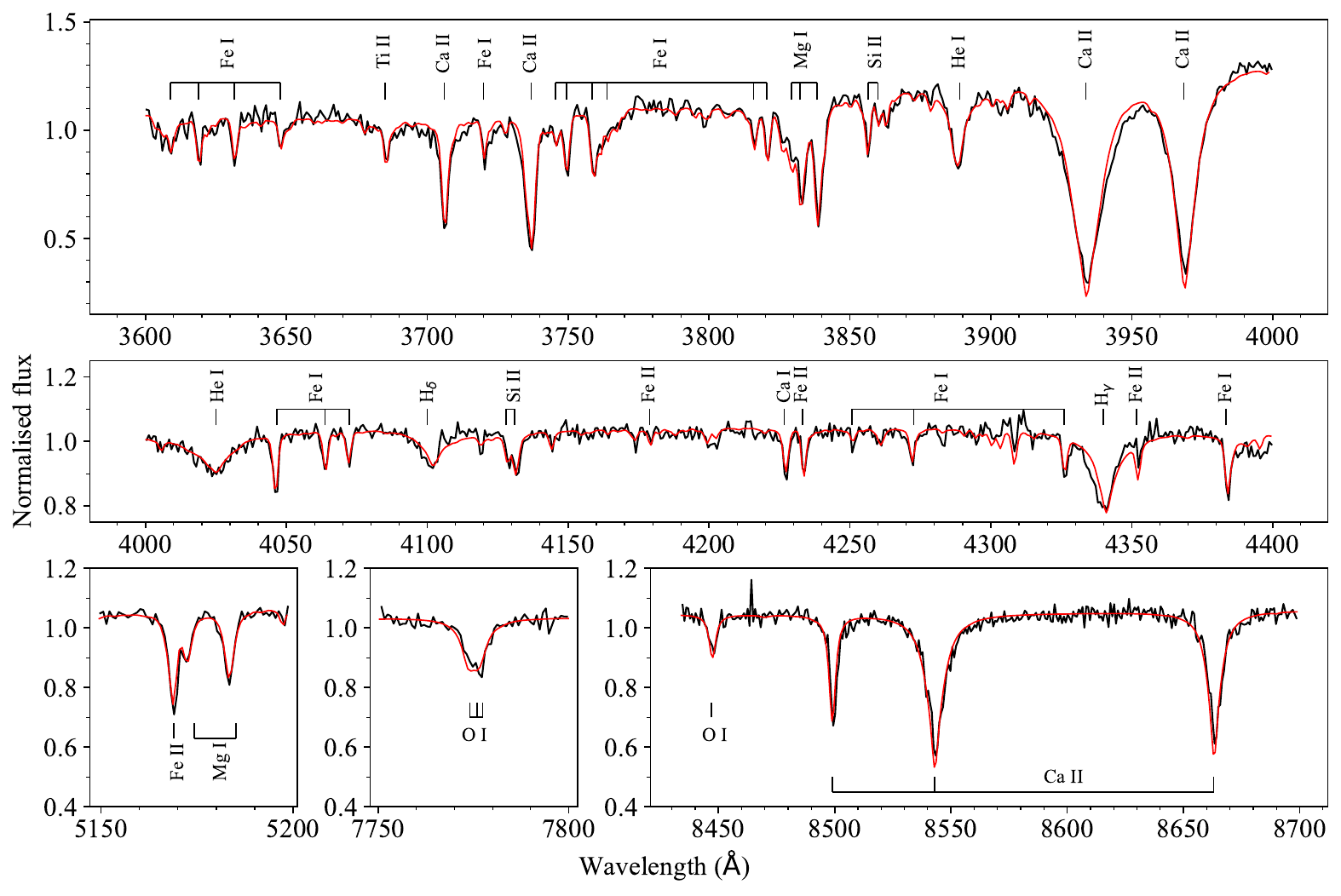}
        \includegraphics[width=0.96\textwidth]{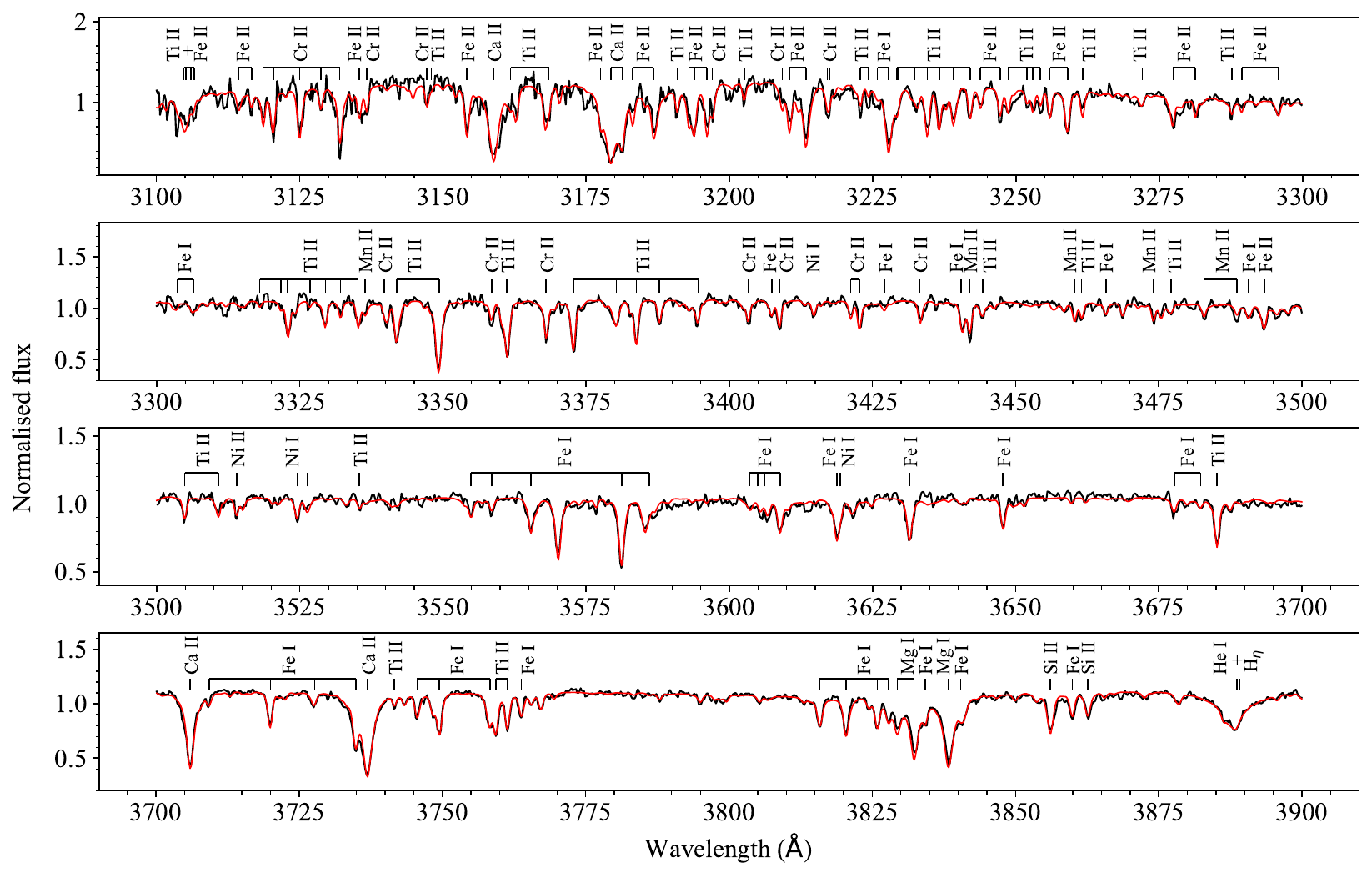}
    \caption{DESI optical coadded spectra (top three rows) and near-UV regions obtained with X-shooter (bottom four panels) of white dwarf 0850+3208. The data and best model (black and red, respectively) have been continuum-normalised by low-order polynomial fits to the regions displayed in each panel. Both datasets display several metal lines among the \protect\Ion{He}{i} and Balmer transitions, but the greater spectral resolution and bluer coverage of X-shooter allows for more identifications.}
    \label{fig:0850_DESIspec_bestfit}
\end{figure*}

The photospheric element abundances of all metals identified in the DESI and X-shooter spectra are listed in Table~\ref{tab:metal_abs}. The quoted uncertainties are statistical, however, based on the comparisons among multiple studies of the same stars, \cite{williams24} concluded that a value of 0.2\,dex is a more conservative and realistic figure.

The spectral energy distribution (SED) of each target are presented in Fig.~\ref{fig:SEDs}, with the best-fit models obtained from the iterative analysis of photometry and spectroscopy (see Section~\ref{sec:iter} for more details).

The best model fits to the 0850+3208 DESI and X-shooter data are shown as an example in Fig.~\ref{fig:0850_DESIspec_bestfit}. The DESI spectrum of this star has a remarkable mean S/N of 46 and displays extreme metal-enrichment that results in numerous overlapping metal transitions, some of them extremely strong. We obtained X-Shooter follow-up observations to increase both the spectral resolution as well as the wavelength range in the ground-based ultraviolet. The data reveal nine different metals: O, Mg, Si, Ca, Ti, Cr, Mn, Fe, and Ni.

We display the DESI and X-Shooter spectra as well as the best-fit models for the other eleven stars in Appendix~\ref{sec:appendix_bestfits}. Even though the models reproduce fairly well all data, there is a shortcoming in the modelling of the Balmer lines for some stars, which seem shifted with respect to both sets of observations (see e.g. Figs.~\ref{fig:1626_bestfit} and \ref{fig:1352_bestfit}). Similar discrepancies have been  previously reported in the analysis of cool He-dominated white dwarfs, and are related to line-broadening mechanisms dominated by interactions with neutral He that are not yet well understood and also possibly related to inadequate treatment of convection \citep[see e.g.][]{koester15,gentile17,tremblay26}. Similar problems are  also noticeable in the fits to the \Ion{He}{i} 4026 or 4472\,\AA\ lines in some stars.

\subsection{Comparison between the DESI and X-Shooter results}

While both data sets share most of their spectral range, the additional near-UV range covered by the X-shooter data, coupled with its superior resolution allows additional metals to be identified. Specifically, the X-shooter spectroscopy confirms the presence of Ti, for which the DESI spectra provide just a hint in the form of two blended lines (\Ion{Ti}{ii}\,3685.2\,\AA\ with \Ion{Fe}{i}\,3686\,\AA; and \Ion{Ti}{ii}\,3759.3+3761.3\,\AA\ with \Ion{Fe}{i}\,3758.2\,\AA) and allows the detection of absorption lines from Cr, Mn and Ni. Ti is an important tracer of crust material, while the relative ratios of Cr, Mn, Ni and Fe help constrain the core fraction of the accreted body. Besides, Ti and Ni also form oxides, and hence their detection improves the accuracy of the O budget (see Section~\ref{sec:water} for more details).

X-shooter spectra extend toward bluer wavelengths than DESI, encompassing more absorption lines from metals identified in both data sets (most notably for Fe and Ti, where most of their transitions lie in that region). We inspected the abundances individually for each element common to both DESI and X-shooter, and found that the differences are similar for all the metals with no exception. Fe enrichment is quite strong in this set of stars, displaying several common absorption lines in DESI and X-shooter data, which can probably explain the compatible estimations. In the case of Ti, the result is less reliable, since just 0850+3208 displays clear isolated features from Ti in both data sets.

Aside from this additional information on the parent body abundances, having two independent data sets for seven stars provides an opportunity to assess systematic uncertainties in the abundances introduced by the use of different instrumentation. 

We calculated the abundance differences for each metal detected in both data sets for each star, using the values of Table~\ref{tab:metal_abs} as input; and computed the weighted mean, with weights proportional to the inverse square of the combined uncertainties. We find a weighted mean of $\bar{\Delta} = 0.007 \pm 0.007$\,dex (being $\Delta$ the difference between DESI and X-shooter $\log{\mathrm{Z/H[e]}}$), corresponding to a 1--2 per cent difference in linear space. This indicates an excellent agreement between our two data sets, showing no significant offset. 


A more detailed assessment of the systematic uncertainties in the parent body abundances determined from debris-enriched white dwarfs will be subject of a future paper but these preliminary assessment shows excellent agreement between the  results from both data sets.

\subsection{Accretion state}
\label{sec:acc_state}

Following the method outlined in Section~\ref{sec:phot_to_pb}, we computed the sinking times, accretion rates and absolute lower limits on the metal masses contained in the CVZ (Table~\ref{tab:sinking_times}). 

\begin{table*}
\setlength{\tabcolsep}{6pt}
\centering
\captionsetup{justification=centering}
\caption{Sinking times $\tau_{\mathrm{z}}$, elemental accretion rates $\dot{M}_{\mathrm{z}}$ assuming the white dwarfs are in steady state, and absolute lower limits on the metal masses contained in the CVZ ($M_{\mathrm{z}}$; Eq.~\ref{eq:m_z}), as there are likely metals that have already sink out or we are not detecting. The H mass is not included in the masses contained in the CVZ since its origin in He-dominated white dwarfs is under debate. The two white dwarfs marked with $^{\star}$ have H-dominated photospheres.}
\label{tab:sinking_times}
\begin{subtable}{\textwidth}
\renewcommand\arraystretch{1.2}
\begin{tabular}{l*{4}{|ccc}}
\hline
& $\tau_{\mathrm{z}}$  & $\dot{M}_{\mathrm{z}}$  & $M_{\mathrm{z}}$
& $\tau_{\mathrm{z}}$ & $\dot{M}_{\mathrm{z}}$ & $M_{\mathrm{z}}$
& $\tau_{\mathrm{z}}$ & $\dot{M}_{\mathrm{z}}$ & $M_{\mathrm{z}}$ 
& $\tau_{\mathrm{z}}$ & $\dot{M}_{\mathrm{z}}$ & $M_{\mathrm{z}}$\\
& ($10^6$ yr) & ($10^8$ g\,s$^{-1}$) & ($10^{20}$ g)
& ($10^6$ yr) & ($10^8$ g\,s$^{-1}$) & ($10^{20}$ g)
& ($10^6$ yr) & ($10^8$ g\,s$^{-1}$) & ($10^{20}$ g)
& ($10^6$ yr) & ($10^8$ g\,s$^{-1}$) & ($10^{20}$ g)\\
\hline
Star && 
\hspace*{-5ex} 1333+3254 \hspace*{-5ex} &&& 
\hspace*{-5ex} 0242+0426 \hspace*{-5ex} &&& 
\hspace*{-5ex} 0850+3208 \hspace*{-5ex} &&& 
\hspace*{-5ex} 1626+3136 \hspace*{-5ex} \\
\hline
H  & $-$ & $-$ & 15 & $-$ & $-$         & 56 & $-$ & $-$ & 1410 & $-$ & $-$ &  172 \\
O  & 3.65 & 58.3 & 6660 & 4.22 & 32.8 & 4320 & 1.48 & 24.6 & 1150 & 1.79 & 61.4 & 3440 \\
Na & $-$ & $-$ & $-$ & $-$ & $-$ & $-$ & $-$ & $-$ & $-$ & $-$ & $-$ & $-$ \\
Mg & 2.80 & 17.9 & 1570 & 3.24 & 9.2 & 928 & 1.13 & 15.9 & 564 & 1.36 & 18.6 & 790 \\
Al & $-$ & $-$ & $-$ & $-$ & $-$ & $-$ & $-$ & $-$ & $-$ & $-$ & $-$ & $-$ \\
Si & 2.60 & 10.2 & 830  & 3.00 & 14.1 & 1320 & 1.05 & 16.1 & 529 & 1.26 & 34.3 & 1350 \\
Ca & 2.06 & 1.48 & 96    & 2.38 & 0.86 & 64 & 0.83 & 2.7 & 71 & 0.99 & 1.01 & 31 \\
Ti & 1.68 & 0.03 & 1.6    & 1.93 & 0.05 & 3.0 & 0.68 & 0.10 & 2.1 & 0.81 & 0.04 & 1.1\\
Cr & 1.61 & 0.22 & $-$ & $-$ & $-$   & $-$ & 0.64 & 0.52 & 11 & $-$ &$-$ & $-$  \\
Mn & $-$ & $-$ & $-$       & $-$ & $-$   &   $-$   & 0.61 & 0.10 & 1.9 & $-$ & $-$ & $-$ \\
Fe & 1.54 & 22.0 & 1060 & 1.77 & 12.5 & 690 & 0.62 & 63.8 & 1240 & 0.74 & 49.6 & 1150 \\
Ni & $-$ & $-$ & $-$ & $-$ & $-$   & $-$  & 0.61 & 3.10 & 59 & $-$ & $-$ & $-$ \\
\hline
Total &  & 110 & 10218 &  & 70 & 7325 &  & 127 & 3628 &  & 165 & 6762 \\
\hline
\end{tabular}
\end{subtable}

\vspace{2em}

\begin{subtable}{\textwidth}
\renewcommand\arraystretch{1.2}
\begin{tabular}{l*{4}{|ccc}}
\hline
& $\tau_{\mathrm{z}}$  & $\dot{M}_{\mathrm{z}}$  & $M_{\mathrm{z}}$
& $\tau_{\mathrm{z}}$ & $\dot{M}_{\mathrm{z}}$ & $M_{\mathrm{z}}$
& $\tau_{\mathrm{z}}$ & $\dot{M}_{\mathrm{z}}$ & $M_{\mathrm{z}}$ 
& $\tau_{\mathrm{z}}$ & $\dot{M}_{\mathrm{z}}$ & $M_{\mathrm{z}}$\\
& ($10^6$ yr) & ($10^8$ g\,s$^{-1}$) & ($10^{20}$ g)
& ($10^6$ yr) & ($10^8$ g\,s$^{-1}$) & ($10^{20}$ g)
& ($10^6$ yr) & ($10^8$ g\,s$^{-1}$) & ($10^{20}$ g)
& ($10^6$ yr) & ($10^8$ g\,s$^{-1}$) & ($10^{20}$ g)\\
\hline
Star &&
\hspace*{-5ex} 0452$-$0214 \hspace*{-5ex} &&&
\hspace*{-5ex} 1352+0323 \hspace*{-5ex} &&&
\hspace*{-5ex} 1252+7352 \hspace*{-5ex} &&&
\hspace*{-5ex} 1756+3816 \hspace*{-5ex} \\
\hline
H  & $-$ & $-$     & 3350  & $-$  & $-$    & 2990  & $-$  & $-$  & 28500 & $-$ & $-$ &  153000 \\
O  & 6.99 & 117 & 25400 & 7.69 & 96.7 & 23200 & $-$  & $-$  & $-$ & $-$  & $-$  & $-$ \\
Na & 5.20 & 0.21   &  34   & 5.64 & 0.05  & 8.6    & $-$  & $-$  & $-$ & 2.89 & 0.27 & 24\\
Mg & 5.28 & 14.8  & 2440  & 5.72 & 4.63   & 828  & 0.96 & 2.02 & 61 & 2.91 & 1.28 & 116 \\
Al & $-$ & $-$ & $-$ & $-$ & $-$ & $-$ & $-$ & $-$ & $-$ & $-$ & $-$ & $-$ \\
Si & 4.88 & 10.2  & 1550  & 5.26  & 5.82  & 956  & 0.88  & 2.55  & 71 & 2.66  & 1.61  & 135 \\
Ca & 3.83 & 0.74   & 88    & 4.08 & 0.18   & 23   & 0.68 & 0.28 & 5.9 & 2.04 & 0.02 & 1.0 \\
Ti & 3.10 & 0.06   & 5.9     & 3.30 & 0.03   & 2.9    & $-$  & $-$  & $-$ & $-$  & $-$  & $-$\\
Cr & $-$  & $-$  & $-$       & 3.13  & 0.13  & 12 & $-$ & $-$ & $-$ & $-$ & $-$ & $-$ \\
Mn & $-$  & $-$  & $-$       & 2.97 & 0.03   & 2.8 & $-$ & $-$ & $-$ & $-$ & $-$ & $-$ \\
Fe & 2.83 & 12.9  & 1140  & 3.30 & 6.71   & 630   & 0.50 & 4.51 & 71 & $-$  & $-$  & $-$ \\
Ni & $-$  & $-$  & $-$       & 2.96 & 0.29   & 26 & $-$ & $-$ & $-$ & $-$ & $-$ & $-$ \\
\hline
Total &  & 156 & 30658 &  & 115 & 25689 &  & 9.36 & 209 &  & 3.18 & 276 \\
\hline
\end{tabular}
\end{subtable}

\vspace{2em}

\begin{subtable}{\textwidth}
\renewcommand\arraystretch{1.2}
\begin{tabular}{l*{4}{|ccc}}
\hline
& $\tau_{\mathrm{z}}$  & $\dot{M}_{\mathrm{z}}$  & $M_{\mathrm{z}}$
& $\tau_{\mathrm{z}}$ & $\dot{M}_{\mathrm{z}}$ & $M_{\mathrm{z}}$
& $\tau_{\mathrm{z}}$ & $\dot{M}_{\mathrm{z}}$ & $M_{\mathrm{z}}$ 
& $\tau_{\mathrm{z}}$ & $\dot{M}_{\mathrm{z}}$ & $M_{\mathrm{z}}$\\
& ($10^6$ yr) & ($10^8$ g\,s$^{-1}$) & ($10^{20}$ g)
& ($10^6$ yr) & ($10^8$ g\,s$^{-1}$) & ($10^{20}$ g)
& ($10^6$ yr) & ($10^8$ g\,s$^{-1}$) & ($10^{20}$ g)
& ($10^6$ yr) & ($10^8$ g\,s$^{-1}$) & ($10^{20}$ g)\\
\hline
Star &&
\hspace*{-5ex} 0255+0237 \hspace*{-5ex} &&&
\hspace*{-5ex} 1336$-$0337$^{\star}$ \hspace*{-5ex} &&&
\hspace*{-5ex} 0922+0103$^{\star}$ \hspace*{-5ex} &&&
\hspace*{-5ex} 2214+0923 \hspace*{-5ex} \\
\hline
H  & $-$ & $-$   & 163000 & $-$  & $-$  & $-$   & $-$  & $-$  & $-$ & $-$ & $-$ &  321 \\
O  & $-$ & $-$ & $-$ & $-$ & $-$ & $-$ & $-$  & $-$  & $-$ & $-$  & $-$  & $-$ \\
Na & $-$  & $-$  & $-$   & 0.04 & 0.02 & 0.02 & $-$  & $-$  & $-$ & 4.34 & 0.01 & 0.7\\
Mg & 1.96 & 0.22 & 13 & 0.04 & 1.27 & 1.47   & 0.05 & 0.47 & 0.76 & 4.32 & 3.85 & 523 \\
Al & $-$ & $-$ & $-$ & 0.03 & 0.08 & 0.09 & $-$ & $-$ & $-$ & $-$ & $-$ & $-$ \\
Si & $-$  & $-$  & $-$   & $-$  & $-$  & $-$   & $-$  & $-$  & $-$ & $-$  & $-$  & $-$ \\
Ca & 1.37 & 0.01 & 0.4 & 0.03 & 0.11 & 0.1  & 0.04 & 0.01 & 0.01 & 2.85 & 0.25 & 23 \\
Ti & 1.11 & 1e-3 & 0.05 & $-$  & $-$  & $-$ & $-$ & $-$ & $-$       & 2.29 & 3e-3 & 0.2 \\
Cr & $-$  & $-$  & $-$  & $-$  & $-$  & $-$ & $-$ & $-$ & $-$        & 2.15 & 0.10 & 6.6 \\
Mn & 1.00 & 3e-4  & 0.04 & $-$  & $-$  & $-$ & $-$ & $-$ & $-$ & $-$ & $-$ & $-$ \\
Fe & 1.00 & 0.69 & 22 & 0.02 & 1.01 & 0.70  & 0.03 & 0.28 & 0.28 & 2.04  & 18.3  & 1170 \\
Ni & 0.98 & 0.06  & 1.7  & $-$  & $-$  & $-$ & $-$ & $-$ & $-$ & $-$ & $-$ & $-$ \\
\hline
Total &  & 0.98 & 36 &  & 2.49 & 2.38 &   & 0.76 & 1.05 &  & 22.5 & 1724 \\
\hline
\end{tabular}
\end{subtable}

\end{table*}


For the He-dominated white dwarfs and cool H-dominated ones in our sample the sinking times are $\tau_{\mathrm{z}} \simeq 10^{4}$--$10^{6}$\,yr, sufficiently long that they are comparable to the estimated duration of accretion events \citep{girven12,cunningham21}. Thus, in this case, the detection of metals does not prove ongoing accretion.

We also checked the available IR photometry for the sources (see Section~\ref{sec:phot_data}) and find no significant excess in that region which could be considered as an indication of active accretion \citep[e.g.][]{zuckerman87,jura03}. However, these white dwarfs are faint and there is growing evidence some dust discs might escape detection due to observational limitations \citep{bergfors14,bonsor17,wilsonetal19}.

\begin{figure}
        \includegraphics[width=\columnwidth]{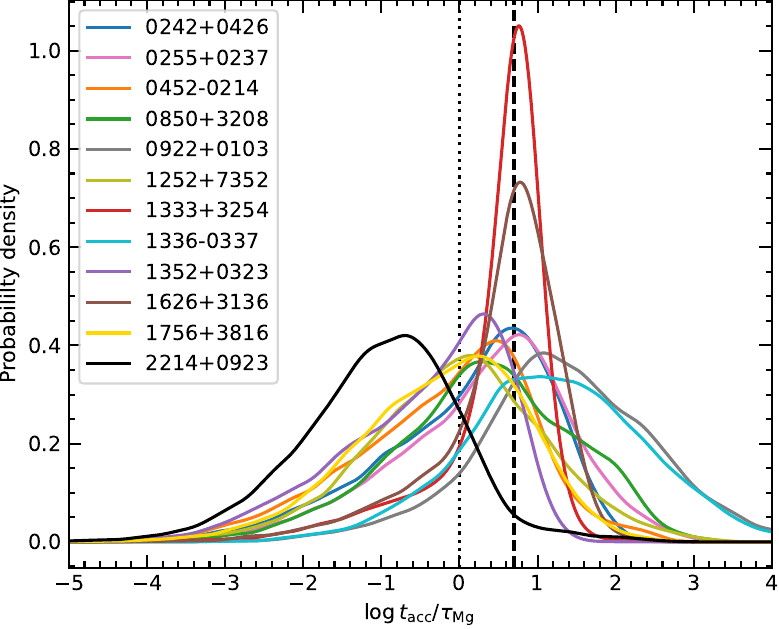}
    \caption{Posterior distributions for the duration of the accretion episode for each star, expressed in terms of sinking timescales for Mg. The posterior probability of a given $\log{t_{\textrm{acc}}/\tau_{\text{Mg}}}$ range is represented by the area under the curve between the range limits. The dotted vertical line at $t_{\text{acc}}=\tau_{\text{Mg}}$ represents the transition out of the increasing phase, where abundances begin to diverge from those in the accreted material. The dashed vertical line at $t_{\text{acc}}=5\tau_{\text{Mg}}$ indicates where abundances have reached a steady state.}
    \label{fig:accretion_duration}
\end{figure}

\begin{figure}
        \includegraphics[width=\columnwidth]{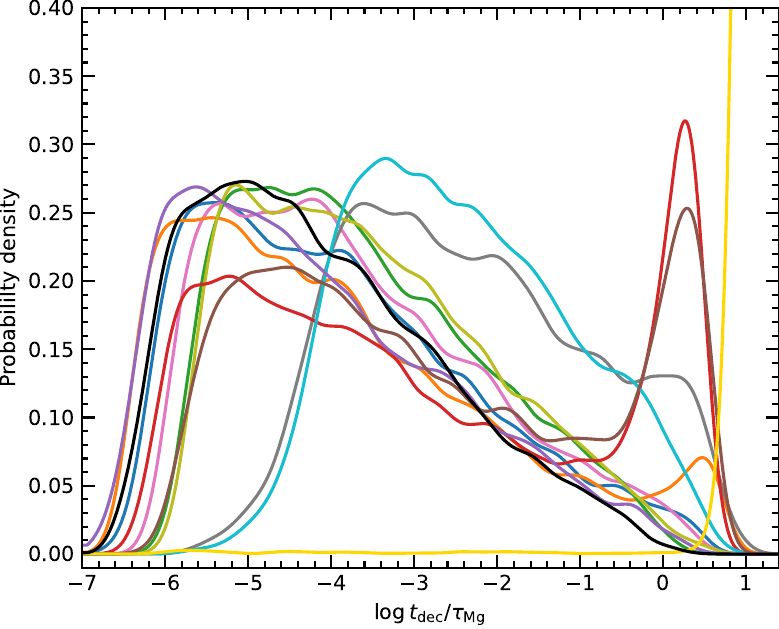}
    \caption{Posterior distributions for the time since accretion ceased, expressed in terms of sinking timescales for Mg. The posterior probability of a given $\log{t_{\text{dec}}/\tau_{\text{Mg}}}$ range is represented by the area under the curve between the range limits. The $y$-axis limits are restricted for readability: the posterior for 1756+3816 is concentrated within a small range and thus has a peak several times higher than the other stars, though we discuss in the text that this result is most likely an artefact caused by insufficient data.}
    \label{fig:time_since_accretion}
\end{figure}

Thus, we used the Bayesian model described in Section~\ref{sec:phot_to_pb} to probe the state of accretion. The abundances and sinking times presented in Tables~\ref{tab:metal_abs} and \ref{tab:sinking_times} were used as input data, with systematic errors of 0.2\,dex (as suggested in Section~\ref{sec:stellar_params}) added in quadrature to the abundance uncertainties presented in those tables to avoid over-fitting. The posterior distributions for the duration of accretion and time since accretion ceased are presented in Figs.~\ref{fig:accretion_duration} and \ref{fig:time_since_accretion}, respectively. Our findings show a posterior median of 0.003 $\tau_{\mathrm{Mg}}$ or less for the time since accretion ceased in all cases except 1756+3816 (where there are insufficient data for a meaningful result), i.e. accretion is most likely ongoing for the stars in our sample. 

However, the tails of the posteriors extend into the decreasing phase for 1333+3254 and 1626+3136, meaning such solutions cannot be ruled out. The time since accretion began is much less constrained. For example, the posterior distribution for 2214+0923 shows a mild preference for the increasing phase, but the 2-$\upsigma$ credible intervals for most of the stars encompass both the increasing phase and steady state.

We note that the small number of detected metals in 0922+0103 and 1756+3816 yielded an unconstrained accretion duration, with the posteriors covering the three different accretion phases. For 0922+0103, the median value places the star in the steady state but for 1756+3816 this analysis is unsuitable due to insufficient data. We therefore refrain from over-interpreting the results for 1756+3816.

The subsequent sections discuss our compositional analysis assuming steady state but, for completeness, alternative accretion scenarios are discussed in Section \ref{sec:declining}.


\subsection{Accretion rates and parent body masses}
\label{sec:mdot_pbmasses}

The accretion rates onto the white dwarfs (computed under the assumption of steady state, Table~\ref{tab:sinking_times}) span three orders of magnitude, reflecting the wide range of photospheric metal abundances, CVZ masses and sinking times. The H-dominated 0922+0103 has the lowest accretion rate, $7.6\times 10^{7}$\,g\,s$^{-1}$, and the He-dominated  1626+3136 has the highest one, $1.65\times 10^{10}$g\,s$^{-1}$. The He-dominated white dwarfs 1333+3254, 0850+3208, 0452$-$0214 and 1352+0323 also show high accretion rates of $\dot{M}_{\mathrm{z}}\simeq10^{10}\mathrm{g\,s}^{-1}$. Such high values have been previously reported for other He-dominated white dwarfs \citep[see e.g. fig. 10 within][]{williams24} but they exceed theoretical predictions of maximum accretion rates of $10^{8} - 10 ^{9}$\,g\,s$^{-1}$ under Poynting-Robertson drag \citep{rafikov11,brouwers22}. The origin of this discrepancy remains under debate, highlighting our limited understanding of accretion processes and, possibly, in the calculation of the accretion rates \citep{bauer19, cunningham19, cresswell25}. \cite{okuya23} demonstrated that the presence of volatiles in the accreted material results in an increased  mass flow through the debris disc, providing one physical avenue that could explain the high accretion rates. Nevertheless, that process would be applicable to both H and He-dominated white dwarfs, and hence not resolving the known discrepancy between the accretion rates determined for these two different types of atmospheres \citep{girven12, bergfors14, williams24}.

The absolute lower limit on the total accreted mass is independent of the accretion phase and was computed excluding H, since its origin may partly be primordial \citep[see e.g.][]{rolland18}. Under this assumption, the inferred accreted masses span a wide range, but are overall comparable with those determined for  other systems \citep[fig.~11 in ][]{williams24}. The largest planetary bodies are accreted by 1333+3254, 0452$-$0214 and 1352+0323 and are comparable in mass to Charon ($1.56\times10^{24}$g). The minimum masses of the parent bodies inferred for 0242+0426, 0850+3208, 1626+3136 and 2214+0923 are more akin to that of the asteroid Vesta ($2.6\times10^{23}$g); while for 1252+7352, 1756+3816, 0255+0237 the objects are more similar in mass to Neptune's moon Proteus ($4.4\times10^{22}$\,g). Finally, the rocky bodies accreted by the two H-dominated white dwarfs are  closer in mass to Janus ($\sim10^{21}$g), a moon of Saturn. 

\subsection{Bulk composition of accreted exoplanets}
\label{sec:steady}

\begin{figure*}
        \includegraphics[width=0.99\hsize]{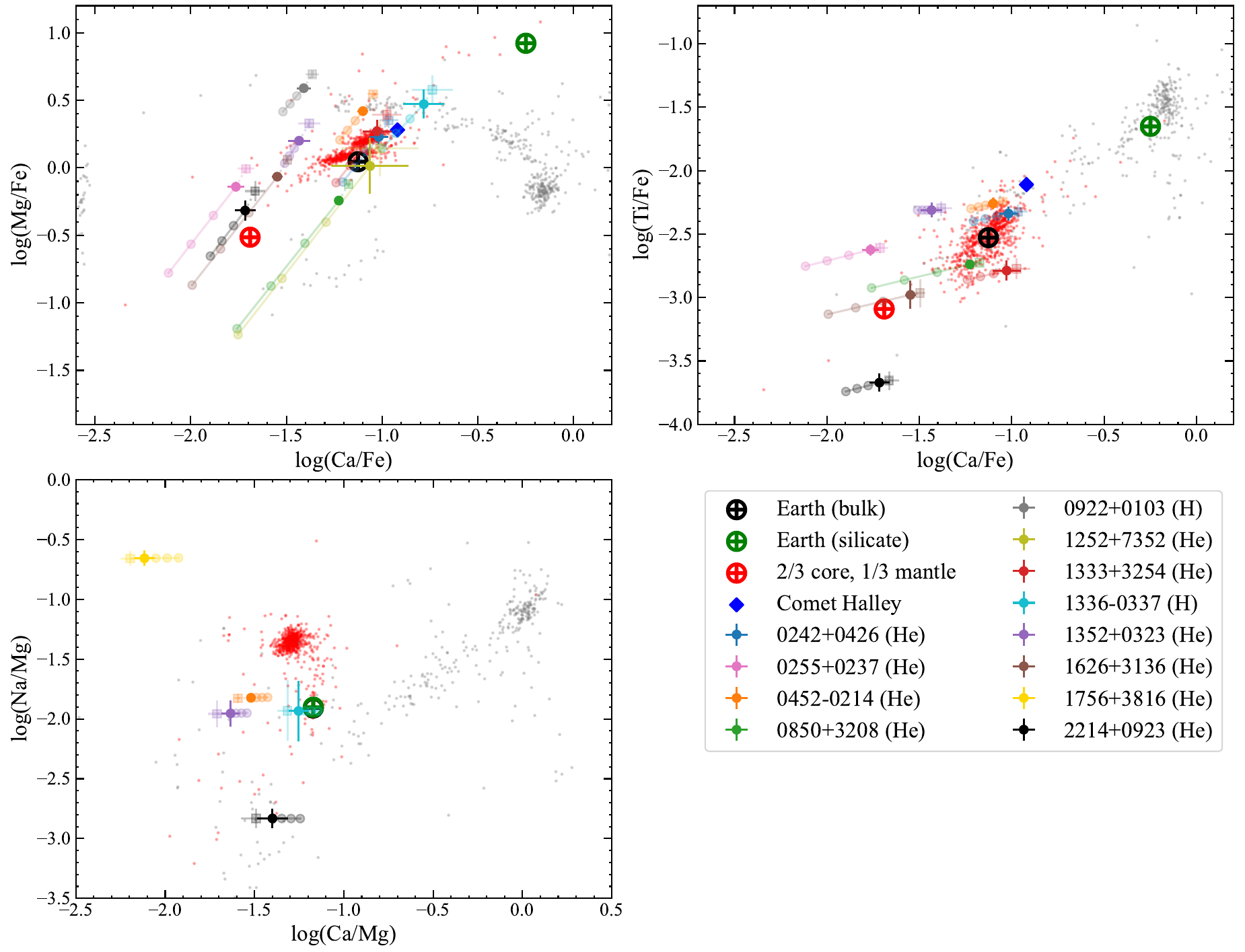}
    \caption{Metal abundance ratios in the accreted parent bodies for the white dwarfs in our sample, along with some Solar System compositions. The circles represent the ratios assuming a steady-state accretion phase, the squares that of increasing phase abundances assuming accretion started $1\times\tau_{\mathrm{Mg}}$ ago, and the translucent tracks represent the decreasing phase, with each semi-transparent point on the track showing the abundance after steps of $10^{6}$\,yr for He-dominated white dwarfs and $10^{4}$\,yr for H-dominated white dwarfs. The small dots in the background of each panel represent abundance ratios of some primitive (red) and processed (grey) meteorites analysed in terrestrial laboratories. Assuming steady state, most of the white dwarfs display accreted material compositions closer to primitive meteorites, with relative abundances similar to those of bulk Earth. There are some exceptions, such as 2214+0923 and 1336$-$0337, which inhabit regions of processed material, hinting towards an accretion of something akin to a core fragment for the former, and similar in composition to the bulk silicate Earth for the latter. In most of the cases, these abundance ratios confirm that the decreasing phase is highly unlikely due to the inferred parent body composition, suggesting exotic and yet unknown planetesimal compositions.}
    \label{fig:ratios_vs_Fe}
\end{figure*}

\begin{figure*}
        \includegraphics[width=0.99\hsize]{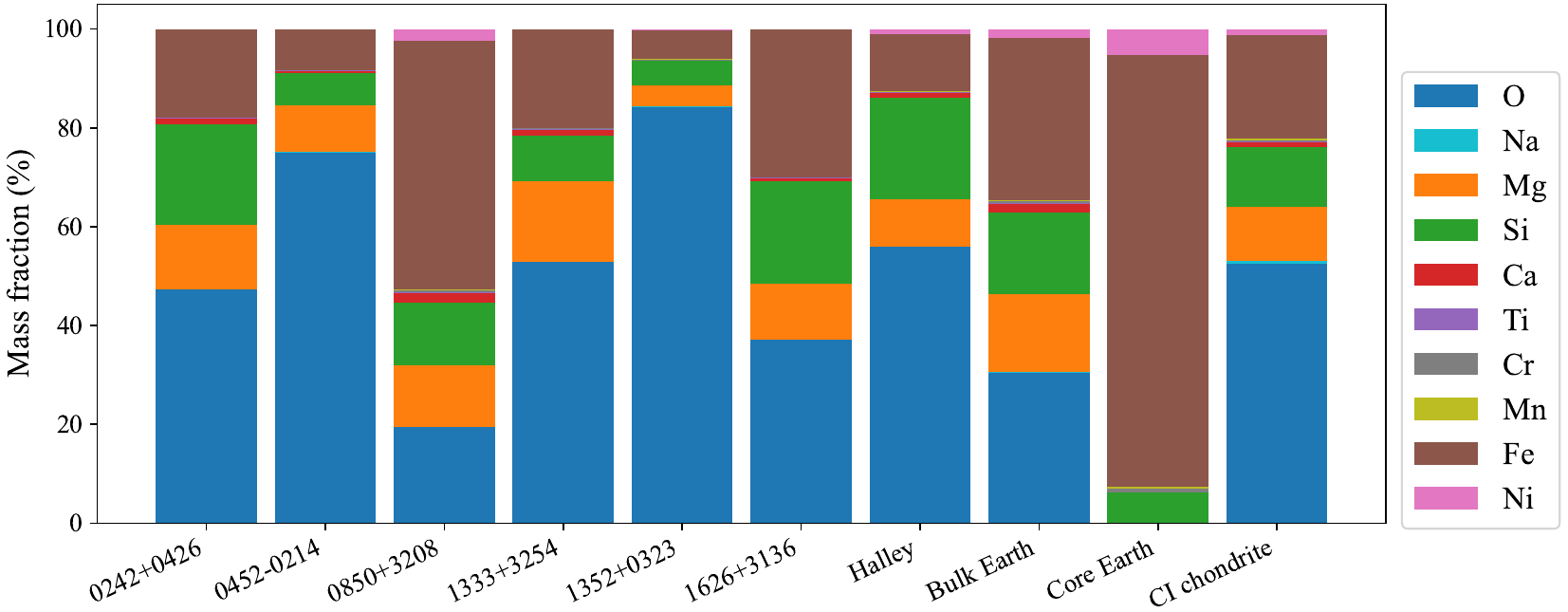}
    \caption{Mass fractions for the six warmer white dwarfs, calculated using the accretion rates for the steady-state accretion phase. We compare these to mass fractions of the comet Halley \citep{jessberger1988}, bulk Earth, core Earth, and CI chondrites \citep{mcdonough00-1,lodders03-1}. Note that the mass fractions of the Solar System bodies are rescaled to only include the detected elements in the studied white dwarfs, with combined mass fractions of the elements \textit{not included} of 2.1 per cent for bulk and core Earth, 12 per cent for CI chondrites and 19 per cent for comet Halley.}
    \label{fig:barcharts_warm}
\end{figure*}

The number element ratios of Ca/Fe, Mg/Fe, Ti/Fe, Na/Mg and Ca/Mg for the accreted bodies derived from our data are shown in Fig.~\ref{fig:ratios_vs_Fe}. Since core Earth has no Mg, it has been represented by 2/3 of the actual core composition and 1/3 of the mantle. The abundance ratios of some primitive meteorites (known as chondrites and unchanged since the solar nebula phase) and some processed ones (whose parent bodies have undergone geological and physical changes) are also displayed in Fig.~\ref{fig:ratios_vs_Fe}.

The relative abundances of some of the major rock-forming elements (Mg, Ca, Fe, Ti, top two panels in Fig.~\ref{fig:ratios_vs_Fe}) in most of the parent bodies in our sample are remarkably similar to those seen in both the bulk Earth and primitive meteorites, suggesting that the accreted parent bodies were rocky and possibly unprocessed. Exceptions are 2214+0923, 0255+0237 and 1626+3136 which are more Fe-rich, placing them closer to core-dominated fragments; and 1336$-$0337, which we discuss below in detail.

The Na/Mg ratio of the material accreted by the white dwarfs in our sample shows a wide range (bottom panel of Fig.~\ref{fig:ratios_vs_Fe}). 0452$-$0214 and 1352$+$0323 are accreting Na/Mg ratios that match bulk Earth. 

The Na/Mg, Al/Mg, and Ca/Mg ratios for the H-dominated 1336$-$0337 are close to those seen in bulk Earth. However, the accreted body is strongly Fe-depleted when compared to the bulk Earth Fe/Mg ratio. These four ratios suggest a similar composition to that seen in the bulk silicate Earth, with a parent body possibly originating from silicate-rich mantle material. 


1756+3816 has an Na/Mg ratio similar to eucrites, i.e. above that of the Sun and nearby stars, and higher than most meteorites in the Solar System. Furthermore, its Ca/Mg ratio is lower than most meteorites. The planetary body accreted by 1756+3816 is likely volatile-rich and refractory-poor indicative of forming in the outer regions of a planetary system. 

2214+0923 on the other hand has refractory abundances that are consistent with the Sun, nearby stars, and meteorites, as seen in all panels of Fig.~\ref{fig:ratios_vs_Fe}, whereas its Na/Mg ratio is low. This is indicative of volatile depletion with the planetary material either forming closer-in to its host star, or experiencing some form of heating during evolution. This low a Na/Mg ratio is found in diogenites and mesosiderites in the Solar System.

We compare the mass fractions of the parent bodies accreted by the six warmer white dwarfs with those of the bulk and core Earth \citep{mcdonough00-1}, CI chondrites \citep{lodders03-1} and comet Halley \citep{nittleretal04-1} in Fig.~\ref{fig:barcharts_warm}. White dwarfs 1252+7352, 1756+3816, 0255+0237, 1336-0337, 0922+0103 and 2214+0923 were omitted as at least one of the main constituents (O, Si, Mg, Fe) is undetected, and hence their mass fractions cannot be accurately represented.

The stars gather in two clear groups (Fig.~\ref{fig:barcharts_warm}), those accreting dry and wet material. White dwarfs 0242+0426, 1333+3254 and 1626+3136, share a remarkable resemblance with bulk Earth or CI chondrites, strengthening the case that chondritic-like compositions dominate among metal-enriched white dwarfs \citep{trierweiler23}. 0850+3208 sits apart in the dry-rock planetesimals accretion due to an Fe enhancement compared to bulk Earth, possibly suggesting that it is accreting a differentiated body, enriched in elements found in planetary cores.

\begin{figure}
        \includegraphics[width=\columnwidth]{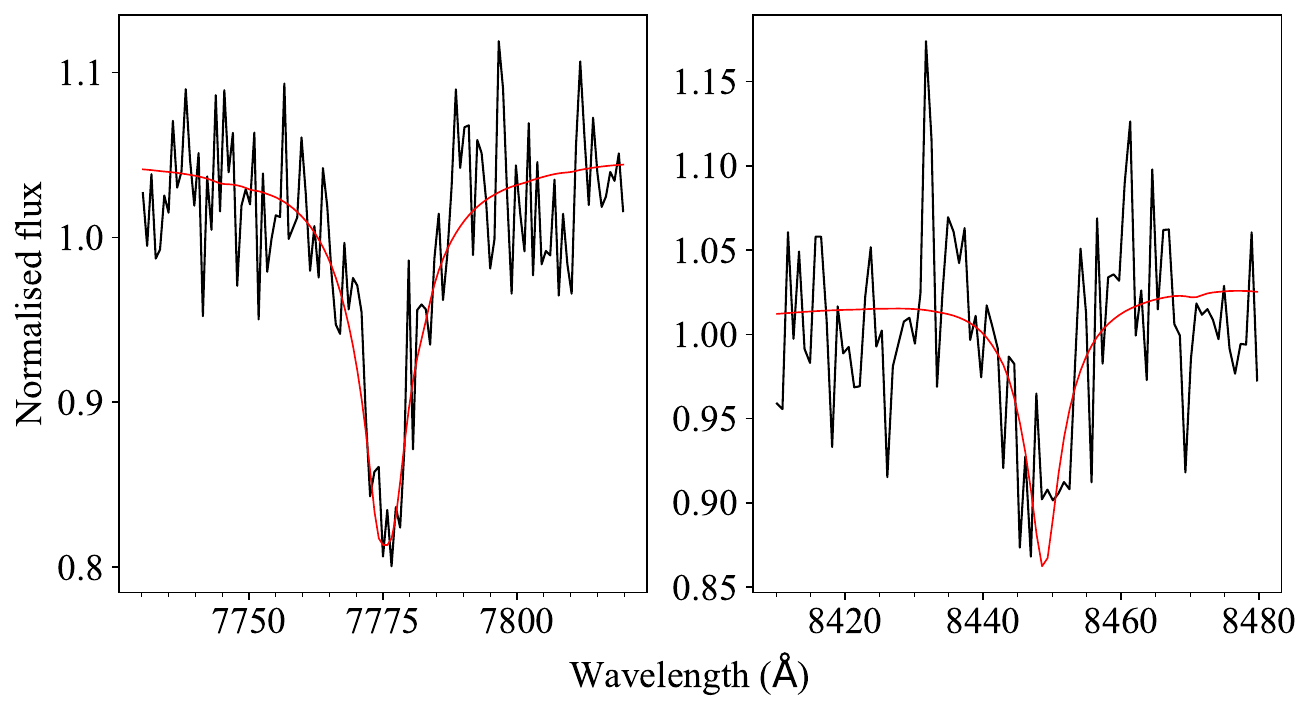}
    \caption{0452$-$0214 DESI data, zoomed in on the oxygen lines. The best fit model is overplotted, with \ztohe{O}$=-4.54$.}
    \label{fig:0452_oxygen}
\end{figure}

\begin{figure}
        \includegraphics[width=\columnwidth]{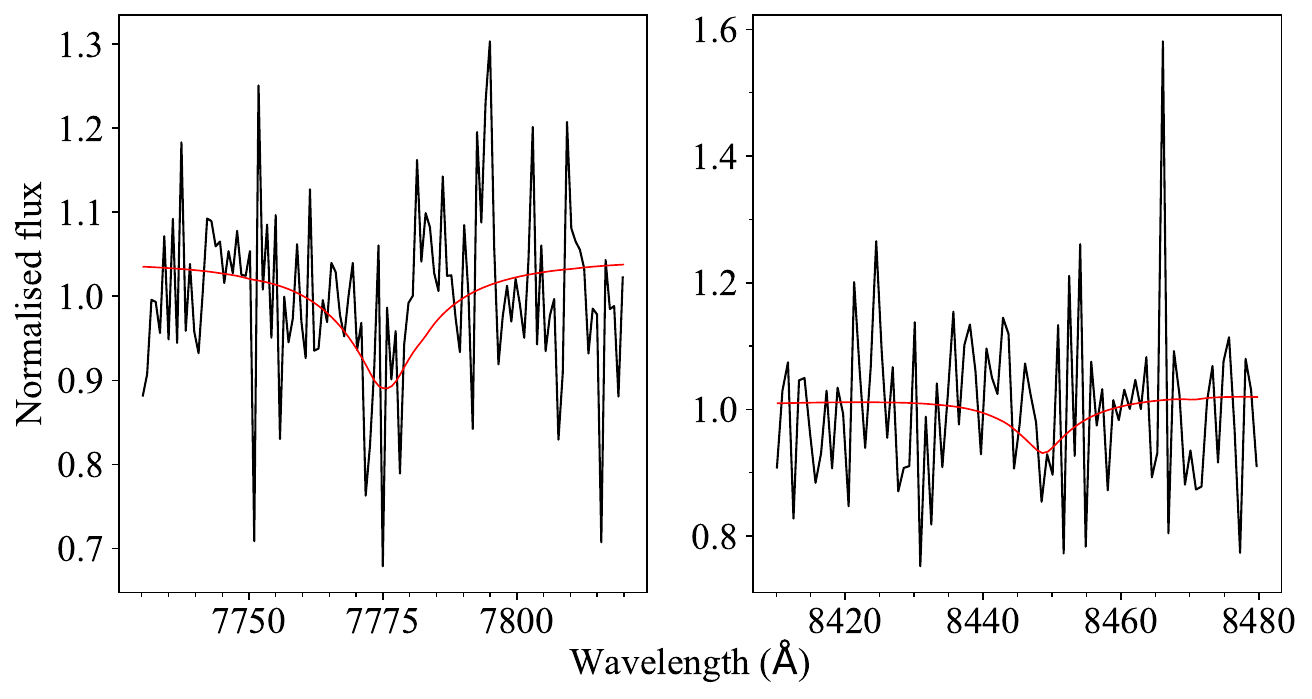}
         \includegraphics[width=\columnwidth]{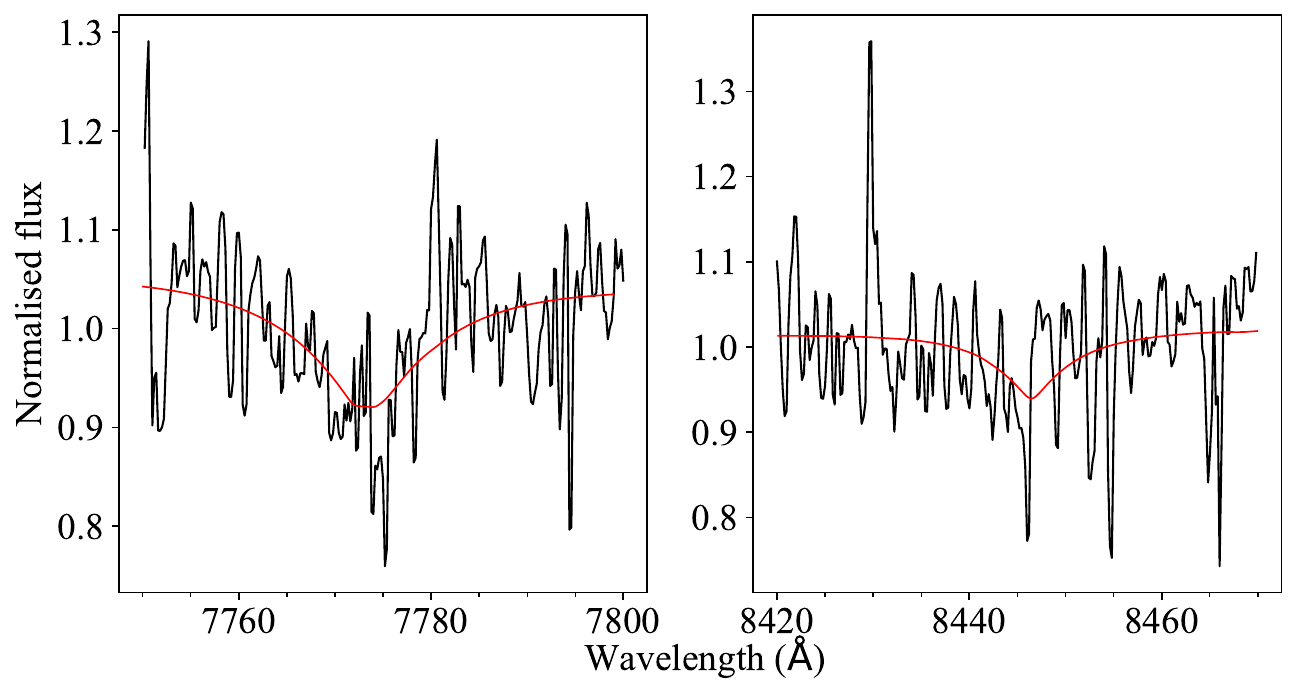}
    \caption{DESI and X-shooter data of 1352+0323 (top and bottom panel, respectively), zoomed in on the oxygen lines. The best fit models are overplot, with \ztohe{O}$=-4.55$ and \ztohe{O}$=-4.65$, respectively.}
    \label{fig:1352_oxygen}
\end{figure}

The stars in the other group, 0452$-$0214 and 1352+0323, display large O mass fractions. In 0452$-$0214, both the \ion{O}{i} 7774\,\AA\ triplet and the 8446.76\,\AA\ lines are well reproduced the best-fit model (Fig.~\ref{fig:0452_oxygen}), suggesting a trustworthy O overabundance that may be linked with the accretion of a water-rich parent body (see Section~\ref{sec:water}). In contrast, the O lines in 1352+0323 are weak and not satisfactorily modelled (Fig.~\ref{fig:1352_oxygen}). Thus the O abundance in this star requires a deeper spectrum for confirmation and should for now be treated with some care. We note that 1352+0323 is the coolest star with an O detection in our sample, which is the result of the relatively high excitation energy of the \ion{O}{i} lines, rather than a genuine abundance pattern. For completeness, all the O lines modelled in this analysis are shown in Figs.~\ref{fig:1333_Olines_DESI}--\ref{fig:1626_Olines_DESI}.

\subsection{Probing for water-rich material}
\label{sec:water}

\subsubsection{Oxygen budget}
\label{sec:o_bgt}

To investigate whether the accreted bodies might contain water, we calculated the O excess or deficit for the six white dwarfs with O detections. We followed the prescription of \citet{klein10}, assuming that O is accreted entirely in the form of the oxides MgO, SiO$_2$, CaO, FeO, Al$_2$O$_3$, NiO, and TiO$_2$. For a bulk Earth composition, a large proportion of Fe and Ni form an oxygen-free alloy in the core. Conversely, depending on the oxygen fugacity, Fe could be in the form of Fe$_2$O$_3$, so the oxygen budget must be considered with these caveats in mind. Elements among the above list of oxides that were not detected in the spectroscopy were included on the calculation of the oxygen budget (O$_{\mathrm{bgt}} = M_{\mathrm{O,oxides}}/ M_{\mathrm{O}}$) with their abundances scaled with respect to the Mg mass fraction of bulk Earth. The error on overall \obgt\ was calculated using a Monte Carlo method, drawing 10\,000 samples from Gaussian distributions whose means and standard deviations were the logarithmic metal abundances and their uncertainties.

An $\obgt>1$ is unphysical because it requires more O than is observed in the white dwarf. Hence some metals must not have fully combined with O in the parent body and were subsequently accreted in the metallic form rather than as oxides. An \obgt\ of 1 suggests all O was locked in oxides within the parent body. Finally an \obgt\ of $<1$ is indicative of an O excess, suggesting an additional oxygen-carrier, the most plausible being H$_2$O. 

\begin{figure*}
        \includegraphics[width=0.99\hsize]{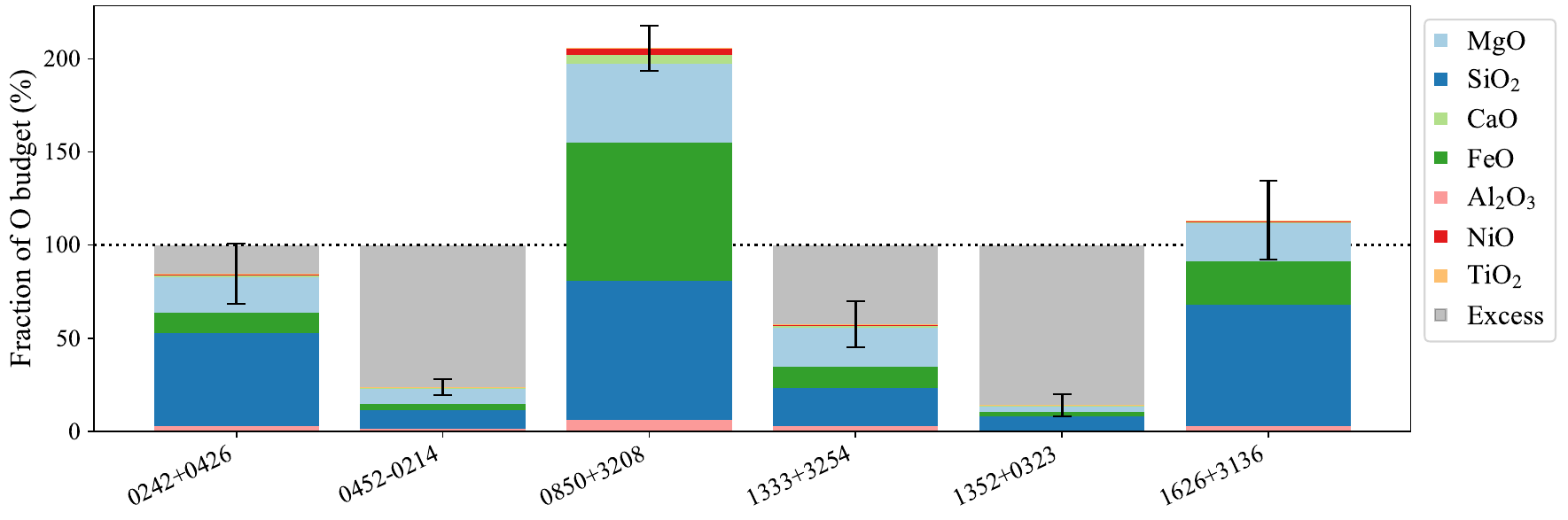}
    \caption{The oxygen budgets for white dwarfs with detected O, assuming steady-state accretion phases. This is calculated by allocating O to the following oxides: MgO, SiO$_2$, CaO, FeO, Al$_2$O$_3$, NiO, and TiO$_2$. The error bar represents the uncertainty on the total O budget. Stars 0452$-$0214, 1333+3254, and 1352+0323 display O excesses (see the main text in Section~\ref{sec:o_bgt} for a detailed discussion on individual objects).}
    \label{fig:Obudget}
\end{figure*}

We present the results of this analysis in Fig.~\ref{fig:Obudget}. 0850+3208 has $\obgt >1$, suggesting the presence of metallic Fe. That could indicate accretion of core-rich material, especially given the high Fe content, but it could also reflect planetesimal formation in a reducing environment \citep{doyle20}. 0242+0426 and 1626+3136 have $\obgt=1$, indicating accretion of dry rocky material. 0452$-$0214, 1352+0323, and (to a lesser extent) 1333+3254 have O excesses, suggesting that water was present in their accreted bodies (but see discussion below regarding 1333+3254).

We also explored the water content of the accreted material assuming that it is similar to Solar System bodies, which relaxes the conservative requirement of maximum oxidation described above. An optional parameter for water content was incorporated into the Bayesian modelling (see Section~\ref{sec:phot_to_pb}). The water fraction of the accreted material was only a free parameter in the wet models, while the dry models assumed exactly zero additional water. Thus, the averaged posterior for the water fraction was conditional on that fraction being non-zero, the probability of which we calculated using the Bayesian evidences. 

\begin{figure}
        \includegraphics[width=\columnwidth]{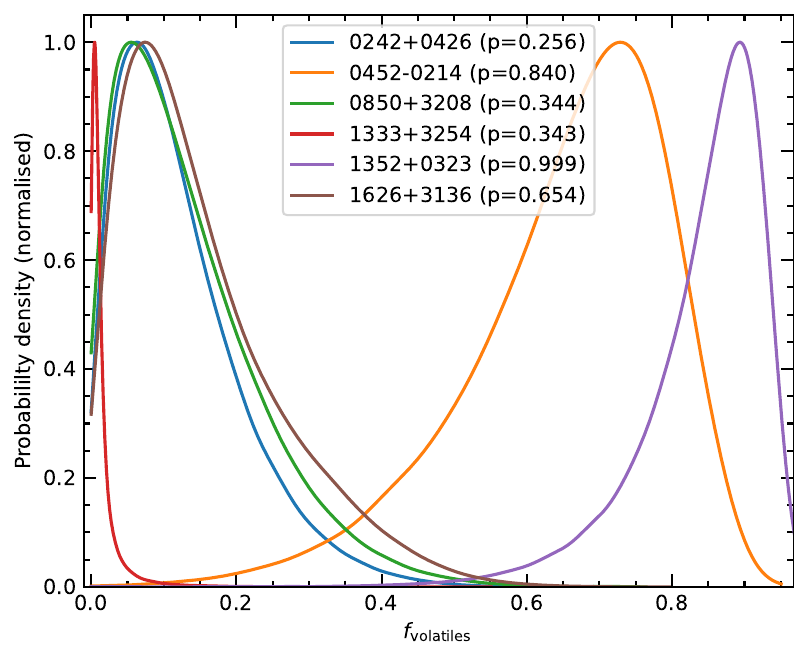}
    \caption{Posterior probability distributions normalised to their maxima for the fraction of water present in the material accreted by each star, \textit{assuming a water-rich model}. The probability that a wet model is a better fit than a dry one is given for each star in the legend in parentheses. For stars 0242+0426, 0850+3208 and 1333+3254 a wet model is disfavoured, while 1626+3136 displays an intermediate case, and stars 1352+0323 and 0452$-$0214 show a high probability of accreting material with a large fraction of water.}
    \label{fig:water_fraction}
\end{figure}

The probability that additional water has been accreted, and the posterior distribution of its quantity in that case, marginalised over the unknown accretion state are shown in Fig.~\ref{fig:water_fraction}. A substantial water fraction (median 0.68) is the preferred model for 0452$-$0214, which serves as a confirmation of the results of the \obgt. Although 1352+0323 is found to be accreting water-rich material at the 3-$\upsigma$ level, with the median of the posterior water fraction being 0.86, we note this result only holds for a reliably-measured O abundance, which may not be the case for this star as discussed above. For 1333+3254, a wet model is disfavoured at about the 1-$\upsigma$ level, and if additional water is present it forms at most a few per cent of the accreted material. For the remainder of the stars, while the data permit a moderate water fraction, they do not decisively rule one in or out.

Both analyses point towards the accretion of volatile-rich material, likely in the form of water, for 0452$-$0214 and 1352+0323, and nearly-dry material in the case of 0850+3208, 1626+3136 and 0242+0426. However, there is a tension between the results for 1333+3254, where the \obgt\ indicates a water fraction of about 25 per cent, while the Bayesian analysis yields a dry composition. The relative over-abundance of O at this star contrasts with a low H abundance (see Table~\ref{tab:metal_abs}), so we now examine the feasibility of H delivery via water.


\subsubsection{Is all the H accreted as H$_2$O?}
\label{sec:hydrogen}

All six white dwarfs with O detections also have trace H in their He-dominated atmospheres, which potentially provides independent information on the possible accretion of water. Unlike metals, H does not sink and can therefore, in principle, be used as a record of the history of H accretion. However, there is an ongoing debate over the extent to which the presence of H in He-dominated white dwarfs is the result of internal processes \citep{koester15,bedard24} versus accretion of water-rich material \citep[see e.g.][]{gentile17}. 



In the following analysis, we assume that (1) \textit{all} H within the white dwarf envelope arises from water contained in the parent body that is currently accreted and that (2) the accretion rate of all individual elements (Table~\ref{tab:sinking_times}) have been constant throughout the ongoing accretion episode. We then explore whether these assumptions are physically possible, and if so, the implications for the parent body masses $M_\mathrm{tot}$ and the duration of the accretion events $t_\mathrm{acc}$. Below, we use \textit{mass} as the total integrated mass accreted throughout the current episode, e.g. $M_\mathrm{H}$, $M_\mathrm{O}$, and $M_\mathrm{Z}$ refer to the total accreted masses of H, O, and that of all elements heavier than He, respectively. We restrict our analysis to the four white dwarfs (1333+3254, 0242+0426, 0452$-$0214, and 1352+0323) that have $\mathrm{O}_\mathrm{bgt} < 1$, i.e. that are likely accreting material containing water. 

We begin by computing  $M_\mathrm{O}$ as the sum of $M_\mathrm{O,H_{2}O}$ and $M_\mathrm{O,oxides}$, i.e. O contained in water and minerals, respectively. $M_\mathrm{O,H_{2}O}$ is given by the H:O mass ratio of 2:16 (1:8) for H$_2$O and $M_\mathrm{H}$ from Table~\ref{tab:sinking_times}. $M_\mathrm{O,oxides}$ is computed from \obgt\ (Table~\ref{tab:water_spec}). 

\begin{equation}
  M_\mathrm{O} = M_\mathrm{O,H_{2}O} + M_\mathrm{O,oxides} = 8M_\mathrm{H} + \obgt M_\mathrm{O}= \frac{8M_\mathrm{H}}{(1-\obgt)}
\end{equation}


\noindent The total mass of accreted material $M_\mathrm{tot}$ is calculated using the O mass fraction,  $f_\mathrm{O} =  M_\mathrm{O}/M_\mathrm{Z} \equiv \dot M_\mathrm{O}/\dot M_\mathrm{Z}$, where $\dot M_\mathrm{O}$ and $\dot M_\mathrm{Z}$ are the measured accretion rates of O and all elements heavier than He (Table~\ref{tab:sinking_times}):

\begin{equation}
    M_\mathrm{tot} = M_\mathrm{Z}+ M_\mathrm{H} =\frac{M_\mathrm{O}}{f_\mathrm{O}} + M_\mathrm{H} = \frac{8M_\mathrm{H}}{(1-\mathrm{O}_\mathrm{bgt})f_\mathrm{O}} + M_\mathrm{H}
    \label{eq:h_mtot}
\end{equation}


\begin{table*}
    \centering
    \caption{Total mass, water mass fraction and total accretion lifetime, required in the hypothetical scenario that \textit{all} of the H in the star was delivered by planetary material with the same abundances as those we have measured. These three quantities are obtained by evaluation of Eqs~\ref{eq:h_mtot}, \ref{eq:h_fwater} and \ref{eq:h_tacc}, respectively. We assumed accretion is in the steady-state phase. We cannot do this calculation for 0850+3208 and 1626+3136 because they have O$_\mathrm{bgt}>1$, which would be unphysical in this scenario.}
    \begin{tabular}{lrrrr}
        \hline
        Star & O$_\mathrm{bgt}$ &  $M_\mathrm{tot}$ ($10^{20}$\,g) & $f_{\mathrm{H_2O}}$ (\%) & $t_\mathrm{acc}$ (Myrs) \\
        \hline
        1333+3254 & $0.574\pm0.124$ & 545.0 & 24.8 & 0.15 \\
        0242+0426 & $0.847\pm0.161$ & 11\,774.0 & 4.3 & 4.8 \\
        0850+3208 & $2.058\pm0.119$ \\
        1626+3136 & $1.134\pm0.211$ \\
        0452$-$0214 & $0.238\pm0.044$ & 50\,163.4 & 60.1 & 9.5 \\
        1352+0323 & $0.141\pm0.060$ & 32\,439.6 & 82.9 & 8.7 \\
        \hline
    \end{tabular}
    \label{tab:water_spec}
\end{table*}

\noindent Under the assumption that all H was accreted as H$_2$O, the total mass of water is $M_\mathrm{H_2O}=9M_\mathrm{H}$, and the water mass fraction becomes 

\begin{equation}
    f_{\mathrm{H_2O}} = \frac{9M_\mathrm{H}}{M_\mathrm{tot}}
    \label{eq:h_fwater}
\end{equation}

\noindent Assuming a constant accretion rate, the total duration of this accretion event is

\begin{equation}
    t_\mathrm{acc} = \frac{M_\mathrm{O}}{\dot{M}_\mathrm{O}} = \frac{8M_\mathrm{H}}{(1-\mathrm{O}_\mathrm{bgt})\dot{M}_\mathrm{O}}
    \label{eq:h_tacc}
\end{equation}

\noindent The results for $M_\mathrm{tot}$ and $t_\mathrm{acc}$ are listed in Table~\ref{tab:water_spec}. Given the above assumptions, all four white dwarfs are accreting objects with masses comparable to those of small Solar System bodies, in the range of moons such as Proteus (1333+3254) to dwarf planets such as Haumea (0452$-$0214), and the duration of the accretion episodes are within the range of predicted disc lifetimes \citep[0.05--30\,Myrs;][]{girven12, cunningham21}. 

We now discuss whether our above analyses yield physically meaningful results for each individual star. 0242+0426 has both a small $M_\mathrm{H}$ and a low water mass fraction that is consistent with the accretion of a relatively dry parent body, and the overall results are compatible with the assumption that the current accretion episode started $\simeq5$\,Myr ago and provided all the H detected in the envelope of the star. 

1333+3254 presents an interesting case. The \obgt\ analysis indicates that the accreted material has a water ice mass fraction of 25\,per cent, within the range predicted for trans-Neptunian objects \citep{arakawa2025} or Ceres \citep{mccord2005}. However, if all H is delivered as H$_2$O then the estimated total mass is less than the measured mass in the white dwarf CVZ (Table~\ref{tab:sinking_times}), which is the result of the very small $M_\mathrm{H}$ and the relatively O-rich material that is currently accreted. This suggests that the assumptions underlying the \obgt\ analysis are not valid for 1333+3254. We speculate that this star may have accreted a highly oxidised object, where O would be delivered by Fe$_2$O$_3$ (instead of FeO as assumed by the \obgt), or other volatiles besides $\mathrm{H_2O}$, such as CO$_2$.

In the cases of 0452$-$0214 and 1352+0323, invoking a highly-oxidised composition cannot explain away the O excess. Both stars have a high H abundance, so the simplest explanation is accretion of water-rich objects. Their water mass fractions are larger than any Solar System dwarf planet or moon, and the inferred duration of the current accretion episodes are long, about 9--10\,Myr. Whereas it cannot be ruled out that 0452$-$0214 and 1352+0323 currently accrete ``water worlds'', it is plausible that the large $M_\mathrm{H}$ are the accumulated results of multiple accretion episodes or the upwelling of H from deeper within the white dwarf.

\subsection{Alternative accretion states}
\label{sec:declining}

The Bayesian analysis presented in Section~\ref{sec:acc_state} demonstrated that ongoing accretion is the most likely scenario for all sources. We thus discussed the parent body compositions assuming steady state (Sections~\ref{sec:mdot_pbmasses}--\ref{sec:water}). Here we discuss the other accretion scenarios: \textit{increasing} or \textit{decreasing} state.


The squares in Fig.~\ref{fig:ratios_vs_Fe} represent the parent body metal abundance ratios in increasing state, assuming accretion started $1\times\tau_{\mathrm{Mg}}$ ago. The inferred element ratios in that figure (Ti/Fe, Na/Mg, Ca/Mg, Mg/Fe and Ca/Fe) differ by less than 0.13\,dex between the increasing and steady scenarios. These lie well within the aforementioned realistic uncertainties \citep[$\simeq 0.2$\,dex;][]{williams24} and therefore we find no significant difference in the composition of the accreted material between the two states, and therefore our adoption of steady state does not affect the interpretation of our results.



In contrast, as soon as accretion stops, i.e. when the system enters the decreasing state, the abundances of each metal drop exponentially on their respective diffusion time scales, causing their abundance ratios to diverge dramatically. In that scenario, our interpretation of the photospheric abundances becomes sensitive to the elapsed time since accretion stopped, and we cannot unambiguously determine the parent body abundances. 

The effect of the different diffusion time scales is illustrated in Fig.~\ref{fig:ratios_vs_Fe}. The faded-coloured lines with round markers show parent body abundances assuming that accretion stopped 1, 2 or 3\,Myr ago for the He-dominated white dwarfs, and $10^4$\,yr ago for those with H-dominated atmospheres, as they have overall shorter diffusion time scales (Table~\ref{tab:sinking_times}). 

As Fe has a diffusion time scale that is shorter than those of Mg, Ca and Ti (by factors of $\simeq0.55$, $\simeq0.75$, and $\simeq0.91$, respectively), the putative parent body compositions become increasingly Fe-rich the longer the system has been in the decreasing phase (top panels of Fig.~\ref{fig:ratios_vs_Fe}). If the systems analysed here were deep into the decreasing phase, most of their parent bodies would be extremely Fe-rich (a few exceptions are discussed in more detail below), with relative Mg/Fe and Ca/Mg abundances unlike any known Solar-system body. In addition to changes in the relative abundance ratios, the inferred parent body masses would \textit{increase} exponentially with the time since accretion stopped, pushing some of them into the regime of Pluto-mass bodies, but accretion of such massive bodies is intrinsically rare. 

Taken together, the above arguments support the results of our Bayesian model that the systems studied are not in the decreasing phase. Exceptions are 1626+3136 and 1333+3254, for which the posterior distribution has a low-probability tail that would admit a post-accretion solution (Fig.~\ref{fig:time_since_accretion}). 

The material accreting onto 1626+3136 exhibits depleted Ca and Ti abundances relative to Mg or Fe in both increasing and steady states (Fig.~\ref{fig:ratios_vs_Fe}). If the system were instead in the decreasing phase, the inferred composition would be Fe-rich and more similar to the Earth's core, while Ca/Mg and Ti/Mg would be closer to bulk Earth or stellar abundances. Given the refractory nature of Ca and Ti, such a composition may be geologically plausible.

The scenario for 1333+3254 is more complex. As already discussed in Section~\ref{sec:hydrogen} this star presents an intriguing case due to its unusual O and H balance, which in the steady state would suggest highly oxidised material or volatiles such as CO$_2$. However, if the system were instead in the decreasing phase, the derived parent body composition would be enriched in transition metals (Fe, Cr and Ti) and to a lesser extent, in Ca, Si and Mg, while the O content would potentially remain equal (\textit{assuming} the elapsed time since accretion stopped is smaller than the O sinking time, $3.65\times 10^{6}$\,yr). This scenario would then enlarge the O-binding material, and thus yield a more typical oxide balance. Ultraviolet observations would be desired to search for photospheric carbon, which would provide an additional O-carrier in the form of CO or CO$_2$ and help disentangle between the two possibilities: highly oxidised object + possible additional CO/CO$_2$, or decreasing phase that enlarges the O-binding material.

In any case, if accretion ceased $2\times\tau_{\mathrm{Mg}}$ ago for 1626+3136 and  1333+3254, both stars would have had to accrete an object of $10^{25}$\,g (approaching the size of the Moon) within the last few Myr. As already mentioned, such events are rare, though not unprecedented \citep{swan23}, but to maintain the observed abundances after differential sinking had removed most of the accreted Fe, both objects would have to resemble planetary cores. These type of accretions are scarcely observed \citep{melis11,gaensicke12,johnson22} but are becoming more common as more metal-enriched white dwarfs are being analysed \citep[see e.g.][]{williams25}.

Finally, we caution that interpretation of photospheric abundances relies on diffusion time scales, which are complex to calculate. Multiple processes in the stellar atmosphere can influence diffusion, and models remain in development and have not yet reached a consensus \citep{buchan25}. Such effects are less of a concern for He-dominated stars like 1333+3254 than for H-dominated stars, but it is nevertheless possible that future refinements of the diffusion time scales could impact our conclusions.

\section{Conclusions}
\label{sec:conclusions}

We analysed the first sample of metal-enriched white dwarfs observed by the Dark Energy Spectroscopic Instrument (DESI). From this data set, we selected 12 of the most heavily enriched white dwarfs, nine of them being new identifications. The majority of these white dwarfs display in their DESI spectra absorptions from O, Mg, Si, Ca, and Fe, some of the most common rock-forming elements in the Solar system. We acquired complementary X-shooter spectra for seven systems, enabling the detection of additional elements: Ti, Cr, Mn and Ni. The excellent agreement between the abundances derived from the two datasets demonstrates that DESI provides reliable and precise measurements of photospheric metal abundances. In eight systems, six or more metals are detected, allowing particularly strong constraints on the composition of the accreted material. 

The photospheric metal abundances reveal a diversity of parent-body compositions. For most of the systems, the material accreted is similar to that seen in primitive chondrites, unprocessed since the era of planet formation. In contrast, the most heavily polluted object in our sample, 0850+3208, exhibits a pronounced enhancement of Fe relative to the silicate-forming elements, suggesting an accreted body more similar to a planetary core. Likewise, the parent body accreted by the H-dominated white dwarf 1336$-$0337 is strongly Fe-depleted relative to the bulk Earth, with a derived composition closer to that of bulk silicate Earth. 0452$-$0214 and 1352+0323 systems show large inferred O mass fractions, which may tentatively indicate the accretion of water-bearing parent bodies, although alternative explanations cannot be excluded. 

Overall, the detection of up to ten different elements in some systems strengthen the need of follow-up observations with more sophisticated spectroscopic instruments to search for additional trace metals to help constrain the bulk compositions of the accreted planetesimals. This study establishes DESI as a powerful instrument for the discovery and characterisation of such systems, enabling the identification of metal-rich targets suitable for detailed follow-up while yielding broad parent compositions. As DESI continues to expand its white dwarf sample, it is expected to play a key role in uncovering new polluted systems and in advancing our understanding of the chemical diversity and internal structure of accreted exoplanetary material.

\section*{Acknowledgements}

This material is based upon work supported by the US Department of Energy (DOE), Office of Science, Office of High-Energy Physics, under Contract No. DE–AC02–05CH11231, and by the National Energy Research Scientific Computing Center, a DOE Office of Science User Facility under the same contract. Additional support for DESI was provided by the US National Science Foundation (NSF), Division of Astronomical Sciences under Contract No. AST-0950945 to the NSF’s National Optical-Infrared Astronomy Research Laboratory; the Science and Technology Facilities Council of the United Kingdom; the Gordon and Betty Moore Foundation; the Heising-Simons Foundation; the French Alternative Energies and Atomic Energy Commission (CEA); the National Council of Humanities, Science and Technology of Mexico (CONAHCYT); the Ministry of Science, Innovation and Universities of Spain (MICIU/AEI/10.13039/501100011033), and by the \href{https://www.desi.lbl.gov/collaborating-institutions}{DESI Member Institutions}. Any opinions, findings, and conclusions or recommendations expressed in this material are those of the author(s) and do not necessarily reflect the views of the US National Science Foundation, the US Department of Energy, or any of the listed funding agencies. 

The authors are honoured to be permitted to conduct scientific research on Iolkam Du’ag (Kitt Peak), a mountain with particular significance to the Tohono O’odham Nation. 

This project has received funding from the European Research Council (ERC) under the European Union’s Horizon 2020 research and innovation programme (Grant agreement No. 101020057 and No. 101221278). LKR is supported by NOIRLab, which is managed by the Association of Universities for Research in Astronomy (AURA) under a cooperative agreement with the U.S. National Science Foundation. D.A. also acknowledges financial support from the Spanish Ministry of Science and Innovation (MICINN) under the 2021 Ramón y Cajal program MICINN RYC2021-032609. Based on observations collected at the European Southern Observatory under ESO programmes 115.28GM.001 and 115.28GM.002.


\section*{Data Availability}

All data used in this work are publicly available. Spectroscopic data were obtained from DESI (\url{https://www.desi.lbl.gov}) and X-shooter (\url{https://www.eso.org/sci/facilities/paranal/instruments/xshooter.html}). Photometric data were retrieved from SDSS (\url{https://www.sdss.org}), Pan-STARRS (\url{https://panstarrs.stsci.edu}) and \textit{Gaia} (\url{https://www.cosmos.esa.int/gaia}). Data on metal-enriched white dwarfs and Solar System bodies were taken from the PEWDD (Polluted White Dwarf Database; \url{https://www.pollutedwhitedwarfs.com}). No new data were generated for this study.

Data used to produce the figures presented in this paper are available in \url{https://zenodo.org/records/18505429}.



\bibliographystyle{mnras}
\bibliography{bib_pau} 



\clearpage

\appendix
\section{Supplementary data}
\label{sec:appendix_phot}

The photometric analysis described in Section~\ref{sec:stellar_mod} employs archival data from SDSS, Pan-STARRS and \textit{Gaia} DR3; which is presented in Table~\ref{tab:phot_points}.

\begin{table*}
\caption{Astrometry and photometry of the 12 white dwarfs. We list the name of the star, the \textit{Gaia} DR3 parallax, and the archival photometry of each target in three different rows per star: the 
point spread function SDSS magnitudes \citep[top row;][]{SDSSpass}, the mean point spread function Pan--STARRS1 magnitudes along with their standard deviations \citep[middle row;][]{PanSTARRS2}, and the broad-band photometry of \textit{Gaia} DR3 \citep[bottom row][]{gaia-edr3}.}
\begin{tabular}{cc|ccccccr}
\hline
  & &  $u$ & $g$ & $r$ & $i$ & $z$ &     & SDSS \\
Star & $\varpi$ (mas)    &      & $g$ & $r$ & $i$ & $z$ & $y$ & PS1 \\
     &       & \mc{$G_\mathrm{BP}$}  & \mc{$G$} & \mc{$G_\mathrm{RP}$} & \textit{Gaia} \\
\hline
& & $18.463\pm0.020$ & $18.344\pm0.012$ & $18.583\pm0.013$ & $18.766\pm0.016$ & $18.967\pm0.043$ \\
0242+0426 & $4.43 \pm 0.21$ & & $18.392\pm0.017$ & $18.588\pm0.013$ & $18.806\pm0.014$ & $19.049\pm0.014$ & $19.195\pm0.018$\\
& & \mc{$18.425\pm0.019$} & \mc{$18.445\pm0.004$} & \mc{$18.444\pm0.027$}\\
\hline
& & $17.656\pm0.015$ & $17.259\pm0.011$ & $17.270\pm0.011$ & $17.315\pm0.012$ & $17.412\pm0.016$\\
0255+0237 & $14.78 \pm 0.08$ &  & $17.267\pm0.010$ & $17.287\pm0.011$ & $17.329 \pm0.011$ & $17.449 \pm 0.011$ & $17.531 \pm0.012$\\
 & & \mc{$17.313\pm0.007$} & \mc{$17.206 \pm 0.003$} & \mc{$17.018 \pm 0.015$}\\
\hline
& & $-$ & $-$ & $-$ & $-$ & $-$\\
0452$-$0214 & $7.11 \pm 0.11$ & & $17.746\pm 0.011$ & $17.907\pm 0.011$ & $18.120\pm 0.011$ & $18.350\pm 0.014$ & $18.422\pm 0.025$\\
& & \mc{$17.839\pm0.009$} & \mc{$17.812 \pm 0.003$} & \mc{$17.815\pm0.017$}\\
\hline
& & $17.080\pm0.012$ & $17.076\pm0.011$ & $17.333\pm0.011$ & $17.527\pm0.012$ & $17.797\pm0.020$\\
0850+3208 & $8.71 \pm 0.08$ & & $17.110\pm0.011$ & $17.326\pm0.011$ & $17.548\pm0.011$  & $17.775\pm0.011$ & $17.919\pm0.016$\\
 & & \mc{$17.154\pm 0.006$} & \mc{$17.182 \pm 0.003$} & \mc{$17.241\pm0.010$}\\
\hline
& & $17.646\pm0.015$ & $16.841\pm0.011$ & $16.496\pm0.011$ & $16.370\pm0.011$ & $16.334\pm0.013$\\
0922+0103 & $30.91\pm0.06$ & & $16.806\pm0.010$  & $16.495\pm0.011$  & $16.381\pm0.011$  & $16.376\pm0.010$  & $16.374\pm0.012$\\
& & \mc{$16.788\pm0.005$} & \mc{$16.499\pm0.003$} & \mc{$16.052 \pm 0.005$}\\
\hline
& & $-$ & $-$ & $-$ & $-$ & $-$\\
1252+7352 & $6.76 \pm 0.15$ & & $18.785 \pm0.012$ & $18.836 \pm0.014$ & $19.000 \pm0.013$ & $19.138 \pm0.027$ & $19.116 \pm0.028$\\
& & \mc{$18.837 \pm 0.019$} & \mc{$18.776 \pm 0.003$} & \mc{$18.729 \pm 0.037$}\\
\hline
& & $18.927 \pm 0.022$ & $18.922 \pm 0.013$ & $19.156 \pm 0.016$ & $19.360 \pm 0.020$  &  $19.558\pm0.060$ \\
1333+3254 & $3.32 \pm 0.19$ & &$18.955\pm 0.014$ & $19.183\pm 0.002$ & $19.395\pm 0.016$  & $19.602 \pm 0.025$ & $19.549\pm 0.113$ \\
& & \mc{$19.019 \pm 0.022$} & \mc{$19.016 \pm 0.003$} & \mc{$19.083 \pm 0.036$} \\
\hline
& & $-$ & $-$ & $-$ & $-$ & $-$\\
1336$-$0337 & $11.54 \pm 0.18$ & &$18.734 \pm0.011$ & $18.448 \pm0.012$ & $18.374 \pm0.012$ & $18.379 \pm0.013$ & $18.378 \pm0.024$\\
& & \mc{$18.717 \pm 0.021$} & \mc{$18.461 \pm 0.003$} & \mc{$18.049 \pm 0.021$}\\
\hline
& & $19.454\pm0.028$ & $19.169\pm0.015$ & $19.268\pm0.016$ & $19.437\pm0.022$ & $19.695\pm0.082$\\
1352+0323 & $4.22 \pm 0.24$ & & $19.137 \pm0.012$ & $19.289 \pm0.013$ & $19.467 \pm0.013$ & $19.681 \pm0.019$ & $19.656 \pm0.062$\\
& & \mc{$19.301 \pm 0.033$} & \mc{$19.187 \pm 0.004$} & \mc{$19.163 \pm 0.041$}\\
\hline
& & $19.278\pm0.029$ & $19.221\pm0.014$ & $19.462\pm0.017$ & $19.699\pm0.023$ & $20.021\pm0.096$\\
1626+3136 & $3.41 \pm 0.22$ & &$19.239\pm0.014$ & $19.462\pm0.011$ & $19.709\pm0.015$ & $19.907\pm0.050$ & $19.861\pm0.024$\\
& & \mc{$19.327\pm0.024$} & \mc{$19.350\pm0.004$} & \mc{$19.409\pm0.056$}\\
\hline
& & $-$ & $-$ & $-$ & $-$ & $-$\\
1756+3816 & $5.88 \pm 0.20$ & &$19.140 \pm 0.018$ & $19.165\pm0.015$ & $19.211\pm0.012$ & $19.358\pm0.023$ & $19.398\pm0.040$\\
& & \mc{$19.192\pm0.037$} & \mc{$19.079\pm0.004$} & \mc{$18.869 \pm 0.042$}\\
\hline
& & $23.101\pm0.421$ & $20.125\pm0.021$ & $19.137\pm0.016$ & $19.063\pm0.019$ & $19.197\pm0.058$\\
2214+0923 & $8.85 \pm 0.31$ & &$20.115\pm0.027$ & $19.171\pm0.039$ & $19.126\pm0.015$ & $19.352\pm0.047$ & $19.151\pm0.011$\\
& & \mc{$20.003\pm0.100$} & \mc{$19.361 \pm 0.005$} & \mc{$18.796 \pm 0.056$}\\
\hline
\end{tabular}
\label{tab:phot_points}
\end{table*}

\clearpage

\section{Best model fits to DESI and X-shooter spectra}
\label{sec:appendix_bestfits}

\begin{figure*}
        \includegraphics[width=0.93\hsize]{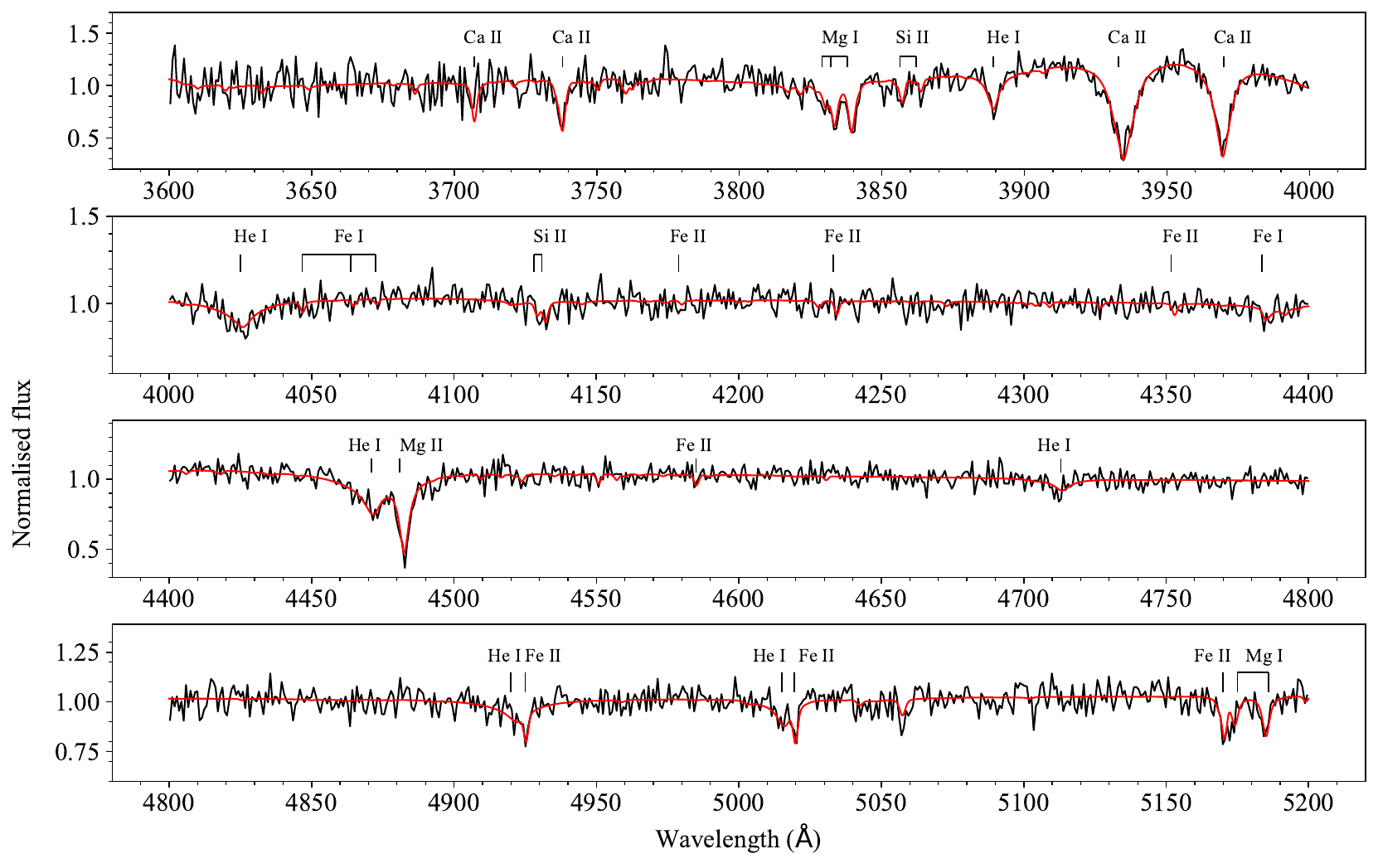}
    \caption{Bluest part of the DESI coadded spectra of 1333+3254 displaying several metal lines along with some \Ion{He}{i} transitions. The data are shown in black with best-fit model overplot in red.}
    \label{fig:1333_bestfit}
\end{figure*}

\begin{figure*}
        \includegraphics[width=0.93\hsize]{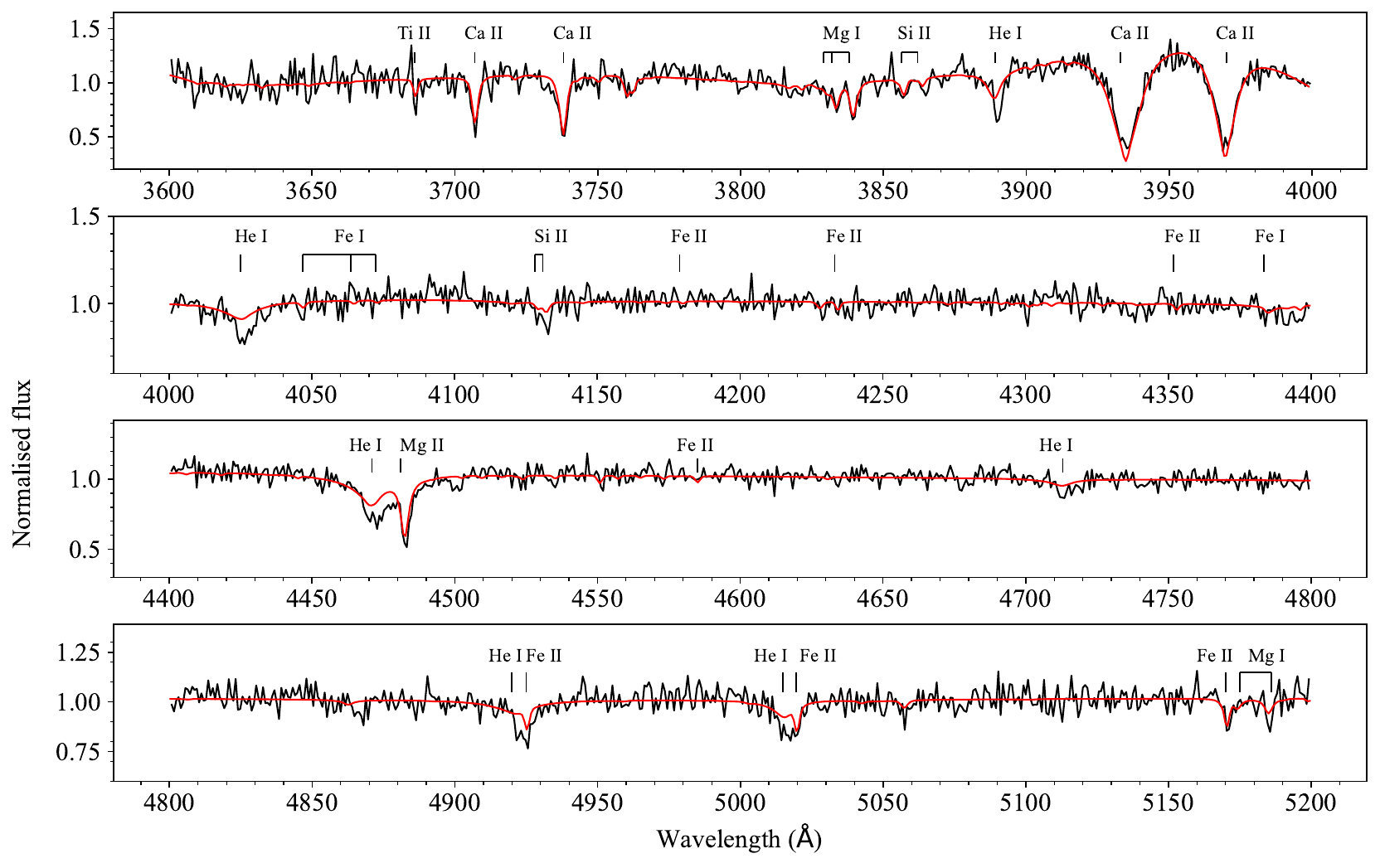}
    \caption{Same as Fig.~\ref{fig:1333_bestfit} but for 0242+0426.}
    \label{fig:0242_bestfit}
\end{figure*}

\begin{figure*}
        \includegraphics[width=0.93\hsize]{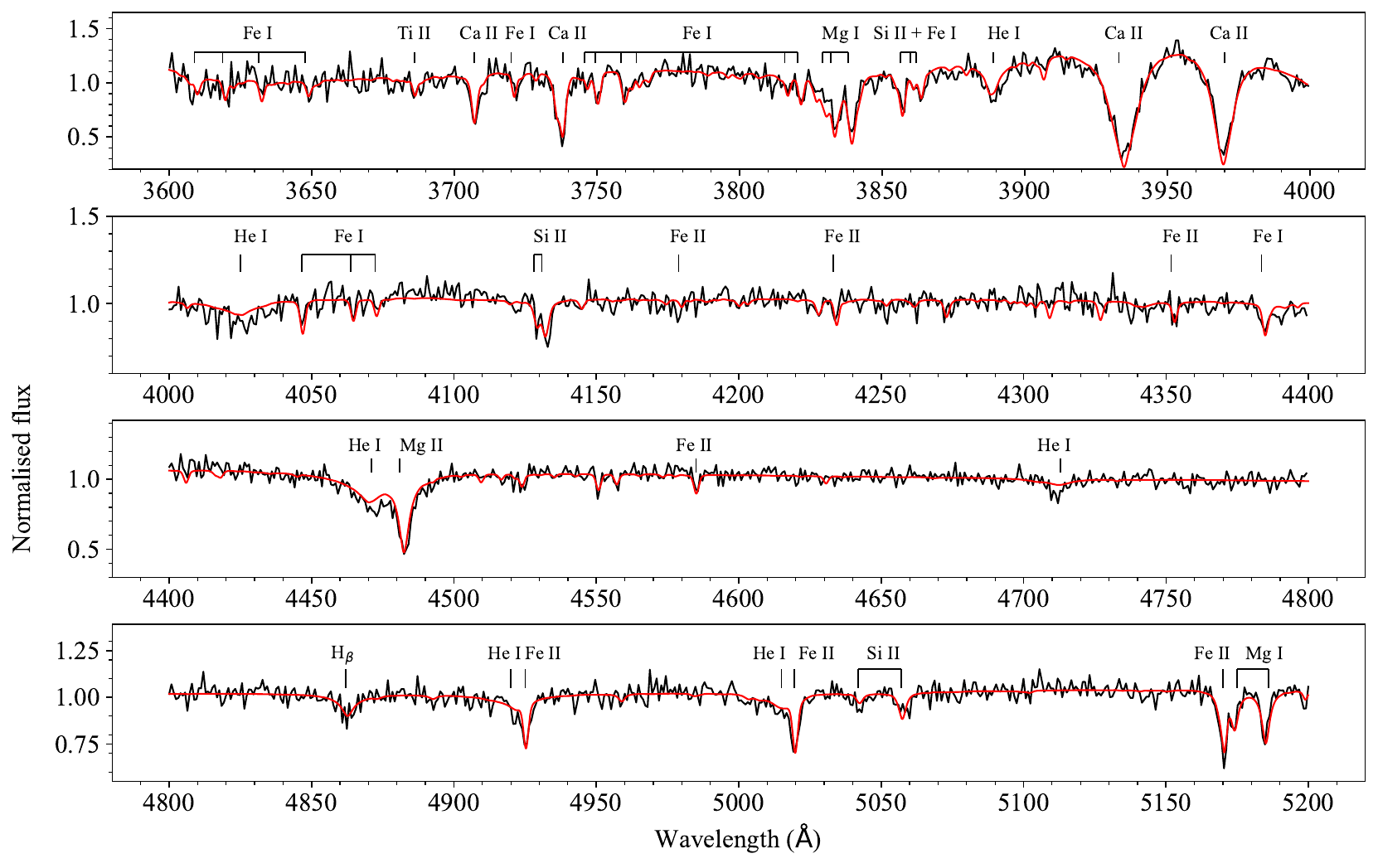}
    \caption{Same as Fig.~\ref{fig:1333_bestfit} but for 1626+3136.}
    \label{fig:1626_bestfit}
\end{figure*}

\begin{figure*}
        \includegraphics[width=0.93\hsize]{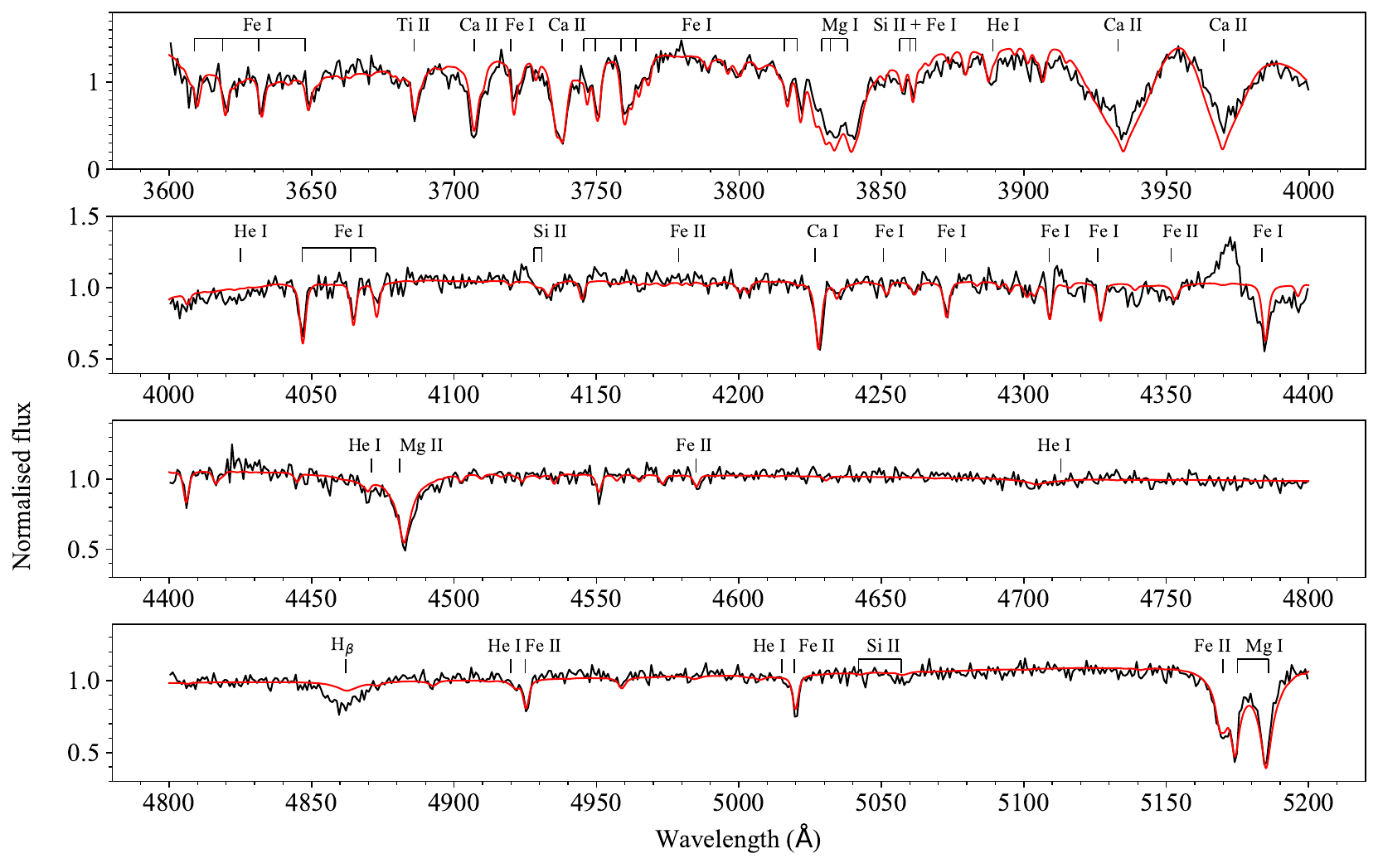}
    \caption{Same as Fig.~\ref{fig:1333_bestfit} but for 0452$-$0214.}
    \label{fig:0452_bestfit}
\end{figure*}

\begin{figure*}
        \includegraphics[width=0.93\hsize]{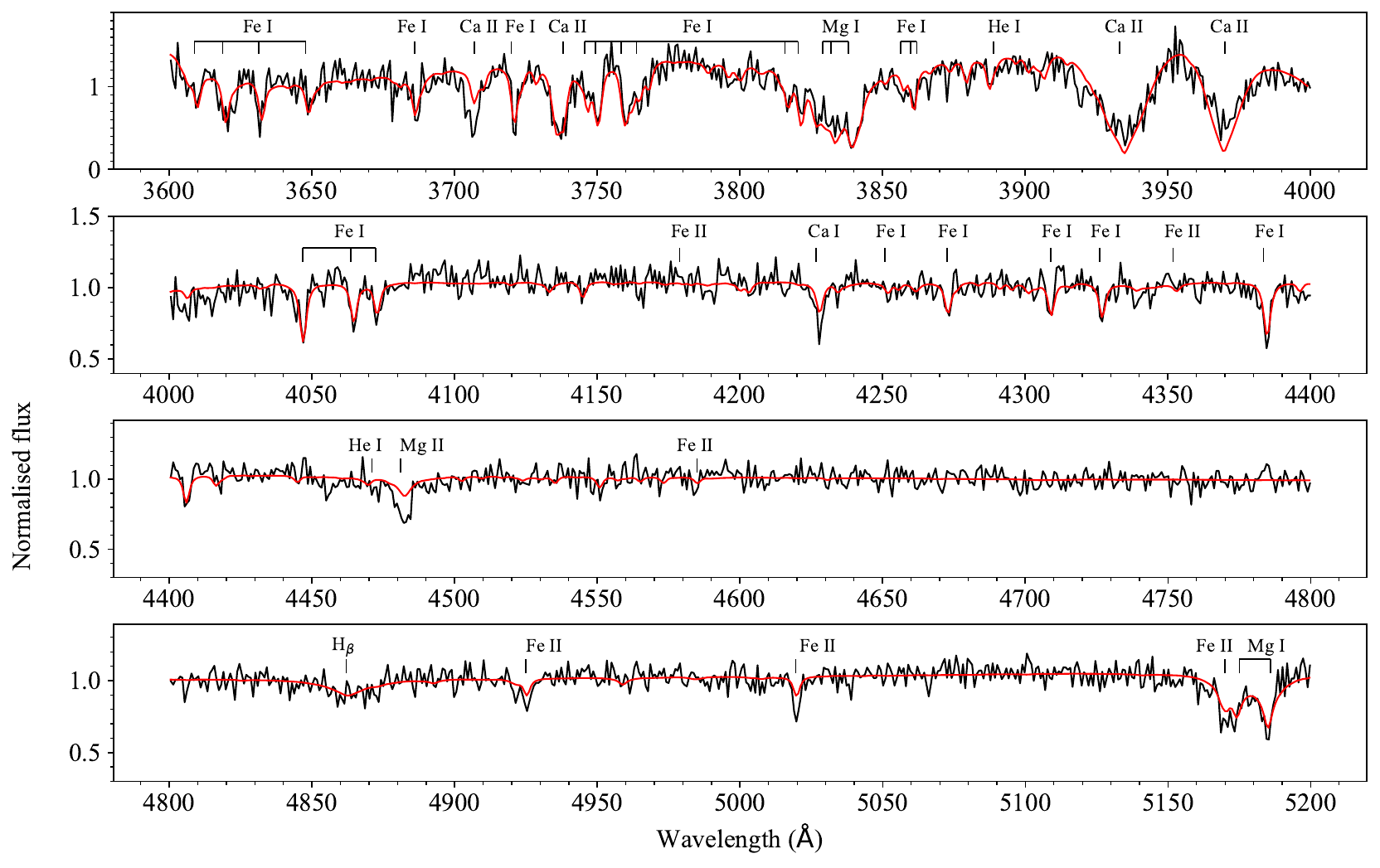}
    \caption{Same as Fig.~\ref{fig:1333_bestfit} but for 1352+0323.}
    \label{fig:1352_bestfit}
\end{figure*}

\begin{figure*}
        \includegraphics[width=0.93\hsize]{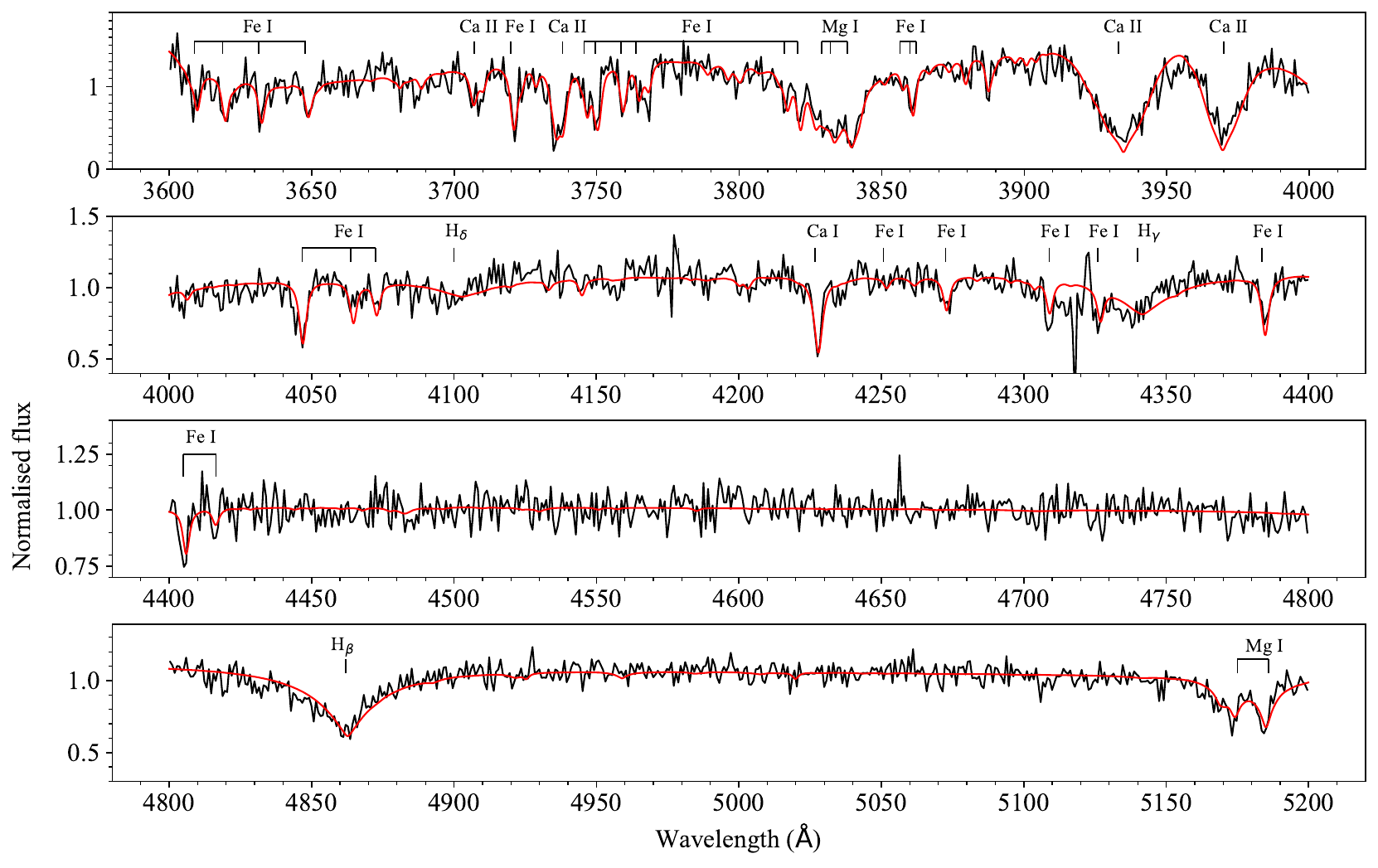}
    \caption{Same as Fig.~\ref{fig:1333_bestfit} but for 1252+7352.}
    \label{fig:1252_bestfit}
\end{figure*}

\begin{figure*}
        \includegraphics[width=0.93\hsize]{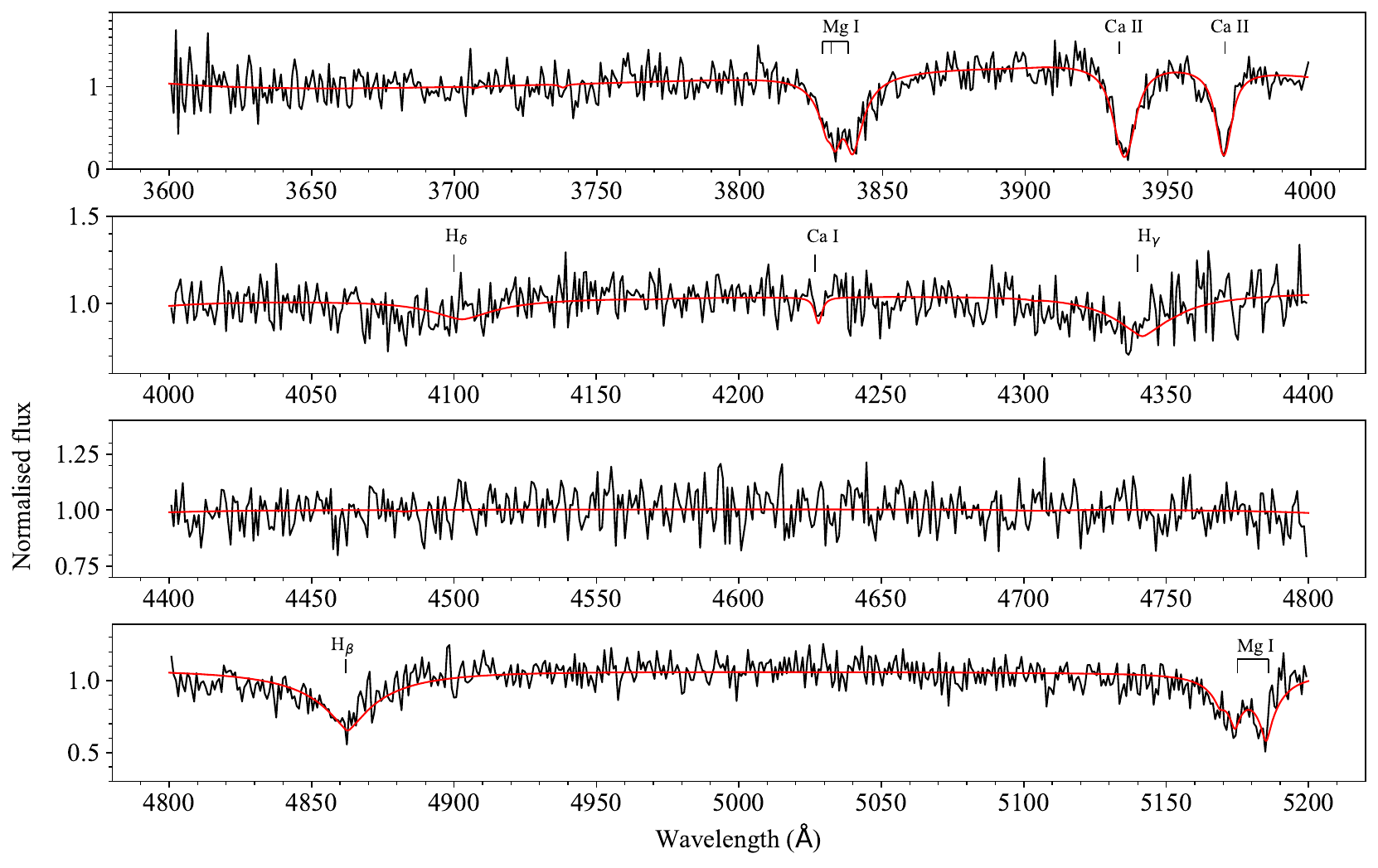}
    \caption{Same as Fig.~\ref{fig:1333_bestfit} but for 1756+3816.}
    \label{fig:1756_bestfit}
\end{figure*}

\begin{figure*}
        \includegraphics[width=0.93\hsize]{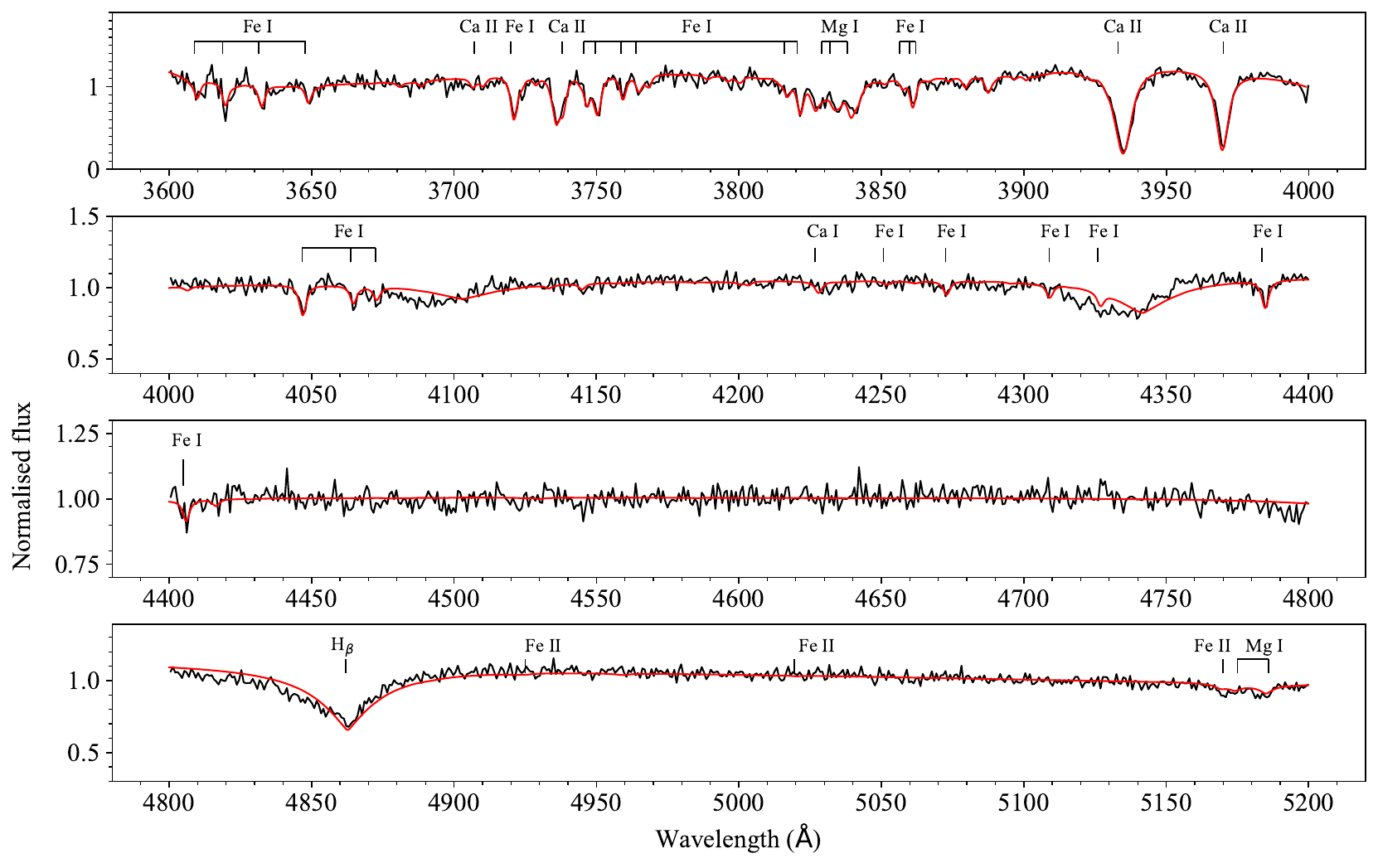}
    \caption{Same as Fig.~\ref{fig:1333_bestfit} but for 0255+0237.}
    \label{fig:0255_bestfit}
\end{figure*}

\begin{figure*}
        \includegraphics[width=0.93\hsize]{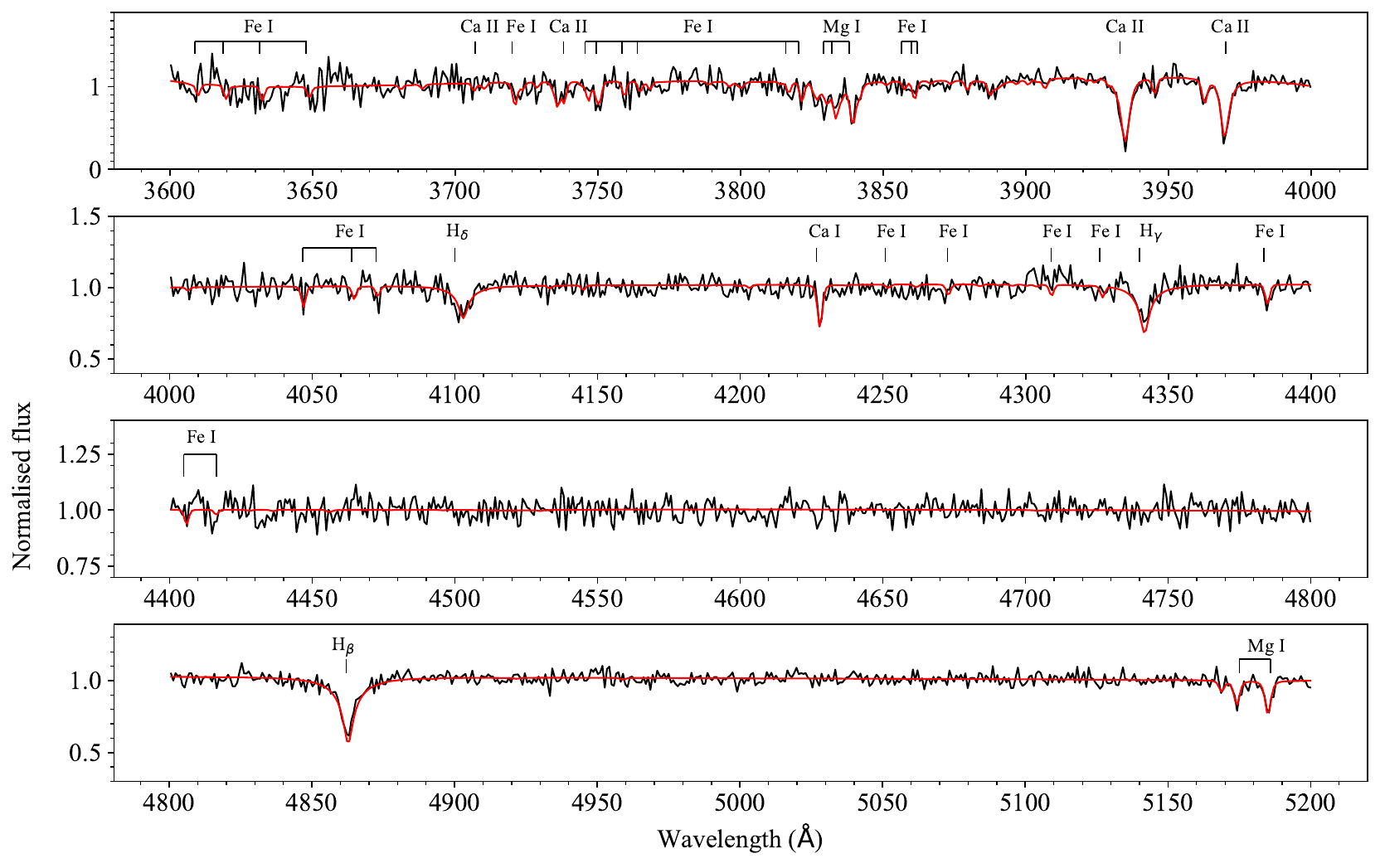}
    \caption{Same as Fig.~\ref{fig:1333_bestfit} but for 1336$-$033733. Note that this is a H-dominated white dwarf and their absorption lines are narrower from those of a warm He-dominated white dwarfs.}
    \label{fig:1336_bestfit}
\end{figure*}

\begin{figure*}
        \includegraphics[width=0.93\hsize]{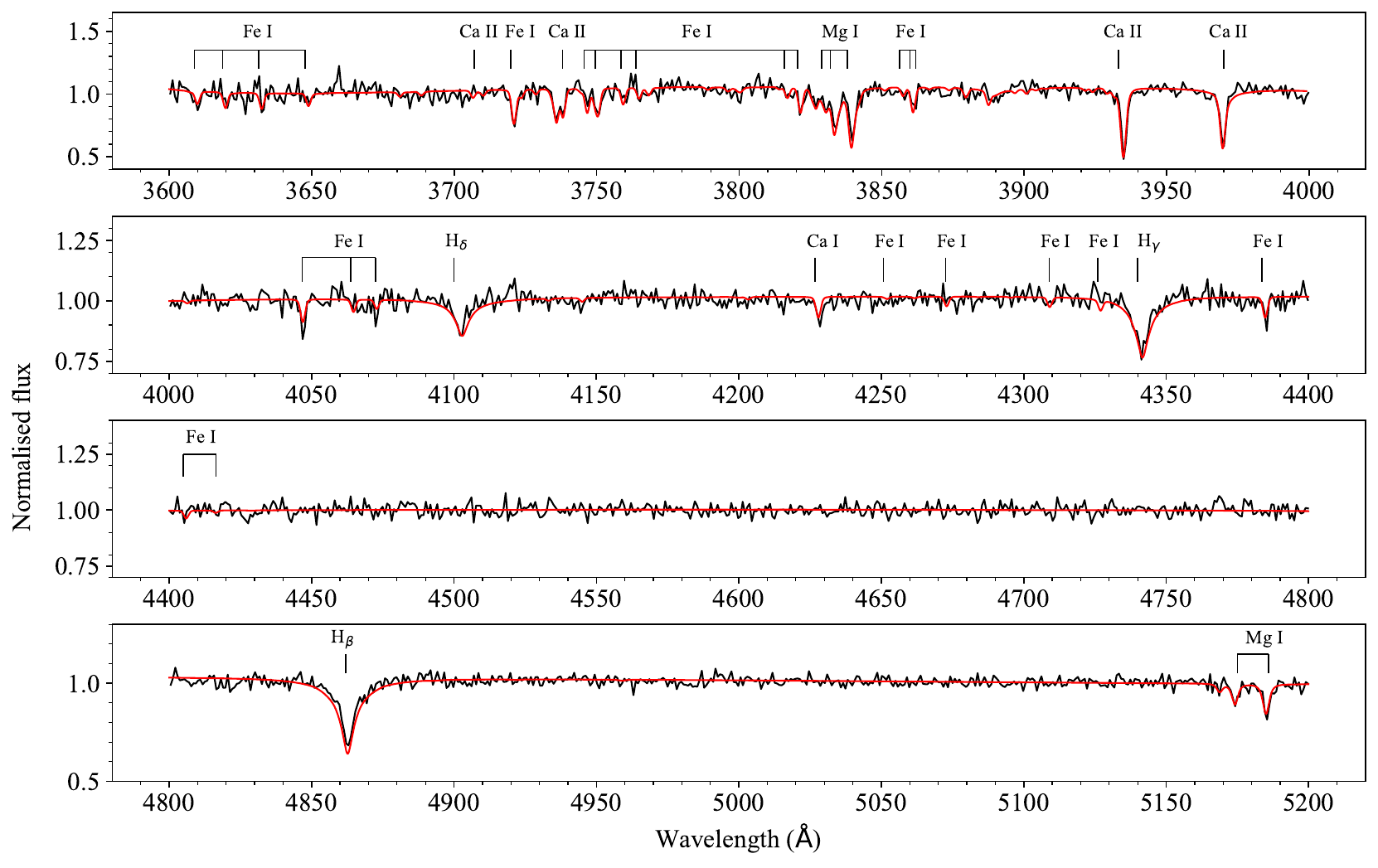}
    \caption{Same as Fig.~\ref{fig:1333_bestfit} but for 0922+0103. Note that this is H-dominated white dwarf and thus its spectral features are remarkably different from those of a warm He-dominated white dwarf.}
    \label{fig:0922_bestfit}
\end{figure*}

\begin{figure*}
        \includegraphics[width=\hsize]{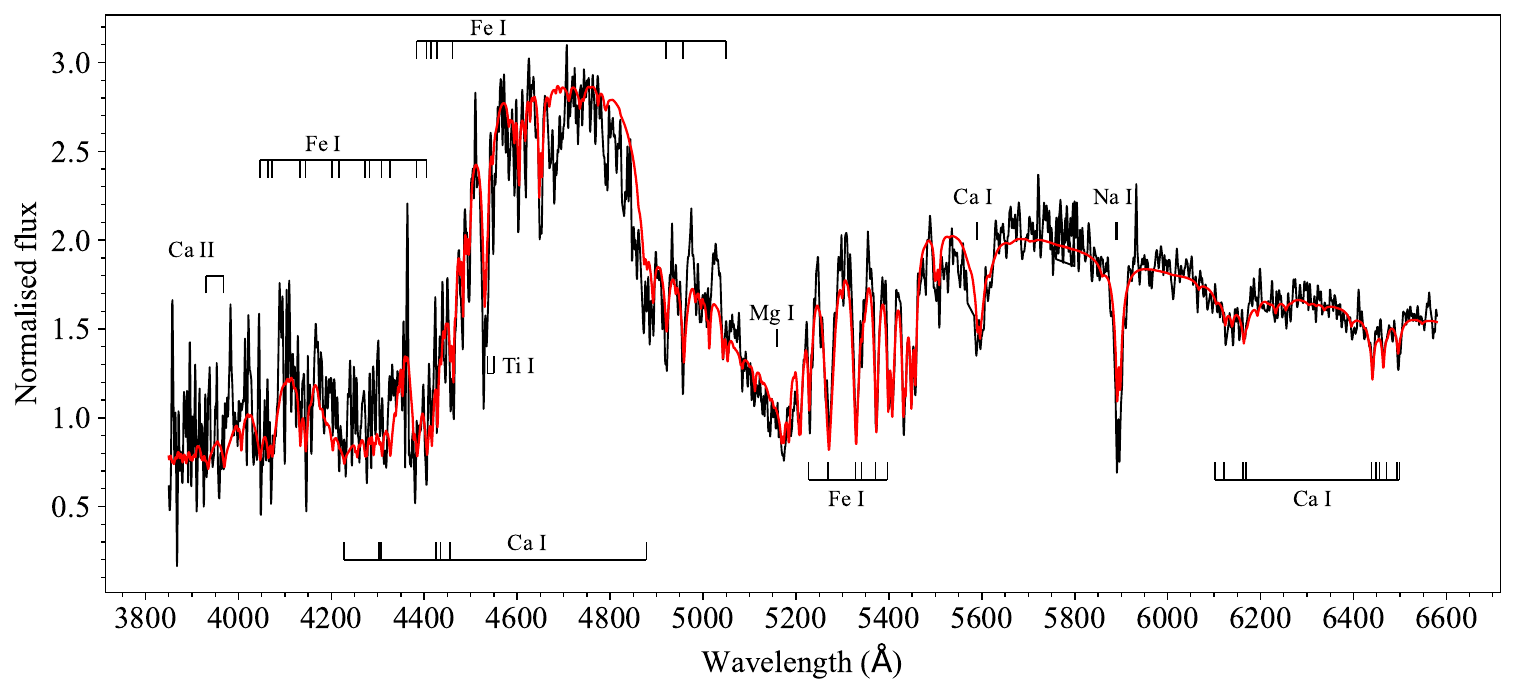}
    \caption{Same as Fig.~\ref{fig:1333_bestfit} but for 2214+0923. Note that this is cool He-dominated white dwarf, known as DZs, and their spectral features are remarkably different from those of warmer He-dominated white dwarfs.}
    \label{fig:2214_bestfit}
\end{figure*}

\begin{figure*}
        \includegraphics[width=\hsize]{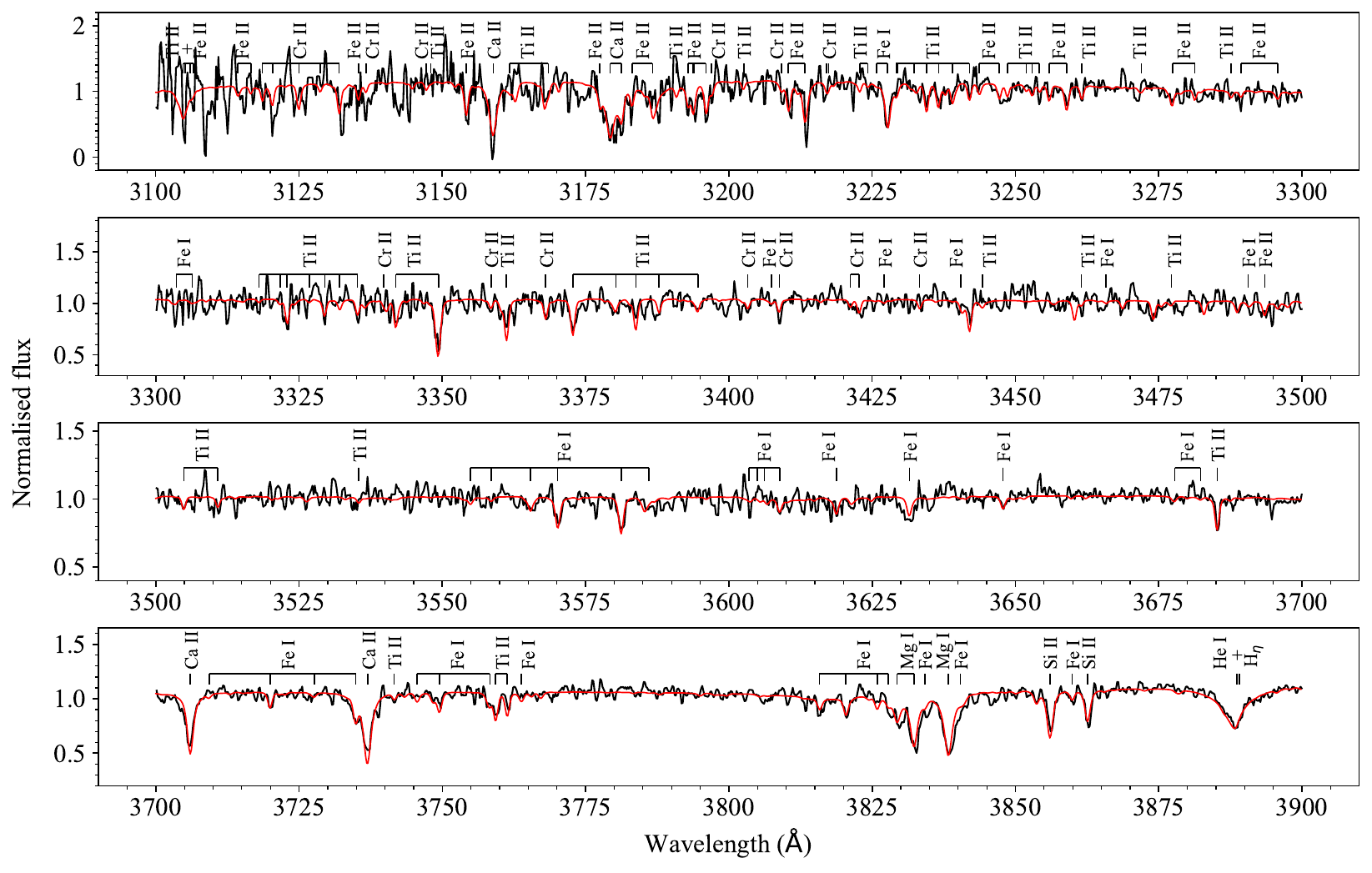}
    \caption{X-shooter spectrum of 1333+3254 in the optical region where most of the metal lines lie. The data, smoothed with a 1D Gaussian filter of standard deviation 1\,\AA, are shown in black with best-fit model overplot in red.}
    \label{fig:1333_bestfit_XS}
\end{figure*}

\begin{figure*}
        \includegraphics[width=0.95\hsize]{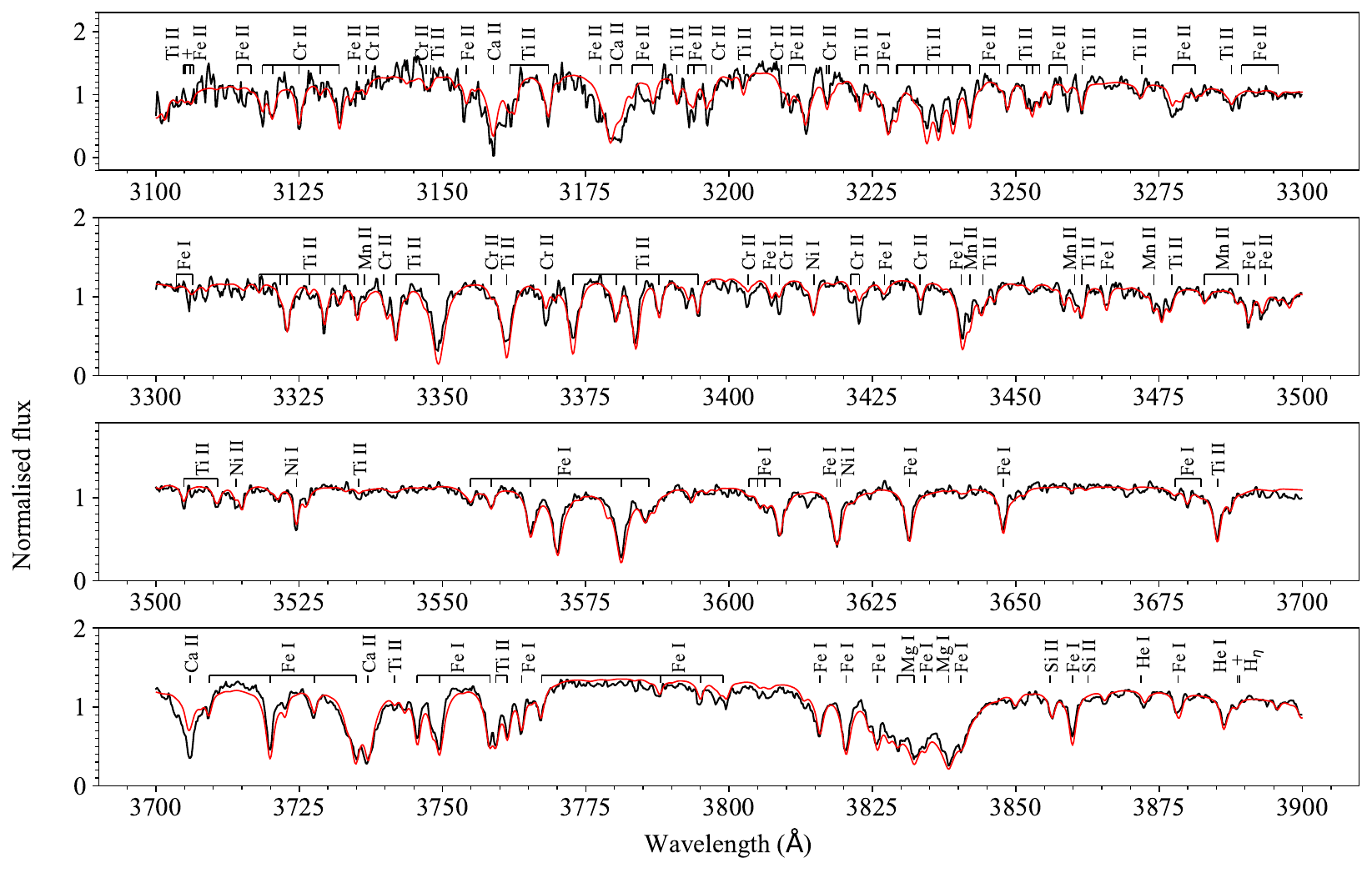}
    \caption{Same as Fig.~\ref{fig:1333_bestfit_XS} but for 1352+0323.}
    \label{fig:1352_bestfit_XS}
\end{figure*}

\begin{figure*}
        \includegraphics[width=0.95\hsize]{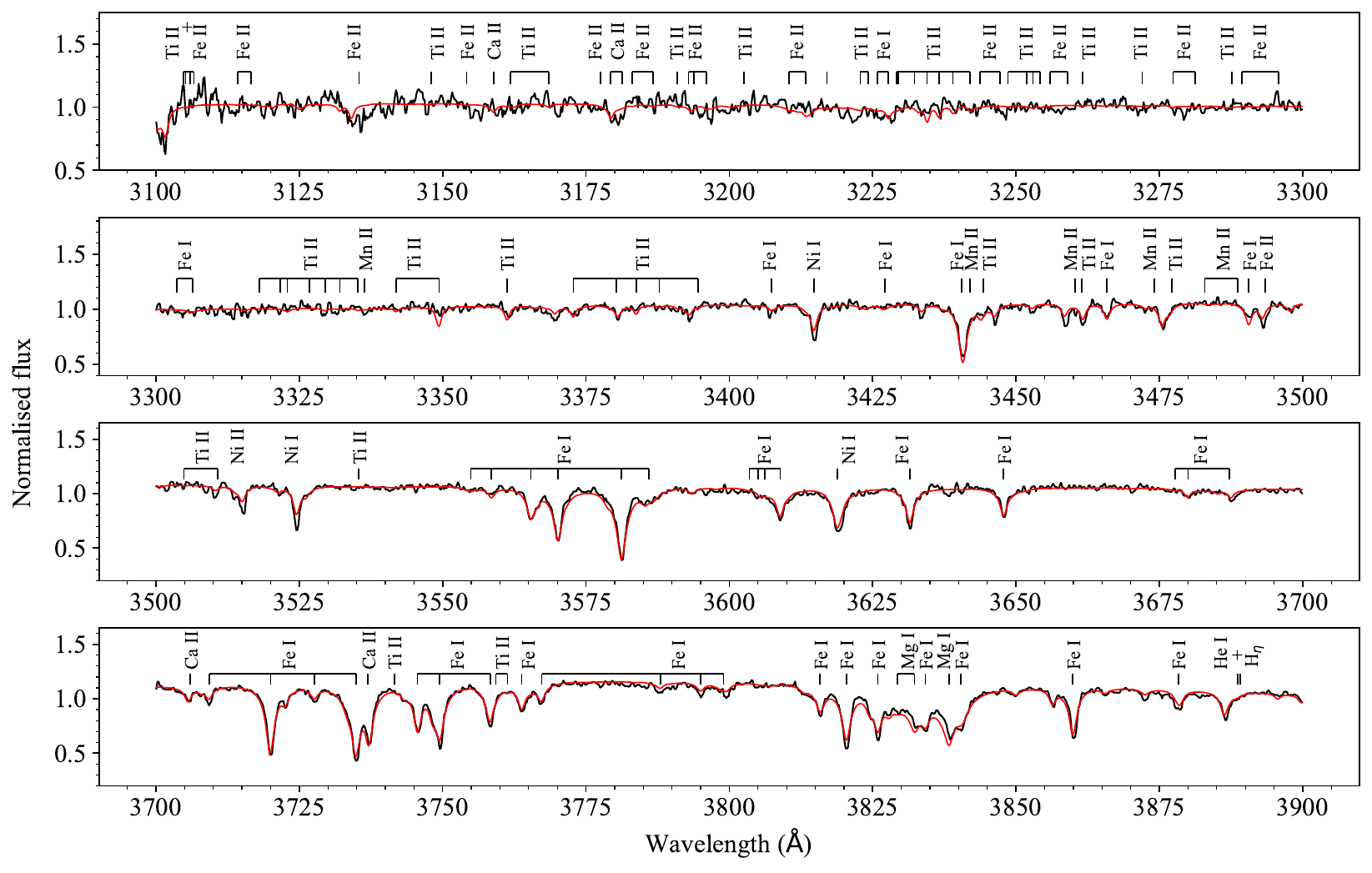}
    \caption{Same as Fig.~\ref{fig:1333_bestfit_XS} but for 0255+0237.}
    \label{fig:0257_bestfit_XS}
\end{figure*}

\begin{figure*}
        \includegraphics[width=0.9\hsize]{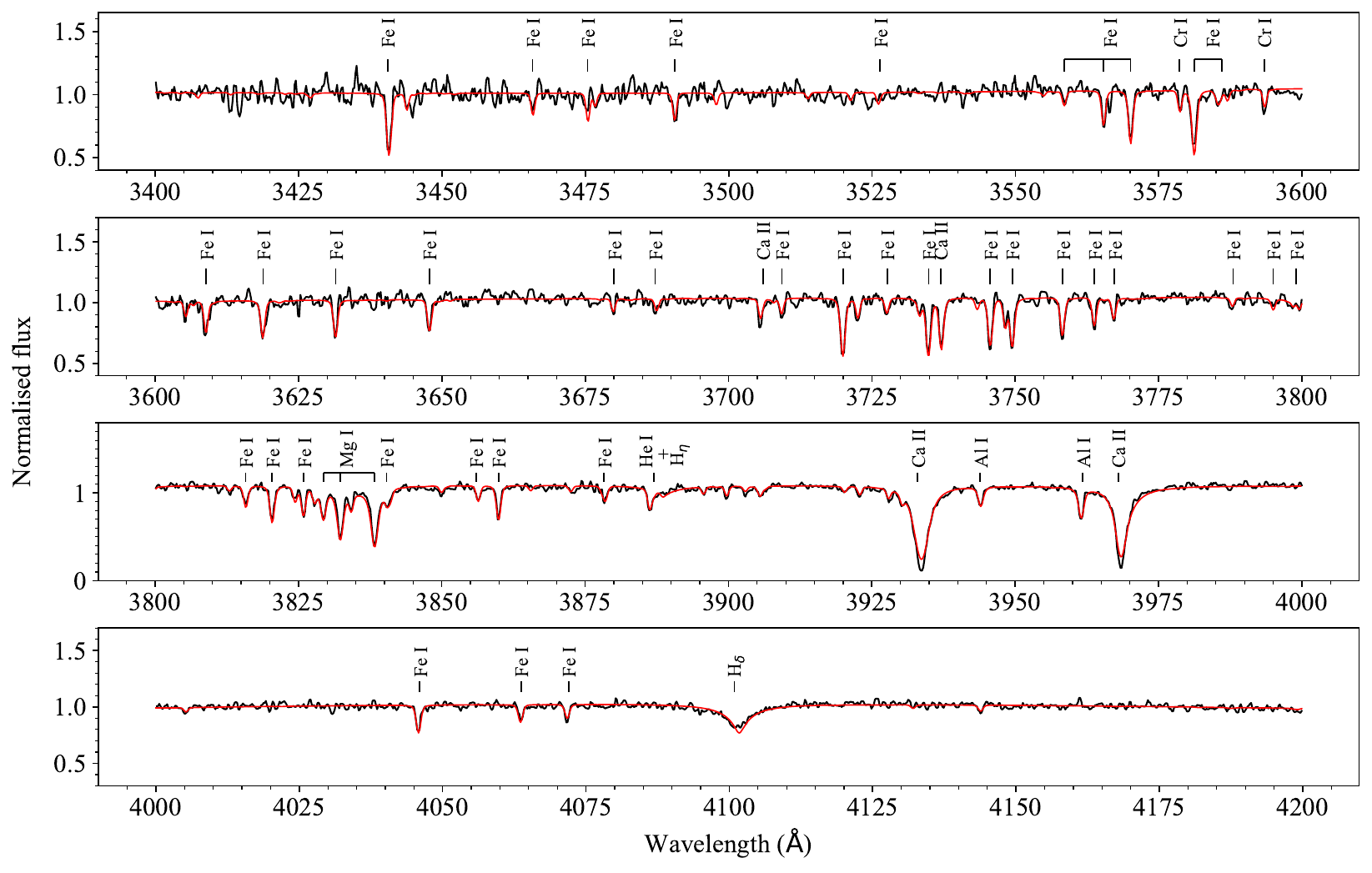}
    \caption{Same as Fig.~\ref{fig:1333_bestfit_XS} but for 1336$-$033733. Note that this is H-dominated white dwarf and thus its spectral features are sharper than those in He-dominated white dwarfs.}
    \label{fig:1336_bestfit_XS}
\end{figure*}

\begin{figure*}
        \includegraphics[width=0.9\hsize]{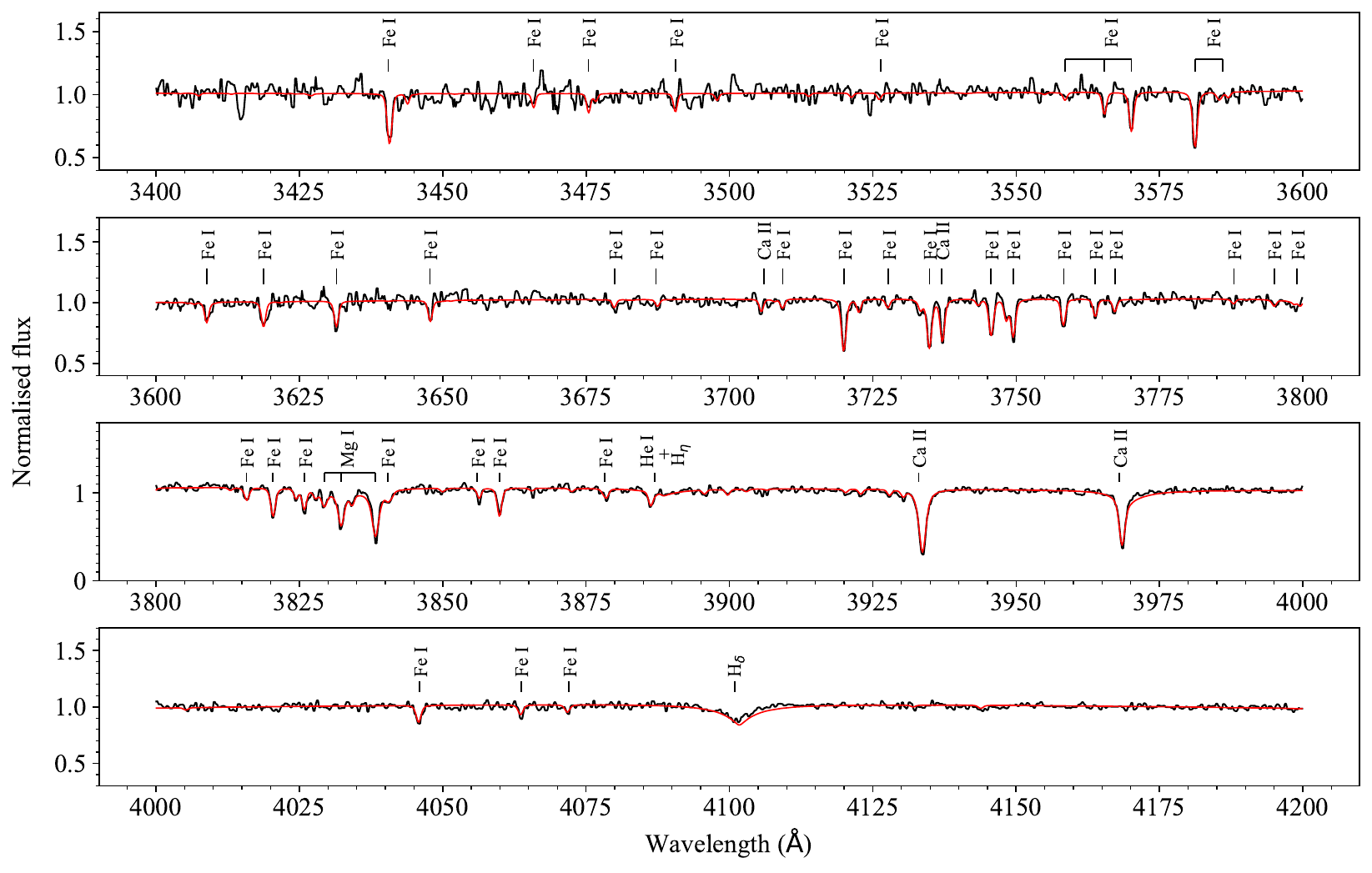}
    \caption{Same as Fig.~\ref{fig:1333_bestfit_XS} but for 0922+0103. Note that this is H-dominated white dwarf and thus its spectral features are sharper than in He-dominated white dwarfs.}
    \label{fig:0922_bestfit_XS}
\end{figure*}

\begin{figure*}
        \includegraphics[width=\hsize]{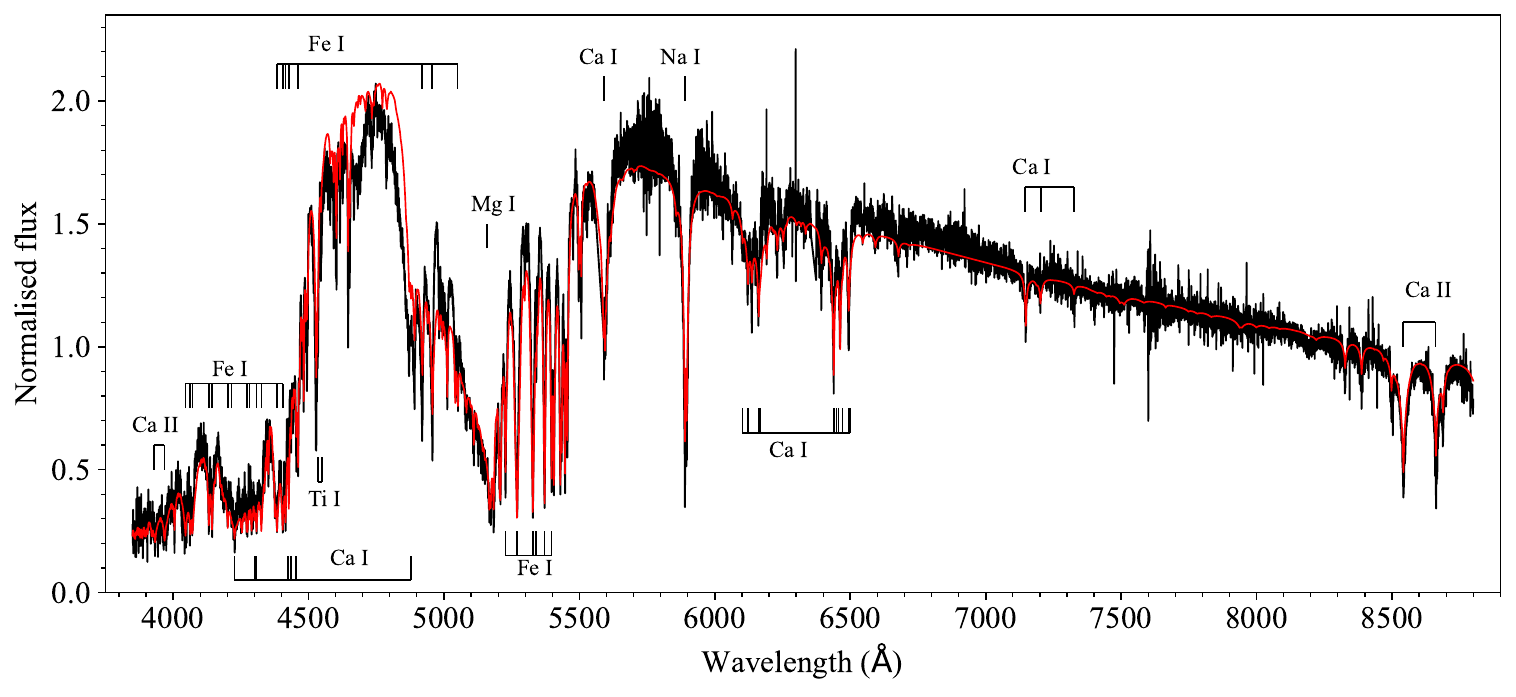}
    \caption{Same as Fig.~\ref{fig:1333_bestfit_XS} but for 2214+0923.}
    \label{fig:2214_bestfit_XS}
\end{figure*}

\begin{figure}
        \includegraphics[width=\hsize]{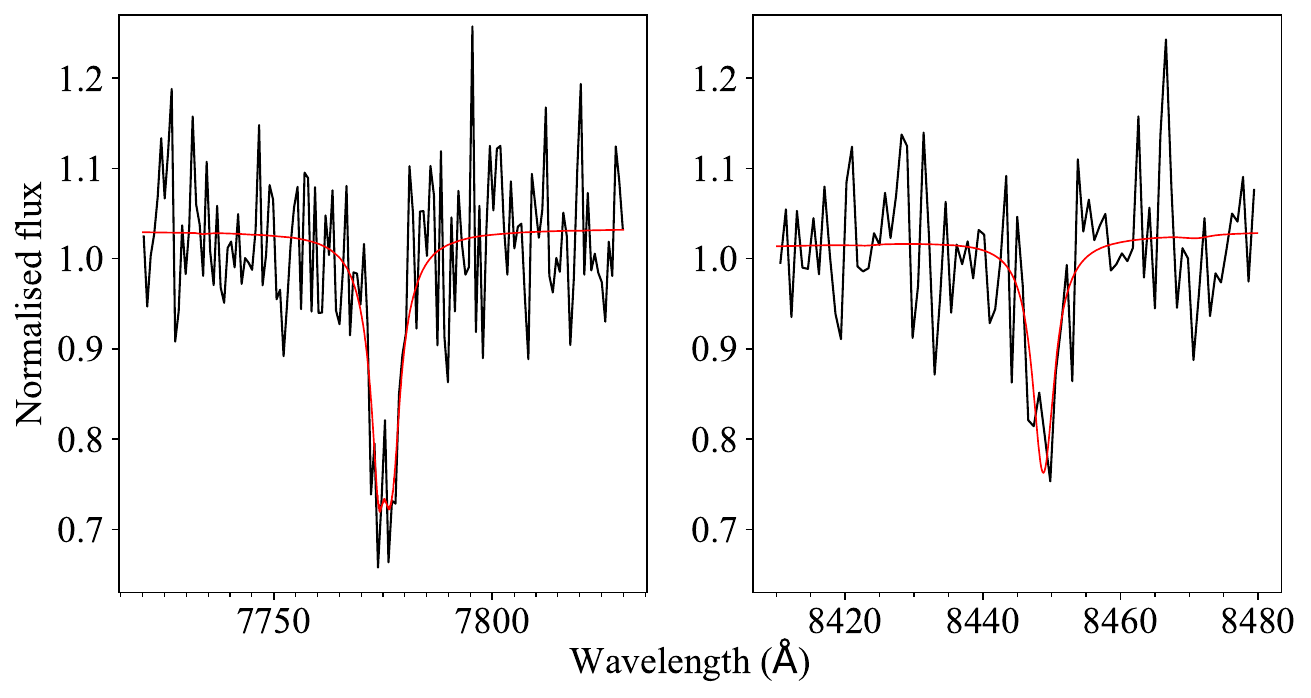}
        \includegraphics[width=\hsize]{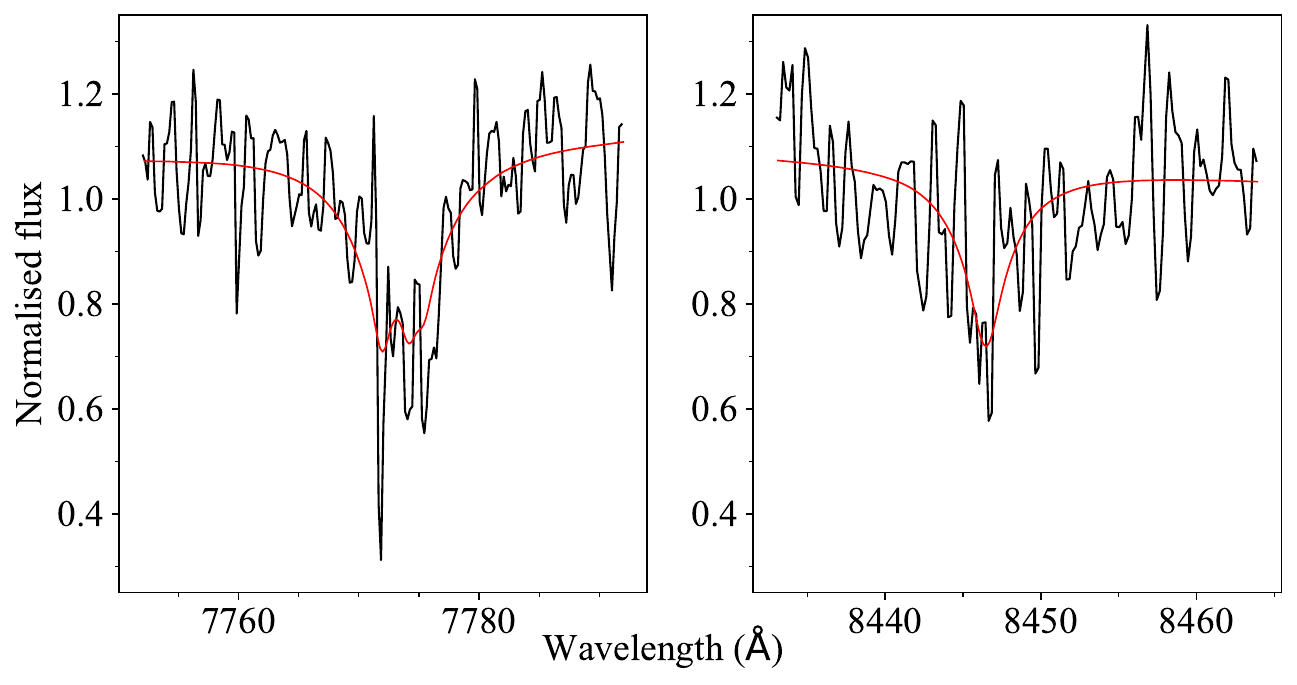}
    \caption{DESI and X-shooter data of 1333+3254 (top and bottom panel, respectively), zoomed in on the oxygen lines. The best model fit is overplot, with $\log{\mathrm{O/He}} = -4.85$.}
    \label{fig:1333_Olines_DESI}
\end{figure}

\begin{figure}
        \includegraphics[width=\hsize]{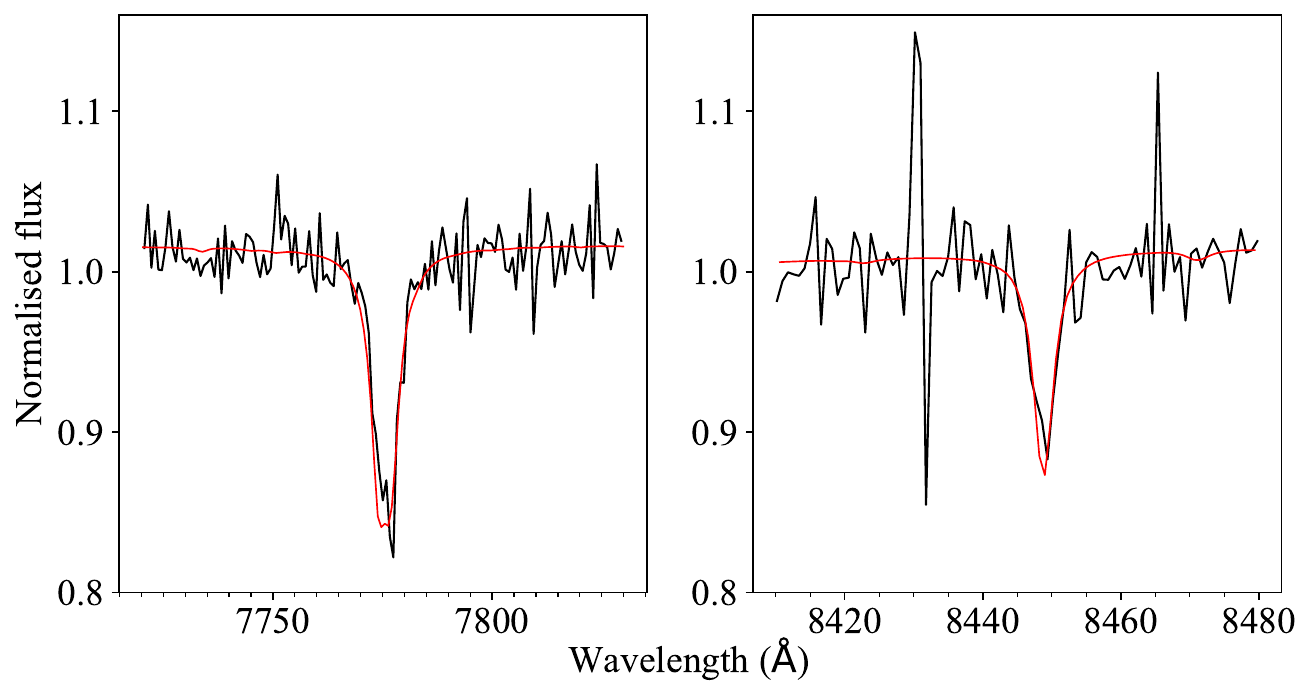}
        \includegraphics[width=\hsize]{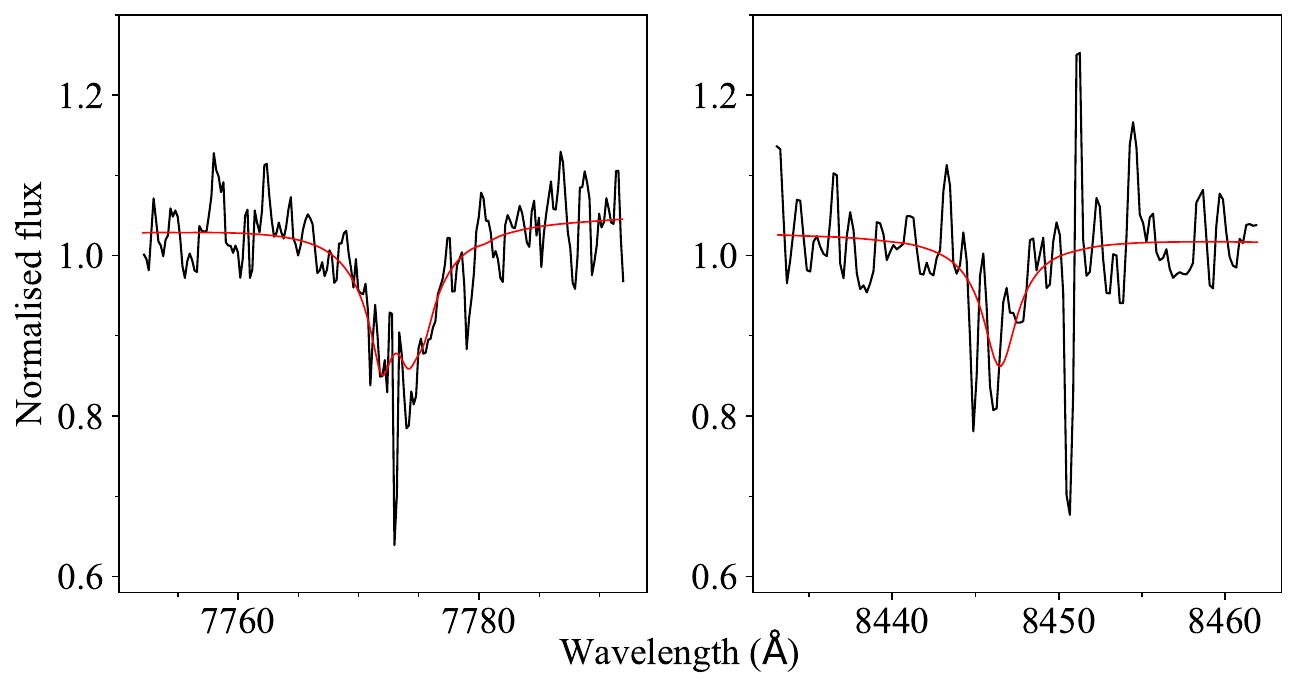}
    \caption{DESI and X-shooter data of 0850+3208 (top and bottom panel, respectively), zoomed in on the oxygen lines. The best model fit is overplot, with $\log{\mathrm{O/He}} = -5.27$.}
    \label{fig:0850_Olines_DESI}
\end{figure}

\begin{figure}
        \includegraphics[width=\hsize]{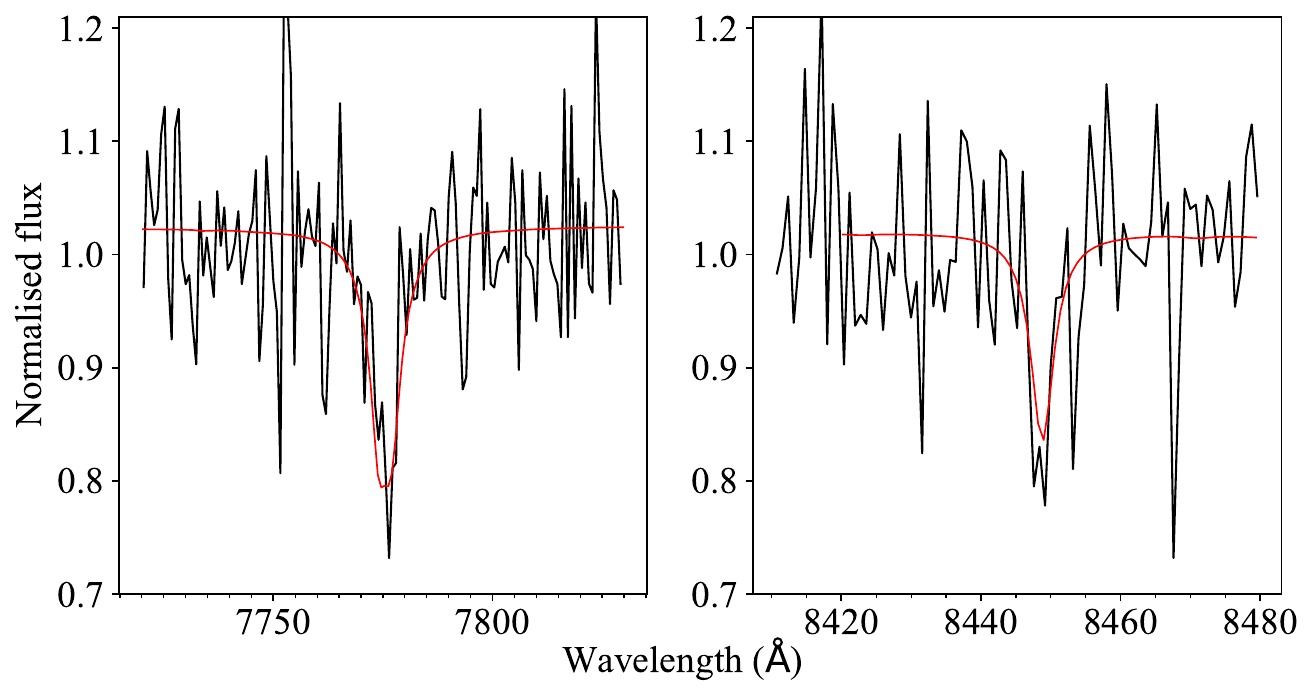}
    \caption{0242+0426 DESI data, zoomed in on the oxygen lines. The best model fit is overplot, with $\log{\mathrm{O/He}} = -5.12$.}
    \label{fig:0242_Olines_DESI}
\end{figure}

\begin{figure}
        \includegraphics[width=\hsize]{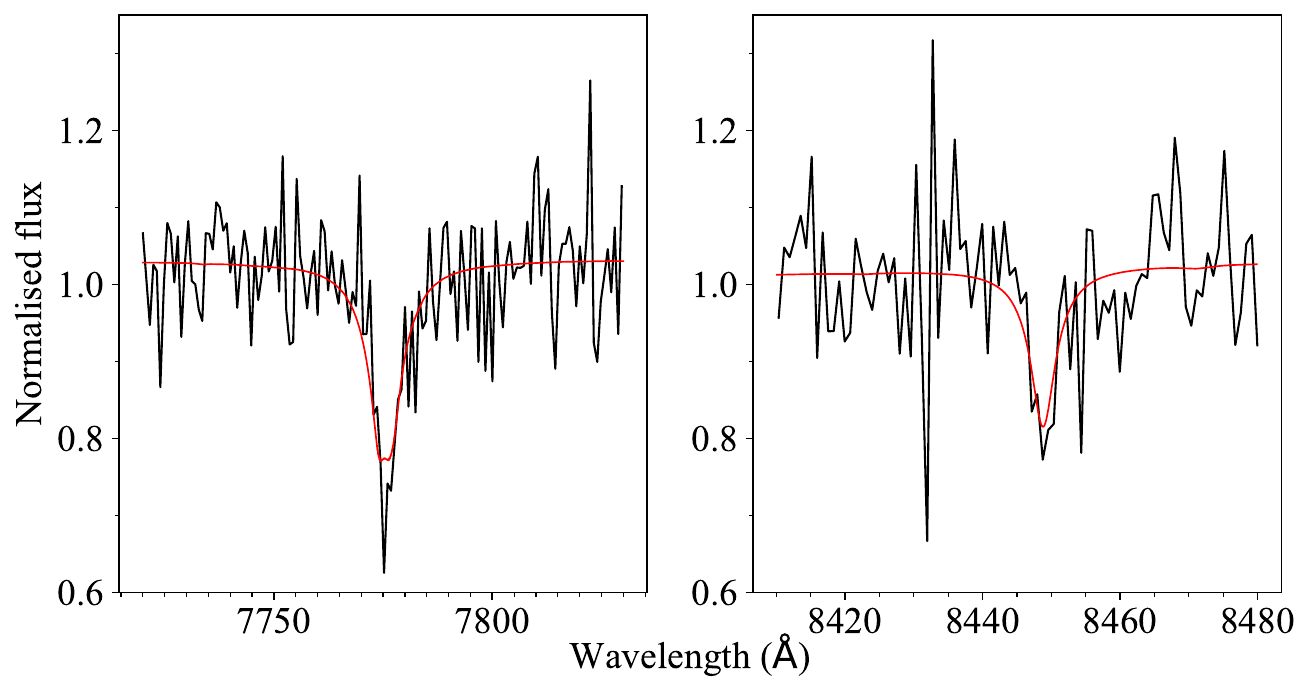}
    \caption{1626+3136 DESI data, zoomed in on the oxygen lines. The best model fit is overplot, with $\log{\mathrm{O/He}} = -4.88$.}
    \label{fig:1626_Olines_DESI}
\end{figure}

The parameters presented in Tables~\ref{tab:stellar_params} and \ref{tab:metal_abs} are derived from the best models that reproduce the available data (spectra and photometry). These best model fits overplot to the various spectra are displayed in Figs.~\ref{fig:1333_bestfit}--\ref{fig:1626_Olines_DESI}.

\clearpage
\section*{Affiliations}

$^{1}$Department of Physics, University of Warwick, Coventry CV4 7AL, UK\\
$^{2}$ Institut f\"{u}r Theoretische Physik und Astrophysik, University of Kiel, 24098 Kiel, Germany\\
$^{3}$ Department of Physics, Universit\`a degli Studi di Trieste, Via A. Valerio 2, I-34127 Trieste, Italy\\
$^{4}$ Gemini Observatory, 670 N. A’ohoku Place, Hilo, HI 96720, USA \\
$^{5}$ Instituto de Astrof\'isica de Canarias, V\'ia L\'actea, La Laguna E-38205, Spain \\
$^{6}$ Lawrence Berkeley National Laboratory, 1 Cyclotron Road, Berkeley, CA 94720, USA \\
$^{7}$ Physics Dept., Boston University, 590 Commonwealth Avenue, Boston, MA 02215, USA \\
$^{8}$ Departamento de Astrof\'isica, Universidad de La Laguna, La Laguna E-38206, Spain \\
$^{9}$ Dipartimento di Fisica “Aldo Pontremoli”, Università degli Studi di Milano, Via Celoria 16, I-20133 Milano, Italy \\
$^{10}$ INAF-Osservatorio Astronomico di Brera, Via Brera 28, 20122 Milano, Italy \\
$^{11}$ Department of Physics and Astronomy, University College London, London WC1E 6BT, UK \\
$^{12}$ Instituto de F\'{\i}sica, Universidad Nacional Aut\'{o}noma de M\'{e}xico,  Cd. de M\'{e}xico  C.P. 04510,  M\'{e}xico \\
$^{13}$ NSF NOIRLab, 950 N. Cherry Ave., Tucson, AZ 85719, USA \\
$^{14}$ Departamento de F\'isica, Universidad de los Andes, Cra. 1 No. 18A-10, Edificio Ip, CP 111711, Bogot\'a, Colombia \\
$^{15}$ Observatorio Astron\'omico, Universidad de los Andes, Cra. 1 No. 18A-10, Edificio H, CP 111711 Bogot\'a, Colombia \\
$^{16}$ Institut d'Estudis Espacials de Catalunya (IEEC), 08034 Barcelona, Spain \\
$^{17}$ Institute of Cosmology \& Gravitation, University of Portsmouth, Dennis Sciama Building, Portsmouth, PO1 3FX, UK \\
$^{18}$ Fermi National Accelerator Laboratory, PO Box 500, Batavia, IL 60510, USA \\
$^{19}$Institute for Astronomy, University of Edinburgh, Royal Observatory, Blackford Hill, Edinburgh EH9 3HJ, UK \\
$^{20}$Institute of Astronomy, University of Cambridge, Madingley Road, Cambridge CB3 0HA, UK \\
$^{21}$Kavli Institute for Cosmology, University of Cambridge, Madingley Road, Cambridge CB3 0HA, UK \\
$^{22}$ Sorbonne Universit\'{e}, CNRS/IN2P3, Laboratoire de Physique Nucl\'{e}aire et de Hautes Energies (LPNHE), FR-75005 Paris, France \\
$^{23}$ Department of Astronomy \& Astrophysics, University of Toronto, Toronto, ON M5S 3H4, Canada \\
$^{24}$ Institut de F\'{i}sica d'Altes Energies (IFAE), The Barcelona Institute of Science and Technology, Campus UAB, 08193 Bellaterra Barcelona, Spain \\
$^{25}$ Instituci\'{o} Catalana de Recerca i Estudis Avan\c{c}ats, Passeig de Llu\'{\i}s Companys, 23, 08010 Barcelona, Spain \\
$^{26}$ Department of Physics and Astronomy, Siena College, 515 Loudon Road, Loudonville, NY 12211, USA \\
$^{27}$ Department of Physics and Astronomy, University of Waterloo, 200 University Ave W, Waterloo, ON N2L 3G1, Canada \\
$^{28}$ Perimeter Institute for Theoretical Physics, 31 Caroline St. North, Waterloo, ON N2L 2Y5, Canada \\
$^{29}$ Waterloo Centre for Astrophysics, University of Waterloo, 200 University Ave W, Waterloo, ON N2L 3G1, Canada \\
$^{30}$ Instituto de Astrof\'{i}sica de Andaluc\'{i}a (CSIC), Glorieta de la Astronom\'{i}a, s/n, E-18008 Granada, Spain \\
$^{31}$ Departament de F\'isica, EEBE, Universitat Polit\`ecnica de Catalunya, c/Eduard Maristany 10, 08930 Barcelona, Spain \\
$^{32}$ Department of Physics and Astronomy, Sejong University, Seoul, 143-747, Korea \\
$^{33}$ CIEMAT, Avenida Complutense 40, E-28040 Madrid, Spain \\
$^{34}$ Department of Physics, University of Michigan, Ann Arbor, MI 48109, USA \\
$^{35}$ National Astronomical Observatories, Chinese Academy of Sciences, A20 Datun Rd., Chaoyang District, Beijing, 100012, P.~R.~China \\


\bsp	
\label{lastpage}
\end{document}